\documentclass[fleqn,usenatbib]{mnras}

\usepackage{newtxtext,newtxmath}
\usepackage[T1]{fontenc}

\DeclareRobustCommand{\VAN}[3]{#2}
\let\VANthebibliography\thebibliography
\def\thebibliography{\DeclareRobustCommand{\VAN}[3]{##3}\VANthebibliography}

\usepackage{graphicx}	% Including figure files
\usepackage{amsmath}	% Advanced maths commands
\usepackage{xspace}
\usepackage{tabularx}
\usepackage{enumitem}
\usepackage{xcolor}
\makeatletter % Workaround to only color the year for cite commands
  \patchcmd{\NAT@citex}
    {\@citea\NAT@hyper@{%
      \NAT@nmfmt{\NAT@nm}%
      \hyper@natlinkbreak{\NAT@aysep\NAT@spacechar}{\@citeb\@extra@b@citeb}%
      \NAT@date}}
    {\@citea\NAT@nmfmt{\NAT@nm}%
    \NAT@aysep\NAT@spacechar\NAT@hyper@{\NAT@date}}{}{}

  \patchcmd{\NAT@citex}
    {\@citea\NAT@hyper@{%
      \NAT@nmfmt{\NAT@nm}%
      \hyper@natlinkbreak{\NAT@spacechar\NAT@@open\if*#1*\else#1\NAT@spacechar\fi}%
        {\@citeb\@extra@b@citeb}%
      \NAT@date}}
    {\@citea\NAT@nmfmt{\NAT@nm}%
    \NAT@spacechar\NAT@@open\if*#1*\else#1\NAT@spacechar\fi\NAT@hyper@{\NAT@date}}
    {}{}
\makeatother

\newcommand{\msun}{{\,\rm M_\odot}}

\newcommand{\kms}{\,{\rm km}\,{\rm s}^{-1}}

\newcommand{\erg}{\,{\rm erg}}
\newcommand{\Gyr}{\,{\rm Gyr}}

\newcommand{\Myr}{\,{\rm Myr}}
\newcommand{\pc}{\,{\rm pc}}
\newcommand{\kpc}{\,{\rm kpc}}
\newcommand{\pkpc}{\,{\rm pkpc}}
\newcommand{\Mpc}{\,{\rm Mpc}}

\newcommand{\mmag}{\,{\rm mag}}

\newcommand{\thesan}{\textsc{thesan}\xspace}
\newcommand{\thesanandhr}{\textsc{thesan(-hr)}\xspace}
\newcommand{\thesanhr}{\textsc{thesan-hr}\xspace}
\newcommand{\thesanone}{\textsc{thesan-1}\xspace}

\def\aap{A\&A}
\def\apj{ApJ}

\def\apjl{ApJ}
\def\mnras{MNRAS}
\def\araa{ARA\&A}
\def\aj{AJ}

\def\na{New Astronomy}

\def\nat{Nature}

\def\apjs{ApJS}

\title[Galaxy sizes at high redshifts in \thesan]{The \thesan project: galaxy sizes during the epoch of reionization}

\author[]{\parbox{17.5cm}{
Xuejian Shen,$^{1}$
Mark Vogelsberger,$^{1,2}$
Josh Borrow,$^{1,3}$
Yongao Hu,$^{1}$
Evan Erickson,$^{1}$
Rahul Kannan,$^{4}$
Aaron Smith,$^{5}$
Enrico Garaldi,$^{6,7}$
Lars Hernquist,$^{8}$
Takahiro Morishita,$^{9}$
Sandro Tacchella,$^{10,11}$
Oliver Zier,$^{1}$
Guochao Sun,$^{12}$
Anna-Christina Eilers$^{1}$
and
Hui Wang$^{1}$
}
\\ \vspace{0.3cm} \\
$^{1}$ Department of Physics \& Kavli Institute for Astrophysics and Space Research, Massachusetts Institute of Technology, Cambridge, MA 02139, USA \\
$^{2}$ The NSF AI Institute for Artificial Intelligence and Fundamental Interactions, Massachusetts Institute of Technology, Cambridge, MA 02139, USA \\
$^{3}$ Department of Physics and Astronomy, University of Pennsylvania, Philadelphia, PA 19104, USA \\
$^{4}$ Department of Physics and Astronomy, York University, 4700 Keele Street, Toronto, ON M3J 1P3, Canada \\
$^{5}$ Department of Physics, The University of Texas at Dallas, SCI10 800 West Campbell Road, Richardson, Texas 75080, USA\\
$^{6}$ Institute for Fundamental Physics of the Universe, via Beirut 2, 34151 Trieste, Italy\\
$^{7}$ Max-Planck Institute for Astrophysics, Karl-Schwarzschild-Str. 1, D-85741 Garching, Germany \\
$^{8}$ Center for Astrophysics | Harvard \& Smithsonian, 60 Garden Street, Cambridge, MA 02138, USA \\
$^{9}$ IPAC, California Institute of Technology, MC 314-6, 1200 E. California Boulevard, Pasadena, CA 91125, USA \\ 
$^{10}$ Kavli Institute for Cosmology, University of Cambridge, Madingley Road, Cambridge, CB3 0HA, UK \\
$^{11}$ Cavendish Laboratory, University of Cambridge, 19 JJ Thomson Avenue, Cambridge, CB3 0HE, UK \\
$^{12}$ CIERA and Department of Physics and Astronomy, Northwestern University, 1800 Sherman Ave, Evanston, IL 60201, USA
}

\date{Accepted XXX. Received YYY; in original form ZZZ}

\pubyear{2024}

\begin{document}
\label{firstpage}
\pagerange{\pageref{firstpage}--\pageref{lastpage}}
\maketitle

\begin{abstract}
We investigate galaxy sizes at redshift $z\gtrsim 6$ with the cosmological radiation-magneto-hydrodynamic simulation suite \thesanandhr. These simulations simultaneously capture reionization of the large-scale intergalactic medium and resolved galaxy properties. The intrinsic sizes ($r^{\ast}_{1/2}$) of simulated galaxies increase moderately with stellar mass at $M_{\ast} \lesssim 10^{8}\msun$ and decrease fast at larger masses, resulting in a hump feature at $M_{\ast}\sim 10^{8}\msun$ that is insensitive to redshift. Low-mass galaxies are in the initial phase of size growth and are better described by a spherical shell model with feedback-driven outflows competing with the cold inflowing gas streams. In contrast, massive galaxies fit better with the disk formation model. They generally experience a phase of rapid compaction and gas depletion, likely driven by internal disk instability rather than external processes. We identify four compact quenched galaxies in the $(95.5\,{\rm cMpc})^{3}$ volume of \thesanone at $z\simeq 6$ and their quenching follows reaching a characteristic stellar surface density akin to the massive compact galaxies at cosmic noon. Compared to observations, we find that the median UV effective radius ($R^{\rm UV}_{\rm eff}$) of simulated galaxies is at least three times larger than the observed ones at $M_{\ast}\lesssim 10^{9}\msun$ or $M_{\rm UV}\gtrsim -20$ at $6 \lesssim z \lesssim 10$. The population of compact galaxies ($R^{\rm UV}_{\rm eff}\lesssim 300\,{\rm pc}$) galaxies at $M_{\ast}\sim 10^{8}\msun$ is missing in our simulations. This inconsistency persists across many other cosmological simulations with different galaxy formation models and demonstrates the potential of using galaxy morphology to constrain physics of galaxy formation at high redshifts.
\end{abstract}


%%%%%%%%%%%%%%%%%%%%%%%%%%%%%%%%%%%%%%%%%%%%%%%%%%

%%%%%%%%%%%%%%%%% BODY OF PAPER %%%%%%%%%%%%%%%%%%

\section{Introduction}

The size of the stellar distribution of galaxies and its evolution with cosmic time provide valuable insights about the formation history of galaxies and the relationship with their dark matter (DM) haloes~\citep[e.g.][]{Mo1998,Kravtsov2013,Somerville2018}. At low redshifts, the two main categories of galaxies (star-forming and quenched) show rather different dependencies between galaxy size and stellar mass/luminosity~\citep[e.g.][]{Shen2003,Kauffmann2003,Marijn2008,vanderWel2014}, which encode their different evolutionary and assembly histories. Galaxy sizes at lower redshifts and their complicated dependencies on various galaxy properties are well reproduced in cosmological hydrodynamic simulations~\citep[e.g.][]{Genel2018,Pillepich2019,Popping2022}.

However, at high redshifts, the classification of galaxy morphology and the measurement of galaxy sizes are challenging due to the limited angular resolution of instruments. Prior to the James Webb Space Telescope (JWST), galaxy size measurements have been mainly driven by the Hubble Space Telescope (HST) with the Advanced Camera for Surveys and the Wide Field Camera 3/IR channel on board~\citep[e.g.][]{Bruce2012,Mosleh2012,vanderWel2014,Morishita2014,Allen2017,Mosleh2020}. This has been pushed for Lyman Break galaxies (LBGs) out to $z\sim 8$ with HST legacy data~\citep[e.g.][]{Oesch2010,Ono2013,Shibuya2015-size,Bouwens2022}. These studies have found stable slopes and scatters of size--mass/luminosity relation at $0\leq z\lesssim 8$ but significantly decreasing average sizes and increased star-formation rate surface density towards high redshifts. These are in good agreement with disk formation theories of star-forming/late-type galaxies~\citep[e.g.][]{Mo1998,Dutton2007,Kravtsov2013}. At $z\sim 3$, there are no significant differences between the size of star-forming versus quiescent galaxies at the massive end~\citep[e.g.][]{Faisst2017,Hill2017,Mowla2019} when the morphological transitions are thought to present.

The early results from the Early Science Release and Cycle-1 observations have already demonstrated the remarkable capabilities of \textit{JWST}. With its infrared sensitivity and angular resolution, \textit{JWST} revealed morphologies of early galaxies at $z\gtrsim 6$ down to scales of $\sim 100\pc$ throughout rest-frame ultraviolet (UV) to optical wavelengths~\citep[e.g.][]{Yang2022,Robertson2023,Tacchella2023,Treu2023,Huertas-Company2023,Ormerod2023}. Notably, \textit{JWST} has revealed a population of extremely compact galaxies at $z\gtrsim 6$~\citep[e.g.][]{Baker2023,Baggen2023,Morishita2023}. These galaxies are more compact than the extrapolations of low-redshift measurements or theoretical model predictions. The formation channel of compact galaxies at very early stages remains unexplored. Many of the compact galaxies show signatures of heavily obscured active galatic nuclei~\citep[AGN; e.g.][]{Greene2023,Harikane2023-AGN,Labbe2023,Kocevski2023,Matthee2023,Maiolino2023,Kokorev2024} with abundance significantly exceeding the extrapolation of UV-selected AGN in the Hubble era~\citep[e.g.][]{Kulkarni2019,Niida2020,Shen2020}. They could bias the constraints on galaxy morphology~\citep[e.g.][]{Harikane2023-AGN,Tacchella2023b} but also pose interesting questions regarding the co-evolution of supermassive black holes (SMBHs) with their host galaxies.

On the other hand, theoretical predictions on galaxy sizes have been made by several numerical simulations, which have the unique power to predict spatially resolved galaxy properties. For example, \citet{Ma2018-size} studied galaxy morphology and sizes at $z>5$ using the FIRE-2 simulations~\citep{Ma2018} and focused on the low-mass end. \citet{Marshall2022} studied galaxy sizes at $7 < z < 12$ using mock images generated from the Bluetides simulations and highlighted the impact of dust. \citet{Roper2022} studied galaxy sizes at $z\geq 5$ using the FLARES simulations~\citep{Lovell2021} and explored physical mechanisms of compact galaxy formation~\citep{Roper2023}. In general, these studies found a negative correlation between intrinsic sizes of galaxies with stellar mass that is contradictory to the observed size--mass/luminosity relation~\citep[e.g.][]{Shibuya2015-size,Kawamata2018}. Concentrated dust attenuation could play a key role in reversing the trend in massive luminous galaxies and reconciling theoretical predictions with observations~\citep[e.g.][]{Marshall2022,Roper2022,Popping2022}. However, it is worth noting that most of these simulations are calibrated based on low-redshift observables and are subject to great uncertainties when extrapolating to high redshifts. For instance, the inhomogeneity of the UV radiation background during the epoch of reionization can affect the star-formation and metal enrichment of low-mass galaxies in non-trivial ways~\citep{Rosdahl2018-Sphinx,Katz2020,Borrow2022} compared to models assuming spatially uniform radiation background. The implementation and calibration of different stellar feedback mechanisms are far from converging between theoretical frameworks and become more uncertain at high redshifts~\citep[e.g.][]{Gnedin2014,Pawlik2017,Pallottini2022,Dekel2023,Ferrara2023}.

In this study, we aim to provide a thorough analysis of galaxy sizes at $z\gtrsim 6$ over a large range of galaxy stellar mass from $\sim 10^{6.5}$ to $10^{10.5}\msun$ using both the flagship \thesan simulations and its high-resolution variants. As a suite of radiation-hydrodynamic simulations, \thesan is able to self-consistently model the reionization process and the galaxies responsible for it with unprecedented physical fidelity. We will first focus on the physical mechanisms driving galaxy size evolution and understanding the causal connection between size and other galaxy properties. We will then compare results with the most recent \textit{JWST} constraints and explore implications for physics models for galaxy formation at high redshifts. The paper is organized as follows. 
In Section~\ref{sec:simulation}, we introduce the simulation suite.
In Section~\ref{sec:method}, we introduce the analysis methods, including galaxy identification, classification, post-processing, and definitions of various galaxy properties.
In Section~\ref{sec:size-mass}, we study the galaxy intrinsic size--mass relation and the correlation of size with other galaxy or environmental properties.
In Section~\ref{sec:physical-driver}, we explore the physical mechanisms that drive galaxy size evolution, especially the compaction of massive galaxies.
In Section~\ref{sec:observation}, we compare the simulation results with observations and discuss various theoretical and observational uncertainties.
In Section~\ref{sec:conclusion}, we present the conclusions of the paper.

\begin{table*}
    \addtolength{\tabcolsep}{6pt}
    \renewcommand{\arraystretch}{1.1}
    \centering
    \caption{ \textbf{Simulations of the \thesan suite. Each column corresponds to the following information:} \newline \hspace{\textwidth}
    (\textbf{1}) Name of the simulation. Simulations labeled with $\ast$ are used only for the numerical convergence study in Appendix~\ref{sec:app-numeric}.
    (\textbf{2}) $L_{\rm box}$: Side-length of the periodic simulation box. The unit is comoving Mpc (cMpc). %\newline \hspace{\textwidth}
    (\textbf{3}) $N_{\rm part}$: Number of particles (cells) in the simulation. In the initial conditions, there are an equal number of DM particles and gas cells. %\newline \hspace{\textwidth}
    (\textbf{4}) $m_{\rm dm}$: Mass of DM particles, which is conserved over time. %\newline \hspace{\textwidth}
    (\textbf{5}) $m_{\rm b}$: Mass of gas cells in the initial conditions as a reference for the baryonic mass resolution. The gas cells are (de-)refined so that the gas mass in each cell is within a factor of two of this target gas mass. Stellar particles stochastically generated out of gas cells have initial masses typically comparable to $m_{\rm b}$ and are subject to mass loss via stellar evolution~\citep{Vogelsberger2013}. %\newline \hspace{\textwidth}
    (\textbf{6}) $\epsilon$: The comoving gravitational softening length for the DM and stellar particles. This is also the minimum gravitational softening length of gas cells, which are adaptively softened. %\newline \hspace{\textwidth}
    (\textbf{7}) $\epsilon^{z=6}_{\rm phy}$: The physical gravitational softening length at $z=6$ for reference. %\newline \hspace{\textwidth}
    (\textbf{8}) $r^{\rm min}_{\rm cell}$: The minimum physical size of gas cells at the end of the simulations ($z\simeq 5.5$).
    }
    \begin{tabular}{lccccccccc}%{p{0.16\textwidth}|p{0.05\textwidth}|p{0.07\textwidth}|p{0.08\textwidth}|p{0.08\textwidth}|p{0.06\textwidth}|p{0.06\textwidth}|p{0.09\textwidth}|p{0.04\textwidth}|p{0.05\textwidth}}
        \hline
        %Simulation & $L_{\rm box}$ & $N_{\rm part}$ & $m_{\rm dm}$ & $m_{\rm baryon}$ & $\epsilon$ & $\epsilon^{z=6}_{\rm phy.}$ & $r^{\rm min}_{\rm cell}$ \\
        %Name & $[{\rm cMpc}]$ &  & $[\msun]$ & $[\msun]$ & $[{\rm ckpc}]$ & $[{\rm pkpc}]$ & $[{\rm pkpc}]$ \\
        Simulation Name & $L_{\rm box}\,[{\rm cMpc}]$ & $N_{\rm part}$ & $m_{\rm dm}\,[\msun]$ & $m_{\rm b}\,[\msun]$ & $\epsilon\,[{\rm ckpc}]$ & $\epsilon^{z=6}_{\rm phy}\,[{\rm pc}]$ & $r^{\rm min}_{\rm cell}\,[{\rm pc}]$ \\
        \hline
        \hline
        \thesanone & 95.5 & $2\times 2100^{3}$ & $3.12\times10^{6}$ & $5.82\times10^{5}$ & $2.2$ & $310$ & $10$ \\
        %\thesantwo$^{\ast}$ & 95.5 & $2\times 1050^{3}$ & $2.49\times 10^{7}$ & $4.66\times 10^{6}$ & $4.1$ & $590$ & $35$\\
        \thesanhr (\textsc{L8-N512}) & 11.8 & $2\times 512^{3}$ & $4.82\times 10^{5}$ & $9.04\times 10^{4}$ & $0.85$ & $120$ & $15$\\
        %(\textsc{L8-N512}) & \\
        \thesanhr (\textsc{L4-N512}$^{\ast}$) & 5.9 & $2\times 512^{3}$ & $6.03\times 10^{4}$ & $1.13\times 10^{4}$ & $0.425$ & $60$ & $8$ \\
        \hline
    \end{tabular}
    \label{tab:sims}
    \renewcommand{\arraystretch}{0.9090909090909090909}
\end{table*}
    
%\citet{Ferreira2022a,Ferreira2022b,Robertson2023,Costantin2022,Dimauro2022,Mowla2022,Lapiner2023,Roberts-Borsani2022,Ono2023,Ormerod2023,Morishita2023,Ito2023,Ward2023}

\section{Simulations}
\label{sec:simulation}

The \thesan project \citep{Kannan2022,Garaldi2022,Smith2022} is a suite of radiation-magneto-hydrodynamic simulations utilizing the moving-mesh hydrodynamics code {\sc AREPO}~\citep{Springel2010, Weinberger2020}. Gravity is solved using the hybrid Tree--PM method \citep{Barnes1986}. The hydrodynamics is solved using the quasi-Lagrangian Godunov method \citep{Godunov1959} on an unstructured Voronoi mesh grid \citep[see][for a review]{Vogelsberger2020NatR}. For self-consistent treatment of ionizing radiation, the \thesan project employs the radiative transfer (RT) extension {\sc Arepo-rt} \citep{Kannan2019}, which solves the first two moments of the RT equation assuming the M1 closure relation \citep{Levermore1984}. The simulation includes the sourcing (from stars and AGNs) and propagation of ionizing photons (in three energy bins relevant for hydrogen and helium photoionization between energy intervals of $[13.6, 24.6, 54.4, \infty)\,{\rm eV}$) as well as a non-equilibrium thermochemistry solver to model the coupling of radiation fields to gas. The luminosity and spectral energy density of stars in \thesan as a complex function of age and metallicity are calculated using the Binary Population and Spectral Synthesis models (BPASS v2.2.1; \citealt{Eldridge2017}). The sub-grid escape fraction of stars was set to be $f_{\rm esc} = 0.37$ in the simulations. This parameter mimics the absorption of Lyman continuum photons happening below the grid scale of the simulation. This parameter was tuned such that the simulated
reionization histories approximately match the observed neutral
fraction evolution in the Universe. For additional details of the simulation methods, we refer to \citet{Kannan2019, Kannan2022}.

In terms of the galaxy formation model, the simulations employ the IllustrisTNG model~\citep{Weinberger2017,Pillepich2018}, which is an update of the Illustris model \citep{Vogelsberger2014,Vogelsberger2014b}. The simulations include (1) density-, temperature-, metallicity- and redshift-dependent cooling of metal-enriched gas~\citep{Smith2008,Wiersma2009}, (2) a two-phase, effective equation of state model for the interstellar medium (ISM) at the sub-resolution level~\citep{Springel2003}, (3) star-formation in dense gas following the empirically defined Kennicutt–Schmidt relation, (4) thermal and mechanical feedback from supernovae and stellar winds, (5) metal enrichment from stellar evolution and supernovae, and (6) SMBH formation, growth, and feedback in the quasar and radio modes as described in \citet{Weinberger2017}. The model has been extensively tested in large-scale simulations and can produce realistic galaxies that match a wide range of observations, in terms of large-scale galaxy clustering~\citep[e.g.][]{Springel2018}, galaxy properties~\citep[e.g.][]{Nelson2018, Genel2018, Pillepich2018b, Naiman2018}, and specifically at high-redshifts~\citep[e.g.][]{Vogelsberger2020, Shen2020, Shen2022, Kannan2023}. 

The initial conditions of the simulations are generated using the \textsc{Gadget4} code~\citep{Springel2021} using the second-order Lagrangian perturbation theory at the initial redshift of $z_{\rm i}=49$. The simulations employ the \citet{Planck2016} cosmology with $h = 0.6774$, $\Omega_{\rm 0} = 0.3089$, and $\Omega_b = 0.0486$, $\sigma_{8} = 0.8159$, $n_{\rm s} = 0.9667$. We will follow these cosmological parameter choices throughout this paper. In the initial conditions, the gas perfectly follows the DM distribution and is assumed to have primordial composition with hydrogen and helium mass fractions of $X=0.76$ and $Y=0.24$, respectively.

The simulations studied in this paper and associated parameters are summarized in Table~\ref{tab:sims} and all of them are made publically available~\citep{Garaldi2023}. Among them, \thesanone is the flagship simulation of the suite with the largest number of resolution elements in a $(95.5\,{\rm cMpc})^{3}$ volume periodic cubic patch of the universe. \thesanhr~\citep{Borrow2022,Shen2024-THR} is a subset of high-resolution small-volume simulations using the same numerical setup and physics inputs as the main \thesan suite, aiming to explore the formation and evolution of low-mass galaxies in the early Universe. The mass resolution of the L8-N512 (L4-N512) run is about $6.4$ ($51.5$) times better than the flagship \thesanone simulation and allows atomic cooling haloes ($M_{\rm halo} \sim 10^{8}\msun$; e.g. \citealt{Bromm2011,Wise2014}) to be properly resolved with $\gtrsim 200$ DM particles, though many important physics mechanisms for haloes below this mass scale have not been included (e.g. explicit treatments of the formation and destruction of molecular hydrogen, Lyman-Werner radiation, and Population-III stars).

\section{Analysis methods}
\label{sec:method}

\begin{figure*}
\raggedright
\includegraphics[width=0.33\linewidth, clip, trim={0.5cm 0 0.5cm 0}]{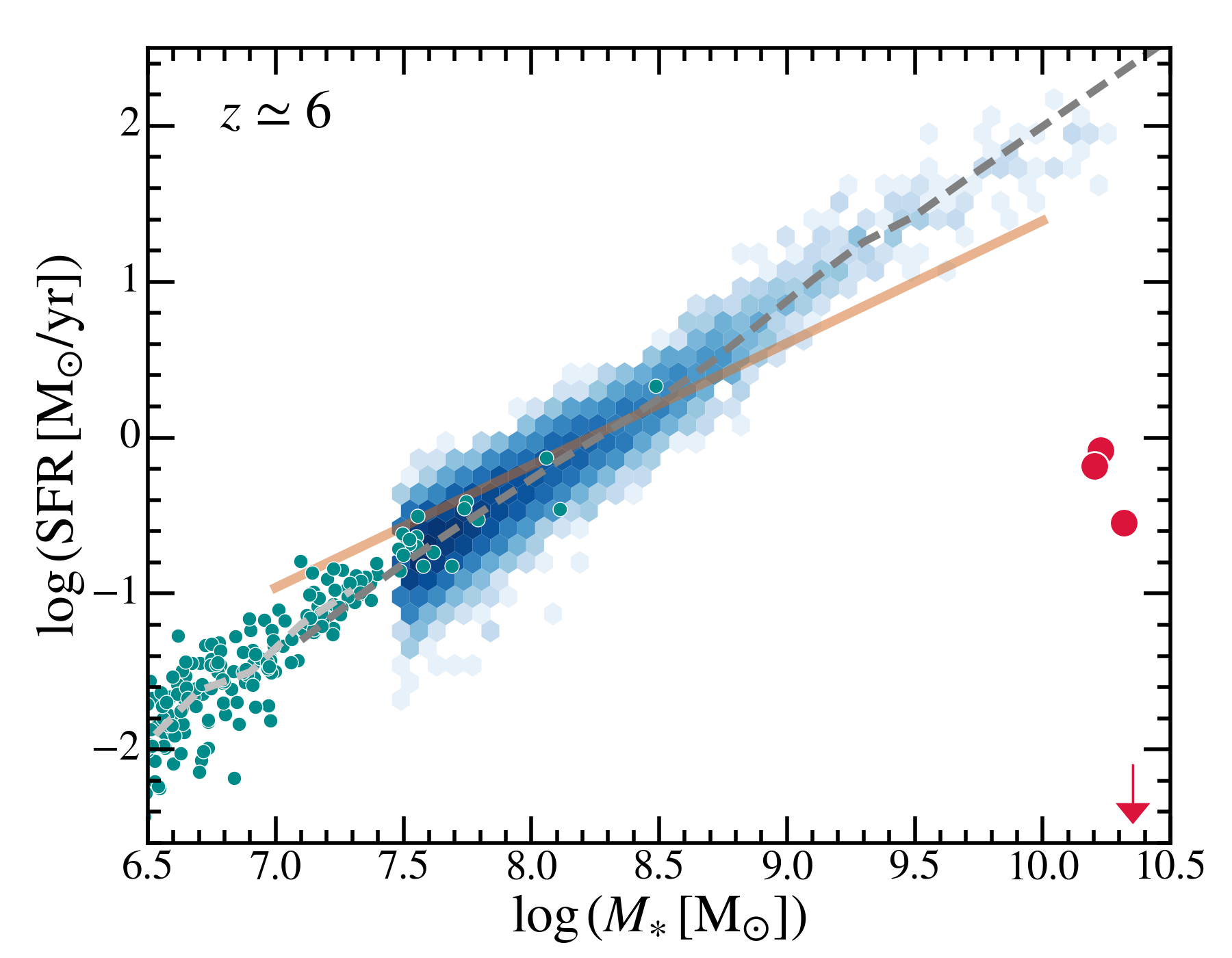}
\includegraphics[width=0.33\linewidth, clip, trim={0.5cm 0 0.5cm 0}]{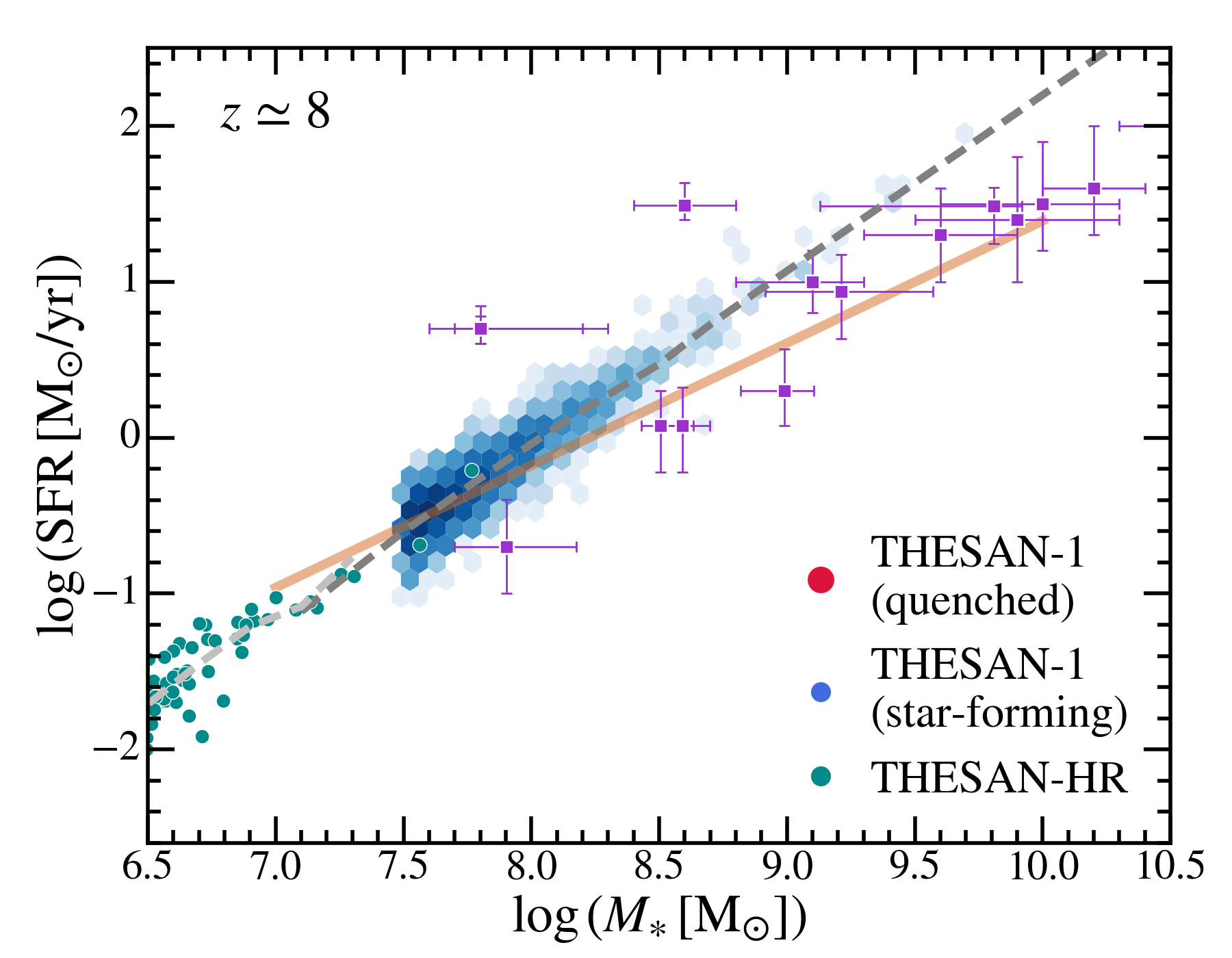}
\includegraphics[width=0.33\linewidth, clip, trim={0.5cm 0 0.5cm 0}]{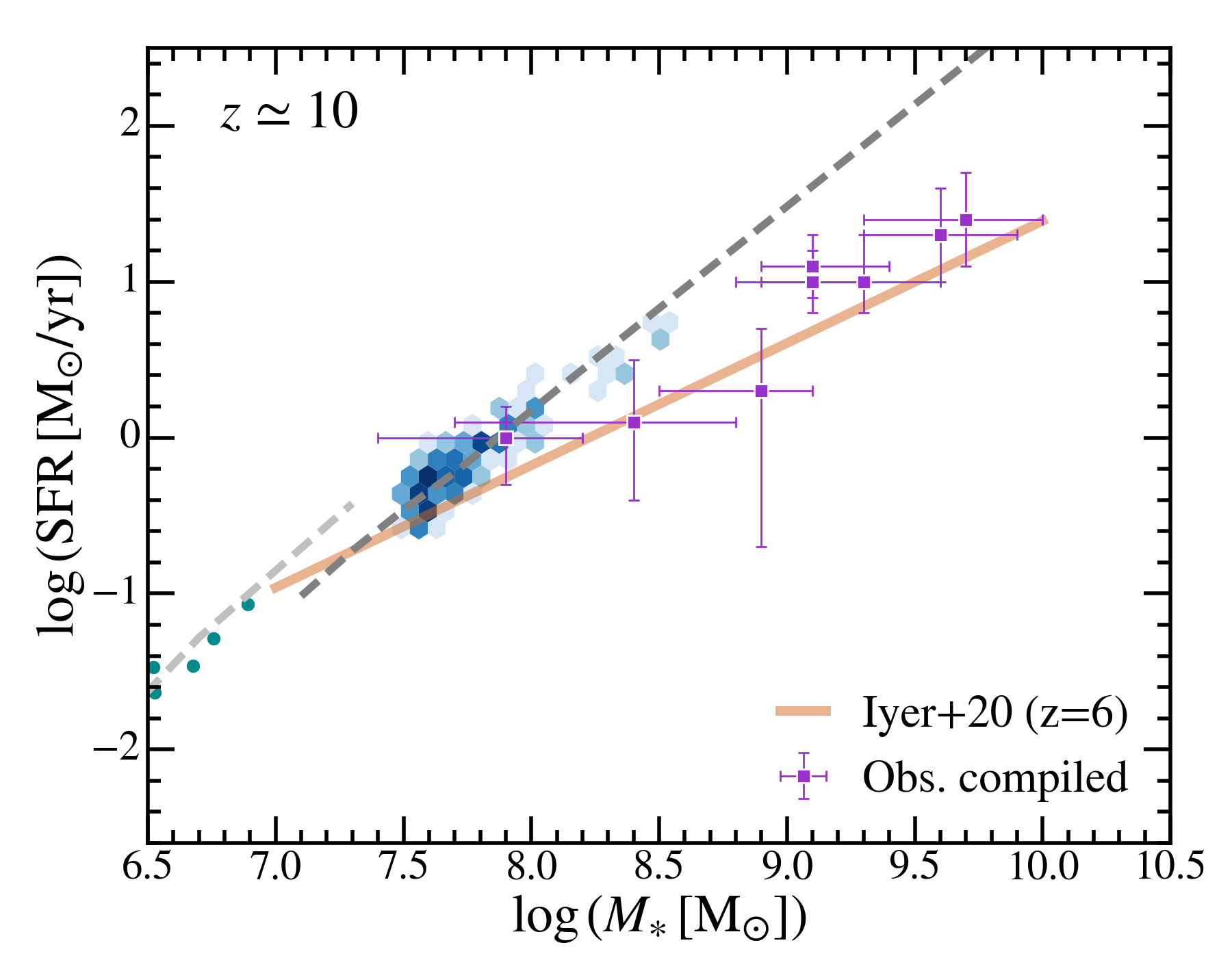}
\caption{The star-formation main sequence (SFMS) of galaxies in \thesanone and \thesanhr at $z\simeq$ 6, 8, and 10. The dashed lines are the median relation derived iteratively as discussed in Section~\ref{sec:method:def-ms}. The four quenched galaxies at $z\simeq 6$ are highlighted with red circles. The orange solid line shows the observation-constrained SFMS at $z\simeq 6$ in \citet{Iyer2020}. The purple data points show the compiled SFR-$M_{\ast}$ of observed $7 \lesssim z \lesssim 13$ galaxies from \citet{Tacchella2022b, Tacchella2023, Robertson2023-NatAst, Fujimoto2023}.}
\label{fig:main_sequence}
\end{figure*}

\begin{table*}
    \renewcommand{\arraystretch}{1.2}
    \centering
    \caption{Definitions of halo and galaxy properties in this paper. Properties marked \textcolor{violet}{purple} are taken directly from the \textsc{Subfind} catalog while others are derived in further post-processing as described.}
    \begin{tabular}{p{0.19\textwidth}p{0.76\textwidth}}
       \hline 
       \hline
       Halo mass (\textcolor{violet}{$M_{\rm halo}$}) & The halo mass of a galaxy is defined as the sum of the mass of all particles gravitationally bound to the subhalo identified by \textsc{Subfind}. We find that the difference of this halo mass compared to the one based on spherical overdensity, $M_{\rm 200,crit}$, is $< 0.1$ ($0.2$) dex for $>95$ ($>99$) percent of the central galaxies in the \thesan simulations at $z\geq 6$. The virial radius of the halo is defined as $R_{\rm vir}\equiv (3\,M_{\rm halo}/4\,\pi\,\Delta_{\rm c}\,\rho_{{\rm crit}, z})^{1/3}$, where $\Delta_{\rm c}=200$ and $\rho_{{\rm crit}, z}$ is the critical density of the Universe at $z$.\\
       \hline
       Stellar mass (\textcolor{violet}{$M_{\ast}$})  &  The stellar mass of a galaxy is defined as the sum of the current mass of all stellar particles within a spherical aperture from the galaxy centre. If not specifically mentioned, the default aperture is twice the stellar half-mass radius (as will be defined below). As shown in \citet{Pillepich2018b}, this stellar mass definition results in good agreement with the galaxy stellar mass functions at $z\lesssim 4$. In practice, we find that it is typically $\sim 0.1$ dex smaller than the total stellar mass gravitationally bound to the subhalo. %In other cases, the stellar mass will be labeled as $M_{\ast}(< r_{0})$, where $r_{0}$ is the aperture size. 
       \\
       \hline
       Star-formation rate (\textcolor{violet}{SFR})  &  To be consistent with the default definition of stellar mass, the SFR of a galaxy is defined as the sum of the instantaneous SFR of gas cells within twice the stellar half-mass radius. The specific star-formation rate (sSFR) is the ratio between SFR and stellar mass defined above. \\
       \hline
       Gas mass (\textcolor{violet}{$M_{\rm gas}$} and \textcolor{violet}{$M^{\rm tot}_{\rm gas}$})  &  Similar to the stellar mass, the gas mass in a galaxy ($M_{\rm gas}$) is defined as the sum of the mass of all gas cells within twice the stellar half-mass radius. Meanwhile, the total halo gas mass ($M^{\rm tot}_{\rm gas}$) is defined as the sum of the mass of all gas cells that are gravitationally bound to the subhalo. This represents the total gas reservoir of a given galaxy. The halo gas fraction ($f_{\rm gas}$) is defined as the ratio between $M^{\rm tot}_{\rm gas}$ and $M_{\rm halo}$.\\
       \hline
       Stellar half-mass radius (\textcolor{violet}{$r^{\ast}_{\rm 1/2}$}) &  We compute the stellar three-dimensional (3D) half-mass radius so that the stellar mass enclosed is half of all the stellar mass gravitationally bound to the subhalo. We note that this could cause a minor inconsistency between our definitions of $M_{\ast}$ and $r^{\ast}_{1/2}$. The $r^{\ast}_{\rm 1/2}$ here should be interpreted as a galaxy size without any pre-defined aperture. \\
       \hline 
       Stellar effective radius ($R_{\rm eff}$) & The effective radius of a galaxy is defined as the two-dimensional (2D) half-mass/light radius. The viewing angle is set to the positive $z$ direction in simulation coordinates. We use $R^{\ast}_{\rm eff}$ to refer to the 2D stellar half-mass radius. We use $R^{\rm x}_{\rm eff}$ to refer to the 2D half-light radius in a given observing band x, measured based on the mock images described in Section~\ref{sec:method:image}. We account for light contributions from all stellar particles without mimicking any surface brightness or signal-to-noise cuts. As a result, our size estimates may be larger than those typically inferred observationally (see Section~\ref{sec:apparent-size-factors} for a detailed discussion). The UV (V-band) sizes are measured in a top-hat filter centering around rest-frame 1500\AA\,(5510\AA) with a width of 100\AA\,(300\AA). \\
       \hline 
       Halo spin ($\lambda^{\prime}$) & The halo spin is defined as $\lambda^{\prime} \equiv J/(\sqrt{2}M_{\rm halo}\,V_{\rm vir}\,R_{\rm vir})$~\citep{Bullock2001}, where $J$ is the total angular momentum of all particles associated with the halo, $V_{\rm vir}\equiv \sqrt{G\,M_{\rm halo}/R_{\rm vir}}$ is the virial velocity. An alternative definition of spin parameter is~\citep{Peebles1969} $\lambda \equiv |J|\, E^{1/2}/G\,M_{\rm halo}^{5/2}$, where $E$ is total energy of the halo. $\lambda^{\prime}$ is related to this as $\lambda^{\prime} \simeq \lambda/\sqrt{f_{\rm c}(c)}$, where $f_{\rm c}$ is a function defined in \citet{Mo1998}. \\
       \hline
       Stellar/gas disk fraction ($f^{\ast}_{\rm disk}$ and $f^{\rm gas}_{\rm disk}$) & We define the face-on direction of a galaxy as the direction of the total angular momentum of stellar particles within twice $r^{\ast}_{\rm 1/2}$. We decompose the stellar content of a galaxy by the circularity parameters of stellar particles, $j_z/j_{\rm tot}$, where $j_z$ is the angular momentum of the particle aligned with the galaxy face-on vector and $j_{\rm tot}$ is the total angular momentum of this particle. Following \citet{Tacchella2019}, we define that a stellar particle belongs to the disk if $j_z/j_{\rm tot}>0.7$. %We note that a purely isotropic system will still contain $\sim 15\%$ of orbits with $j_z/j_{\rm tot}>0.7$ according to this definition. 
       The stellar disk fraction is the mass fraction of the disk component to the total stellar mass. The gas disk fraction is defined in the same way as the stellar content, except that the face-on direction is instead defined by the total angular momentum of gas cells within twice $r^{\ast}_{\rm 1/2}$.\\
       \hline
       \hline
    \end{tabular}
    \renewcommand{\arraystretch}{0.9090909090909090909}
    \label{tab:prop}
\end{table*}

\subsection{Galaxy identification and properties}
\label{sec:method:galprop}

We identify dark matter haloes using the Friends-of-Friends (FoF; \citealt{Davis1985,Springel2005}) algorithm and further identify inhabiting subhaloes via the \textsc{Subfind} \citep[first described in][]{Springel2001} algorithm. Gravitationally bound resolution elements are associated with each subhalo, and by definition, exclude elements bound to satellite galaxies. In this paper, galaxies are subhaloes with stellar masses larger than $50$ times the baryonic mass resolution ($m_{\rm b}$) of the corresponding simulation, including both central and satellite galaxies. A comparison of the sizes of centrals and satellites is presented in Appendix~\ref{sec:app-satellite} and displays no significant offsets. Unless otherwise stated, we study both central and satellite galaxies without distinction. The centre position of a galaxy is taken from the particle with the minimum gravitational potential energy. The derived galaxy properties that will be used in this paper are summarized in Table~\ref{tab:prop}.

\subsection{Classification of star-forming and quenched galaxies}
\label{sec:method:def-ms}

Following common practice in the observational community, we label our galaxies as star-forming versus quenched by recursively searching for the ridge line of the star-formation main sequence (SFMS, defined as the median sSFR of the star-forming galaxies). Following \citet{Pillepich2019}, we use $0.2$ dex bins of stellar mass and reject quenched galaxies as those whose logarithmic distance from the SFMS ridge line at their respective stellar mass lie in the range $\Delta \log_{10}{\rm sSFR} \leq -1.0$. We evaluate the SFMS in each stellar mass bin until the number of galaxies in the bin drops below $20$ and, in other bins, the reference locus of the SFMS is linearly extrapolated. Star-forming galaxies are those with $\Delta \log_{10}{\rm sSFR} > -1.0$. This sSFR-based classification is consistent at the $1-10$ percent level with commonly adopted selections in the UVJ diagram \citep[e.g.][]{Fang2018, Donnari2019}. Due to the dependence of sSFR on both stellar mass and redshift, this selection of star-forming galaxies will be different from a straight cut in sSFR. Furthermore, as discussed in Appendix~\ref{sec:app-numeric}, the lower envelope of the SFR-$M_{\ast}$ distribution could be subject to numerical effects at the low-mass end. Therefore, we conservatively remove the population of spurious quenched galaxies at the low-mass end from our analysis. In Figure~\ref{fig:main_sequence}, we show the identified SFMS at $z=6-10$ for \thesanone and \thesanhr, respectively. Four massive quenched galaxies are found at $z=6$ and are highlighted as red circles.

\subsection{Mock galaxy images}
\label{sec:method:image}

We generate images and datacubes (full spectra in pixels) for simulated galaxies. These images will be used for the comparison with observational results in Section~\ref{sec:observation}. Following our definition of galaxies in Section~\ref{sec:method:galprop}, we generate images only for galaxies with stellar mass larger than $50\,m_{\rm b}$, and specifically for \thesanone at $z\simeq 6$ and 8, we increase the stellar mass limit of galaxies to $175\,m_{\rm b} \sim 10^{8}\msun$ (for \thesanone) to save computational cost.

Following \citet{Vogelsberger2020,Shen2020,Kannan2022-LIM}, we use the Monte Carlo dust radiative transfer code {\sc Skirt}~\citep[version 8;][]{Baes2011,Camps2015} to generate images of simulated galaxies. We assigned intrinsic emission to stellar particles in the simulations according to their ages and metallicities using the stellar population synthesis method. All stellar particles within a $20\pkpc$ (physical $\kpc$) aperture are included in this calculation. We adopt the Flexible Stellar Population Synthesis (\textsc{Fsps}) model~\citep{Conroy2009} for continuum stellar emission\footnote{We note that the use of FSPS in post-processing here can introduce some inconsistencies with the BPASS model used for on-the-fly radiative transfer. The choice here is inherited from the pipeline developed in \citet{Vogelsberger2020}, which has been calibrated to the observed UV luminosity functions at $z = 4-10$ and successfully applied to \thesan in \citet{Kannan2022}.}. For the nebular emission originating from the birth cloud of young star clusters, we use the \citet{Byler2017} model built in \textsc{Fsps}, where line luminosities are generated with \textsc{Cloudy} and tabulated for grids of age, metallicity, gas density, and ionization parameter of the birth cloud. Following \citet{Byler2017}, we assume an ionization parameter~\footnote{Although \thesan can self-consistently model the radiative transfer of ionizing photons from young massive stars and the non-equilibrium coupling to gas, H{\small II} regions are not fully resolved. Therefore, the effective treatment of nebular emission is still required. We note that the galaxy sizes studied in this paper are not sensitive to the choice of the detailed treatment of nebular emissions.} $\log_{10}{U} = -2$, which agrees well with the ionization properties of recently \textit{JWST}-identified high-redshift galaxies~\citep[e.g.][]{Reddy2023}. The gas-phase metallicity of the H{\small II} regions is chosen to be the same as the initial metallicity of the stellar particles, which is inherited from the gas cell from which a stellar particle is created. Photon packages are randomly released based on the smoothed source distribution characterized by the positions and SEDs of stellar particles. Each stellar particle has been smoothed with smoothing length determined by the distance to the $16$th nearest neighbour. Radiative transfer is conducted on a wavelength grid with $657$ points spanning from $0.05$ to $2\,\mu{\rm m}$, refined around emission lines, as designed in \citet{Kannan2022-LIM}.

The emitted photon packages will further interact with the dust in the ISM. We assume that dust is traced by metals in the cold, star-forming gas and turn the metal mass distribution into the dust mass distribution with a constant, averaged dust-to-metal ratio of all galaxies at a fixed redshift. This dust-to-metal ratio depends on redshift as $0.9 \times (z/2)^{-1.92}$, which has been calibrated based on the observed UV luminosity functions at $z=2-10$~\citep{Vogelsberger2020} and resulted in consistent predictions when applied to \thesan~\citep{Kannan2022-LIM}. The radiative transfer calculations are performed on an adaptively refined octree grid~\citep{Saftly2014} with the minimum and maximum refinement levels set to $3$ and $10$. The grid spans $20\pkpc$ on each dimension and the resulting minimum Octree cell size is about $40\pc$. Dust emission is modelled assuming thermal equilibrium with the local radiation field. We adopt the \citet{Draine2007} dust mix, which includes a composition of graphite, silicate, and polycyclic aromatic hydrocarbon (PAH) grains. This model reproduces the average Milky Way extinction curve and is widely used~\citep[e.g.][]{Jonsson2010,Remy-Ruyer2014}. 

For each wavelength on the wavelength grid, $N_{\rm p}$ photon packets are launched isotropically. Compared to \citet{Kannan2022-LIM}, we increase $N_{\rm p}$ to $3\times 10^{5}$ to reduce the Poisson noises in mock images. The photon packets then propagate through the resolved gas (dust) distribution in the ISM and interact with the dust cells randomly before they are finally collected by a pre-defined photon detector, which is placed at a distance of $10\Mpc$ from the galaxy centre in the positive $z$ direction. The pixel-wise galaxy spectra are then recorded, which can be processed to integrated SED, imaging, or integrated field unit data. For rest-frame broadband photometry, galaxy SEDs are convolved with the transmission curves using the \textsc{Sedpy} code. For the calculation of apparent band magnitudes, the rest-frame flux is redshifted, corrected for intergalactic medium absorption~\citep{Madau1995}, and converted to the observed spectra. The final images have a field-of-view of $15\pkpc$ and $512\times 512$ pixels. Convolution with the point spread function (PSF) of observational instruments or background noises has not been considered here. The pixel size is $\sim 30\pc$ (physical) and is several times smaller than the observational limit of galaxy size ($\sim 120 - 160 \pc$) estimated from the half of the FWHM of NIRCam in \citet{Morishita2023}.

\begin{figure}
    \centering
    \includegraphics[width=1\linewidth]{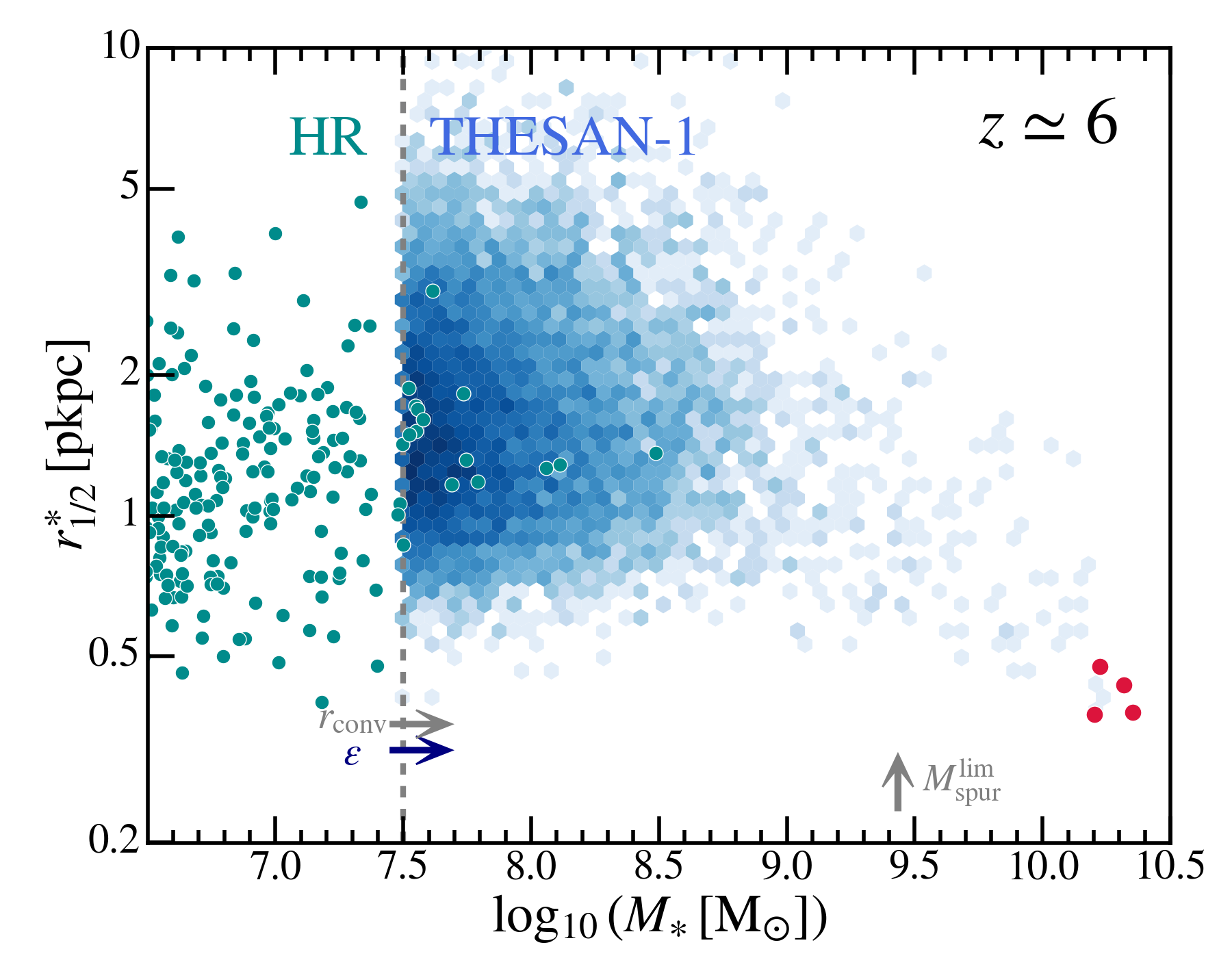}
    \includegraphics[width=1\linewidth]{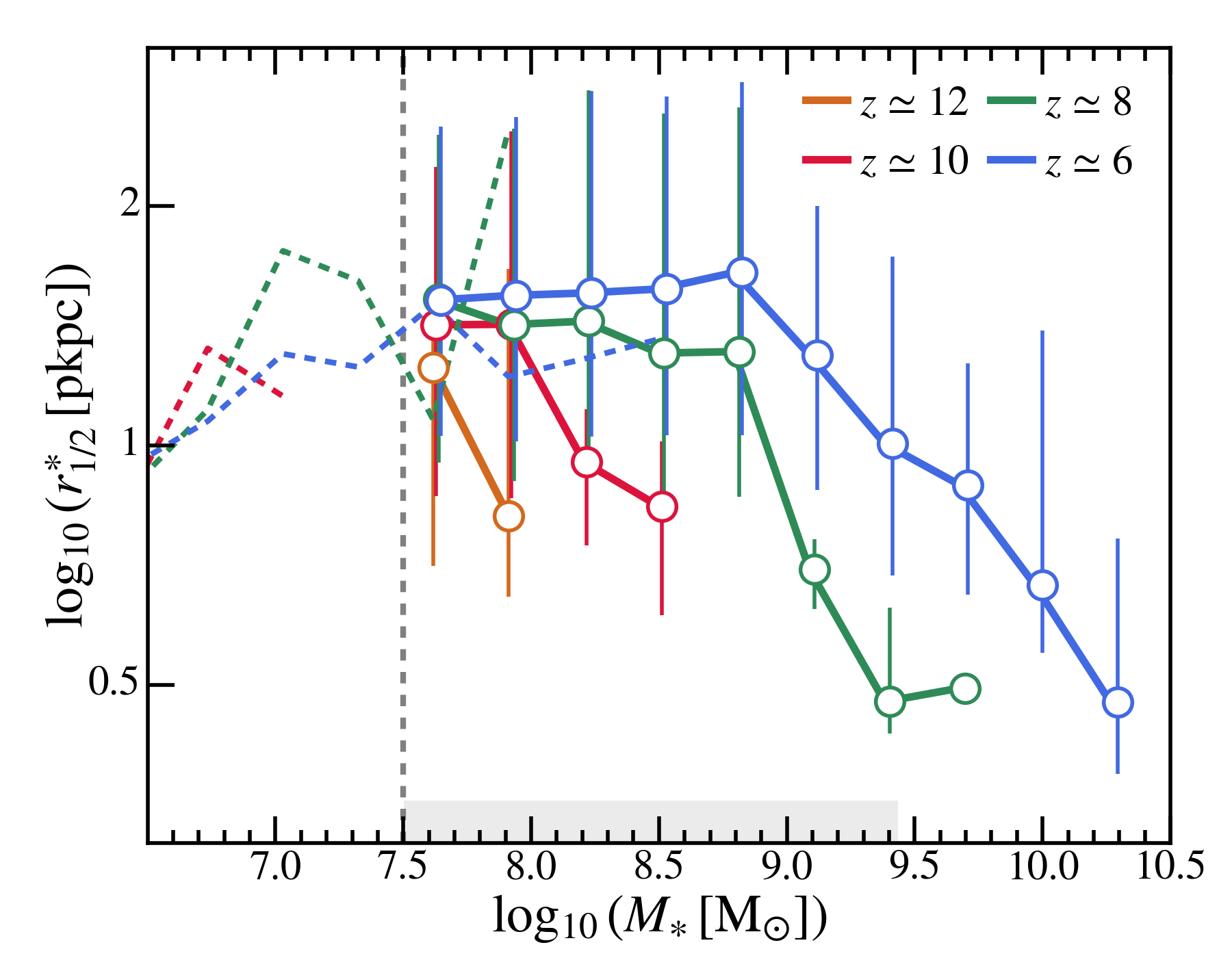}
    \caption{\textit{Top:} Galaxy intrinsic size ($r^{\ast}_{1/2}$) versus stellar mass at $z\simeq 6$ in \thesanone (blue cloud) and \thesanhr (cyan circles representing individual galaxies). The four quenched galaxies identified in \thesanone are marked with red circles. The vertical dashed line indicates the stellar mass limit we choose for \thesanone. The blue arrow shows the physical gravitational softening length of \thesanone. The gray arrows show the convergence radius ($r_{\rm conv}$) and the spurious heating mass limit ($M^{\rm lim}_{\rm spur}$) of \thesanone (see Appendix~\ref{sec:app-numeric} for details). In the massive end, we find that $r^{\ast}_{1/2}$ anti-correlates with galaxy stellar mass at $M_{\ast}\gtrsim 10^{8}\msun$. Four compact quenched galaxies are identified in \thesanone. \textit{Bottom:} Redshift evolution of galaxy size--mass relation from $z\simeq12$ to $z\simeq 6$. The error bars show the median and $1$-$\sigma$ dispersion of $r^{\ast}_{1/2}$ in each bin. The gray shaded region shows the regime closed by $r_{\rm conv}$ and $M^{\rm lim}_{\rm spur}$, where numerical effects may prevent the formation of compact galaxies. Negative slopes of the intrinsic size--mass relation are consistently found at these redshifts. Through cosmic time, there is limited evolution at the low-mass end of the size--mass relation while the most massive galaxies continuously become more compact.}
    \label{fig:size-mass}
\end{figure}

\section{Galaxy intrinsic size--mass relation}
\label{sec:size-mass}

The intrinsic size of a galaxy is represented by the 3D stellar half-mass radius $r^{\ast}_{1/2}$ (as defined in Table~\ref{tab:prop}). This is the preferred quantity for our analysis of the physical evolution of galaxy sizes, as it is less affected by geometrical/projection effects and tracer biases (e.g. using light in certain bands as a tracer for stellar mass). Therefore, the analysis will begin with $r^{\ast}_{1/2}$ and we leave the discussions on observed galaxy sizes to Section~\ref{sec:apparent-size}.

In the top panel of Figure~\ref{fig:size-mass}, we show the intrinsic size--mass relation of galaxies in the \thesan simulations. The intrinsic sizes of galaxies show a mild increase with stellar mass at the low-mass end ($M_{\ast} \lesssim 10^{8}\msun$) while displaying a negative correlation with stellar mass towards the massive end. Notably, all galaxies at the massive end ($M_{\ast} \gtrsim 10^{10}\msun$) are compact ($r^{\ast}_{1/2} \lesssim 1 \pkpc$). Four massive compact galaxies in \thesanone are quenched at $z\simeq 6$. The scatter of galaxy sizes at the low-mass end is substantial. A population of low-mass compact galaxies exists at the same time with extended galaxies with sizes as large as $\gtrsim 5 \pkpc$. Due to the limited mass and spatial resolution of the simulations, numerical effects can affect galaxy sizes in non-trivial ways~\citep[e.g.][]{Power2003, Ludlow2019-size,Ludlow2023}. The choice of gravitational softening lengths of the \thesan simulations is motivated to minimize these effects. This is discussed in detail in Appendix~\ref{sec:app-numeric}. In Figure~\ref{fig:size-mass}, we show two relevant spatial scales for numerical effects, the convergence radius ($r_{\rm conv}$) and the physical gravitational softening length ($\epsilon$) of stars as well as the mass limit below which galaxy size can be affected by spurious heating of DM particles. In this regime, we could miss a population of low-mass compact galaxies in our simulations. However, in terms of the median relation, the numerical convergence is decent when comparing \thesanone and \thesanhr results. We refer readers to Appendix~\ref{sec:app-numeric} for a more thorough discussion.

\begin{figure}
    \centering 
    \includegraphics[width=1\linewidth]{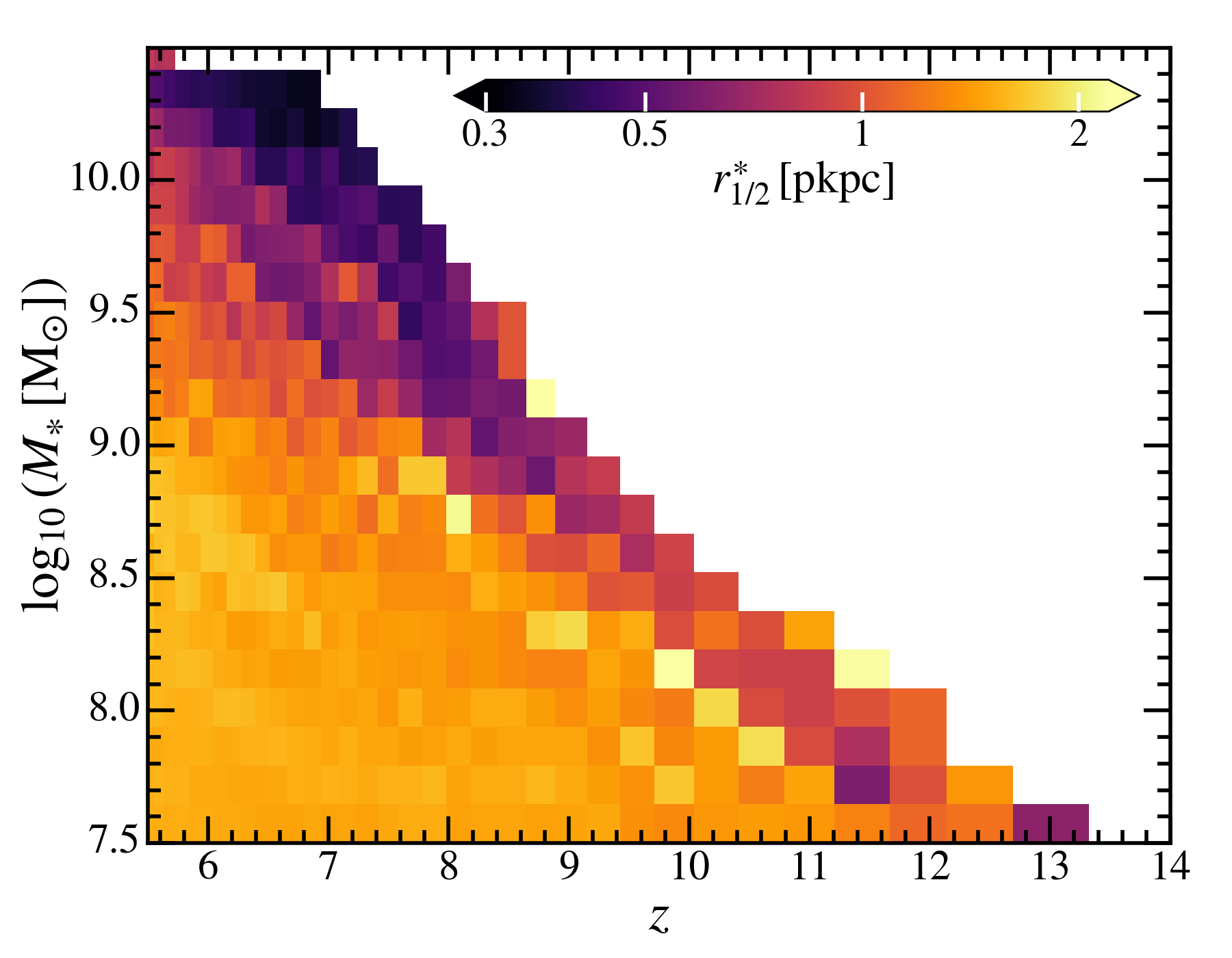}
    \caption{The evolution median galaxy size as a function of stellar mass and redshift in the \thesanone simulation. The sizes of most massive galaxies become increasingly more compact through time starting at $z\sim 10$. Slightly lower-mass galaxies follow a similar track but with compaction starting at lower redshifts. On the other hand, the low-mass galaxies have little size evolution with $r^{\ast}_{1/2}$ consistently above $\sim 1\pkpc$. }
    \label{fig:size-mass-colormap}
\end{figure}

In the bottom panel of Figure~\ref{fig:size-mass}, we show the evolution of the intrinsic size--mass relation from $z\simeq 12$ to $z\simeq 6$. The shaded regions show the regime affected by spurious heating at $z\simeq 6$ (as characterized by $r_{\rm conv}$ and the spurious heating mass limit). At all redshifts, galaxies at the massive end have their sizes negatively correlated with stellar mass. The massive compact tail continuously evolves towards the massive compact end. As we will discuss later, the tail is represented by galaxies that consistently stay in the compact end of the galaxy population and some of them are quenched at $z\simeq 6$. However, at the low-mass end, there is limited evolution. The scatters of the size--mass relation increase with cosmic time. In Figure~\ref{fig:size-mass-colormap}, we present a more thorough view of galaxy size as a function of stellar mass and redshift. The upper edge of this colormap can be interpreted as the main progenitor mass growth history of the most massive galaxies in the \thesanone simulation. These massive galaxies become increasingly more compact through time since $z\sim 10$. The compaction phase is similar for slightly lower-mass galaxies but with a delayed starting time. At fixed stellar mass and at $M_{\ast} \gtrsim 10^{8.5}\msun$, galaxy sizes are larger at lower redshift but this is primarily due to their different phases on the track of compaction. On the other hand, the sizes of low-mass galaxies have little evolution and are consistently above $\sim 1\pkpc$, suggesting fundamentally different physical mechanisms controlling galaxy sizes at different mass scales.

The negative slope of the size--mass relation at the massive end is consistent with findings in previous theoretical studies using simulations~\citep[e.g.][]{Roper2022,Popping2022,Costantin2022,Marshall2022}. However, it is contradictory to the positive correlations found in observations at low redshifts~\citep[though at sightly larger stellar masses; e.g.][]{Mosleh2012,vanderWel2014,Lange2015} and some more recent constraints at $z\gtrsim 6$~\citep[e.g.][]{Shibuya2015-size,Kawamata2018,Ormerod2023,Morishita2023}. The contradiction could be due to the concentrated dust attenuation in massive galaxies, resulting in less cuspy surface brightness profiles and larger apparent sizes~\citep[e.g.][]{Roper2022,Marshall2022,Popping2022}. In addition, the intrinsic 3D sizes we find in \thesan are generally larger than the effective radii reported for observed galaxies. A combination of projection effects, tracer biases, and dust attenuation could lead to this discrepancy. This will be discussed in detail in Section~\ref{sec:apparent-size-factors}.

\begin{figure}
    \includegraphics[width=1\linewidth]{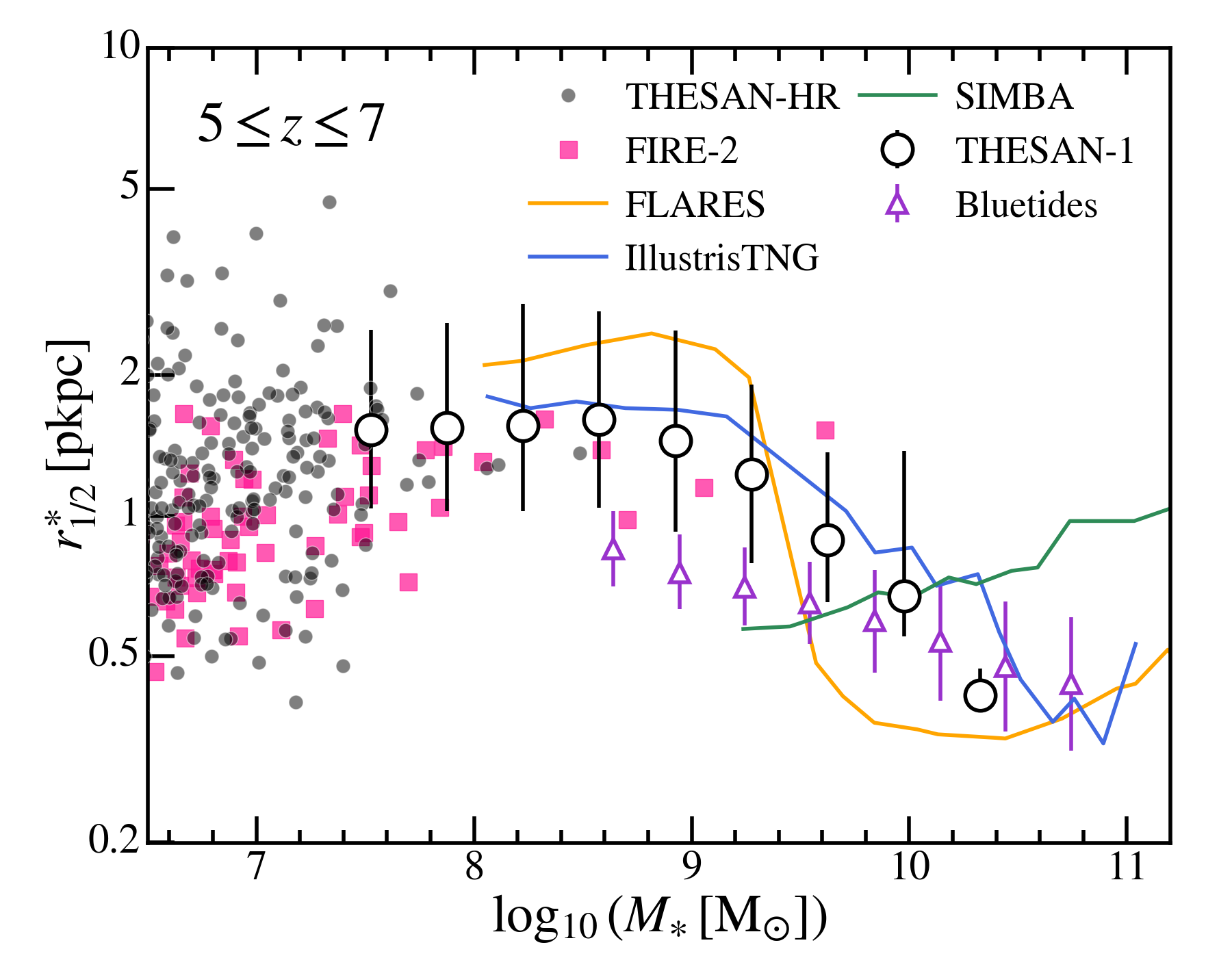}
    \includegraphics[width=1\linewidth]{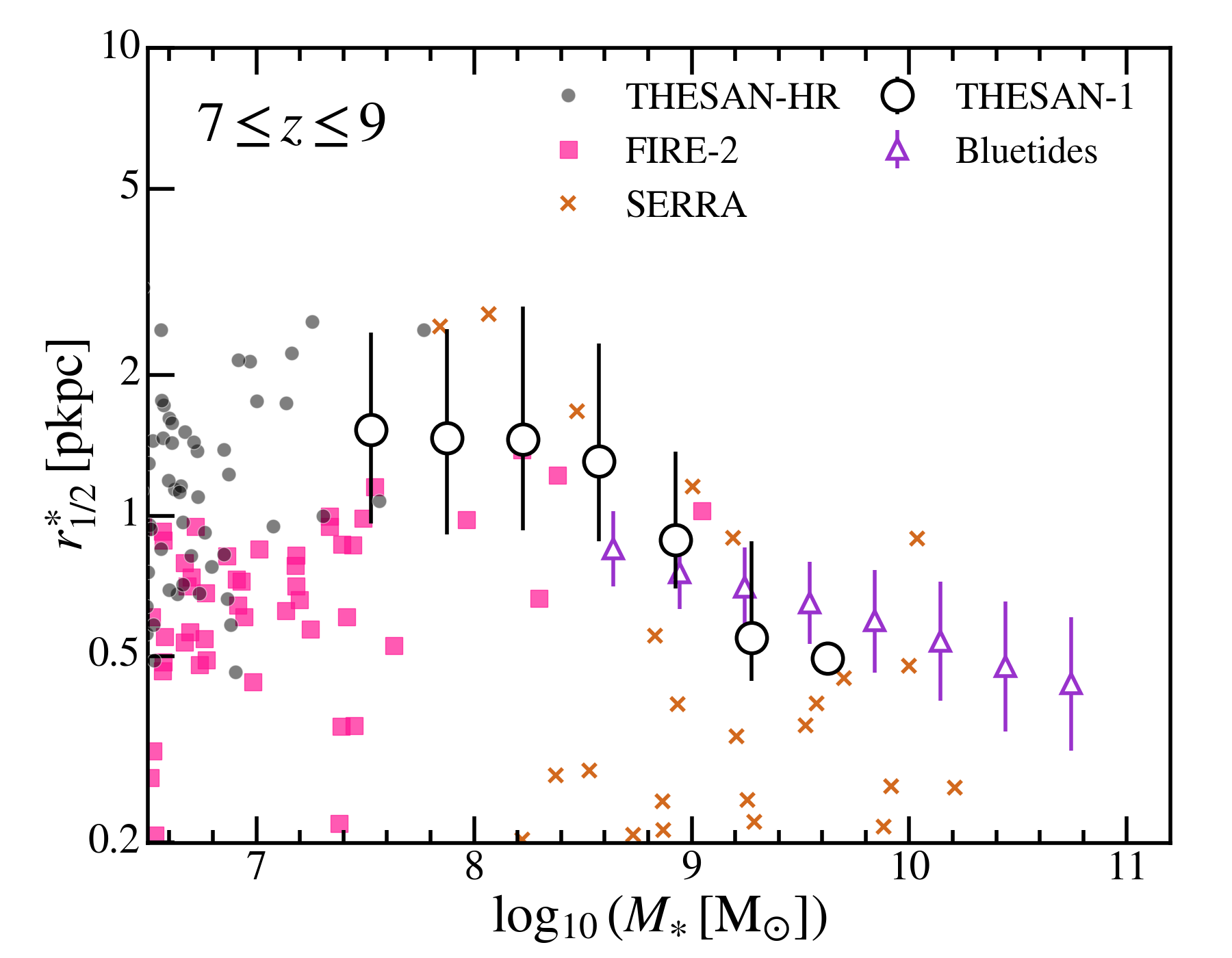}
    \caption{Galaxy intrinsic size--mass relation from a compilation of cosmological hydrodynamic simulations. The \thesanone and \thesanhr predictions are shown in black open and solid circles, respectively. For comparison, we show results from the FIRE-2 simulations at $z=6,8$~\citep{Ma2018-size}, the FLARES simulation at $z=5$~\citep{Roper2022,Roper2023}, the IllustrisTNG simulation~\citep{Pillepich2018} at $z=5$, the SIMBA simulation~\citep{Romeel2019} at $z=6$, the Bluetides simulations at $z=7$~\citep{Marshall2022}, and the SERRA simulations~\citep{Pallottini2022}. The \thesan results are in general consistent with these simulations in shared dynamical range, despite rather different galaxy formation physics models and numerical resolutions of them. An exception is the SERRA simulations which produce increasingly more compact galaxies at $M_{\ast}\lesssim 10^{9}\msun$.}
    \label{fig:size-mass-comparision}
\end{figure}

\subsection{Comparison to other simulations}
\label{sec:size-mass-theory-compare}

In Figure~\ref{fig:size-mass-comparision}, we compare the \thesan results with other cosmological hydrodynamic simulations using diverse numerical setups and galaxy formation physics models. We include predictions from the FLARES simulation~\citep{Vijayan2021,Lovell2021} at $z=5$, the IllustrisTNG simulation~\citep[specifically TNG100-1;][]{Pillepich2018} at $z=5$, the SIMBA simulation~\citep[m100n1024;][]{Romeel2019} at $z=5$, as shown and compared in \citet{Roper2023}. In addition, we include predictions of the 2D projected half-stellar-mass radius from the FIRE-2 simulations~\citep{Ma2018,Ma2018-size} and the Bluetides simulations~\citep{Marshall2022}. The 2D half-mass radius is converted to the 3D half-mass radius using a constant multiplication factor, $\sim 1.3$, as will be motivated in Section~\ref{sec:apparent-size-factors}. We also include the UV effective radius predictions of central galaxies from the SERRA simulations~\citep{Pallottini2022}, for which we adopt the same multiplication factor and effectively assume that the UV light is a perfect tracer of stellar mass.

The \thesan predictions are in good agreement with IllustrisTNG as expected due to the same galaxy formation model employed. Compared to FLARES, \thesanone predicts roughly the same normalization and negative dependence of size versus mass, but \thesanone does not predict the sharp galaxy size transition at $M_{\ast}\sim 10^{9.5}\msun$ as seen in FLARES. Compared to Bluetides, \thesanone results are in good agreement with it at the massive end while giving about a factor of two larger sizes at $M_{\ast}\lesssim 10^{9}\msun$ at $z\sim 6$. Predictions from SIMBA exhibit a positive slope of size--mass relation, in contrast with the results of all other simulations. This could be due to the poorer mass resolution of the m100n1024 run of SIMBA~\citep{Roper2023} but also details in the feedback model.

Compared to FIRE-2, which is a suite of high-resolution ($\epsilon \sim \pc$) zoom-in simulations, the results from \thesanone and \thesanhr are consistent with it at $z\sim 6$. This consistency is non-trivial given the drastically different simulation strategy, numerical resolution, star-formation, and feedback model employed. However, FIRE-2 did not predict the relatively extended galaxy population ($r^{\ast}_{1/2}\gtrsim 2\pkpc$) at the low-mass end and this discrepancy becomes more apparent at $z\sim 8$. For IllustrisTNG, the effective ISM model does not resolve the cold molecular phase, and the spatial and time clustering of star-formation may be suppressed compared to simulations employing a high-density threshold and more explicit criteria for star-formation~\citep[e.g.][]{Pontzen2014}. Therefore, the \thesan simulations may miss the temporary extremely compact phase of galaxies, especially at the low-mass end where the stellar component of a galaxy can be dominated by a single compact star cluster~\citep[e.g.][]{Yajima2017,Ma2018-size}. 

Even within high-resolution zoom-in simulations employing resolved ISM physics, predicted sizes can differ substantially due to the detailed implementation of star-formation and feedback. An example is the outlier in our size comparison in Figure~\ref{fig:size-mass-comparision}, the SERRA simulations~\citep{Pallottini2022}, which predicts increasingly more compact galaxies at $M_{\ast}\lesssim 10^{9}\msun$ than all other simulations. As will be discussed later in Section~\ref{sec:observation}, this result turns out to agree better with the observational constraints. These comparisons here highlight the potential of galaxy sizes at high redshifts as a channel to constrain physics models of galaxy formation and evolution.

We note that the simulations compared here together indicate a hump feature in the intrinsic size mass relation of galaxies at $z\gtrsim 6$ in the stellar mass range of $10^{8}$ to $10^{9}\msun$. In the mass range, the majority of the galaxies have $r^{\ast}_{1/2}\gtrsim 1\pkpc$. This means that a single power-law relation is not adequate to describe the connection between $r^{\ast}_{1/2}$ and $M_{\ast}$ at $z\gtrsim 6$, in contrast to the common practice in observational studies~\citep[e.g.][]{vanderWel2014,Shibuya2015-size}. Meanwhile, as will be demonstrated in Section~\ref{sec:apparent-size}, the extended galaxy sizes around the hump feature are in tension with the latest observational constraints, including the ones based on \textit{JWST} spectroscopically confirmed galaxy samples.

\begin{figure*}
    \raggedright
    \includegraphics[width=0.49\linewidth]{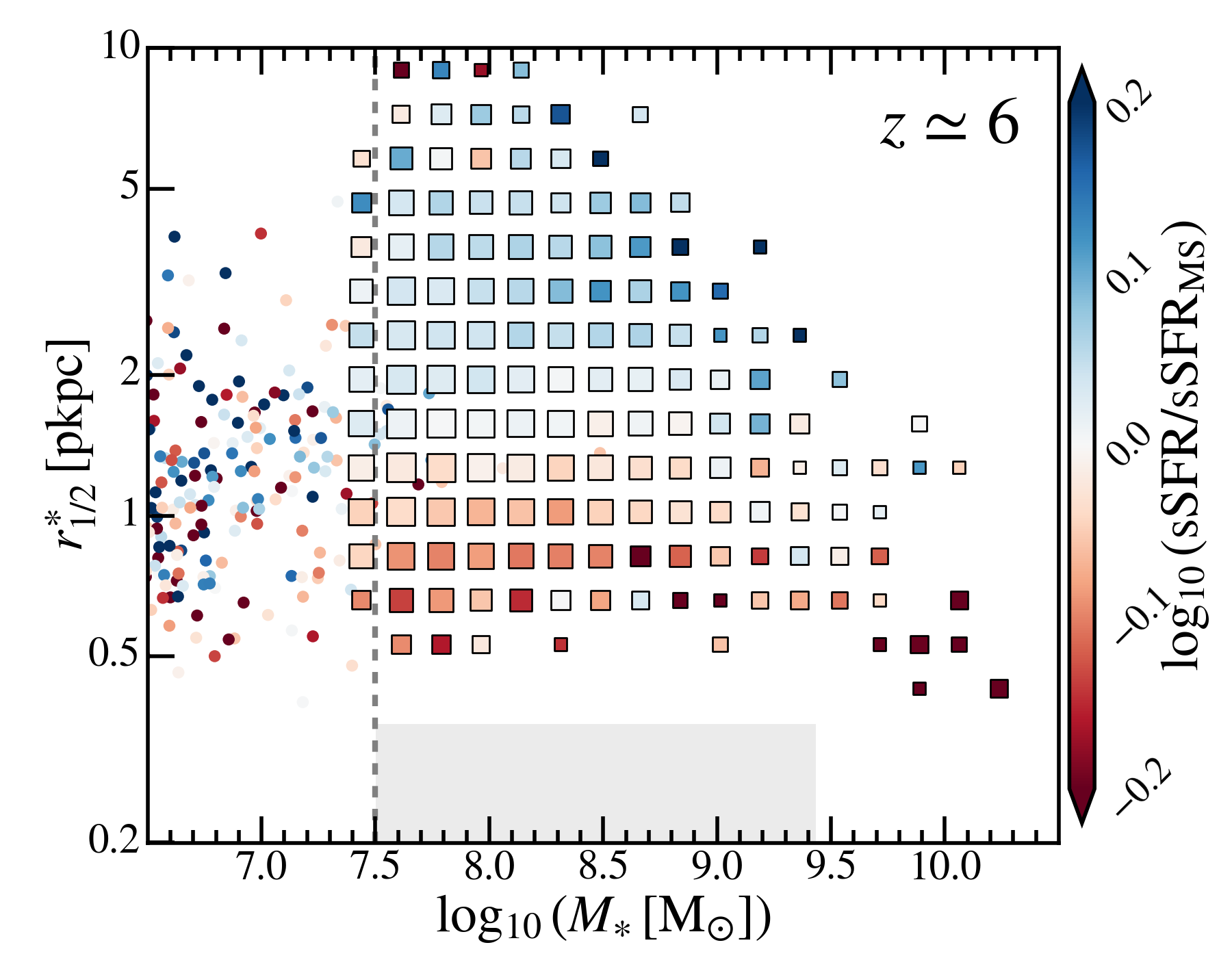}
    \includegraphics[width=0.49\linewidth]{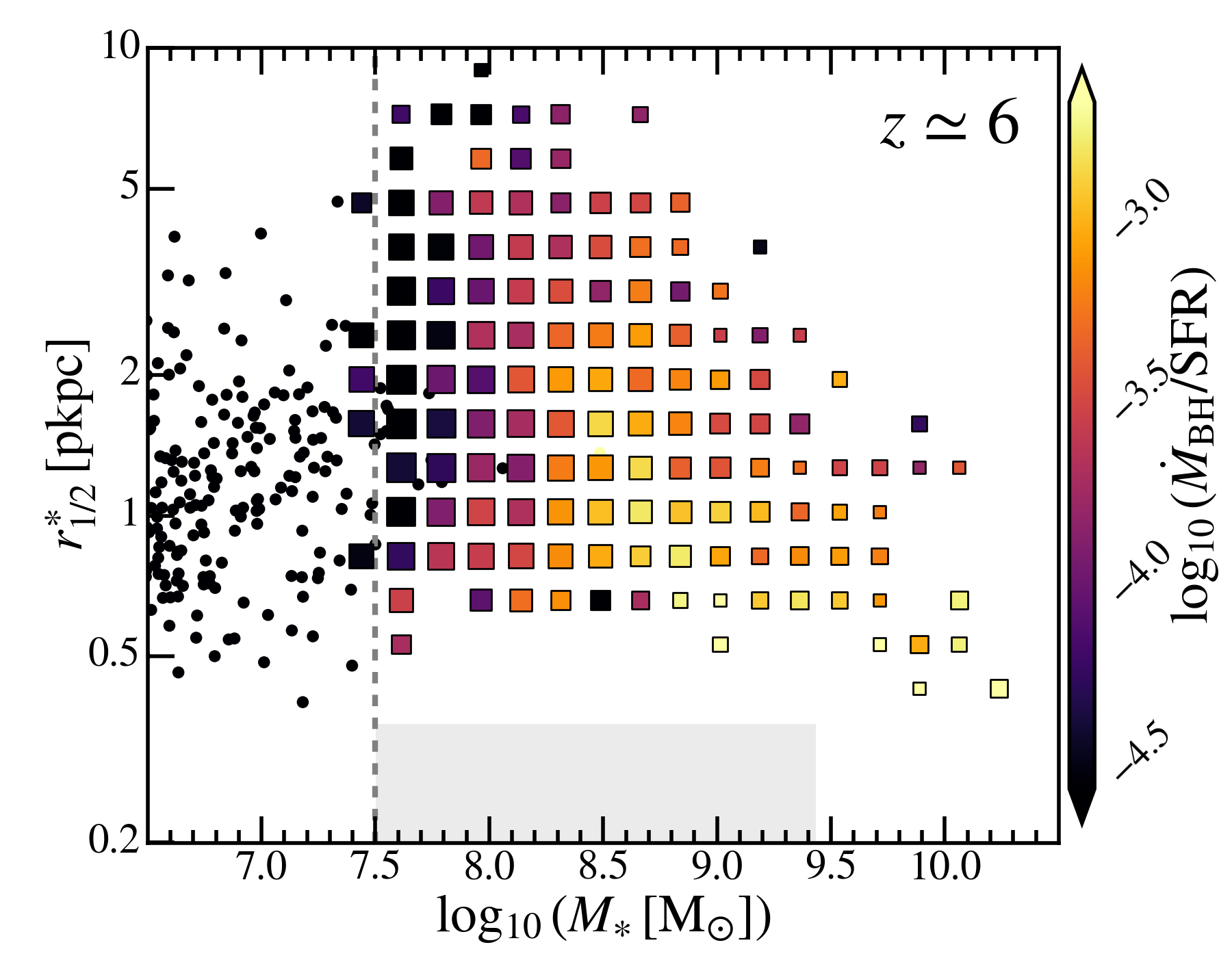}
    \includegraphics[width=0.49\linewidth]{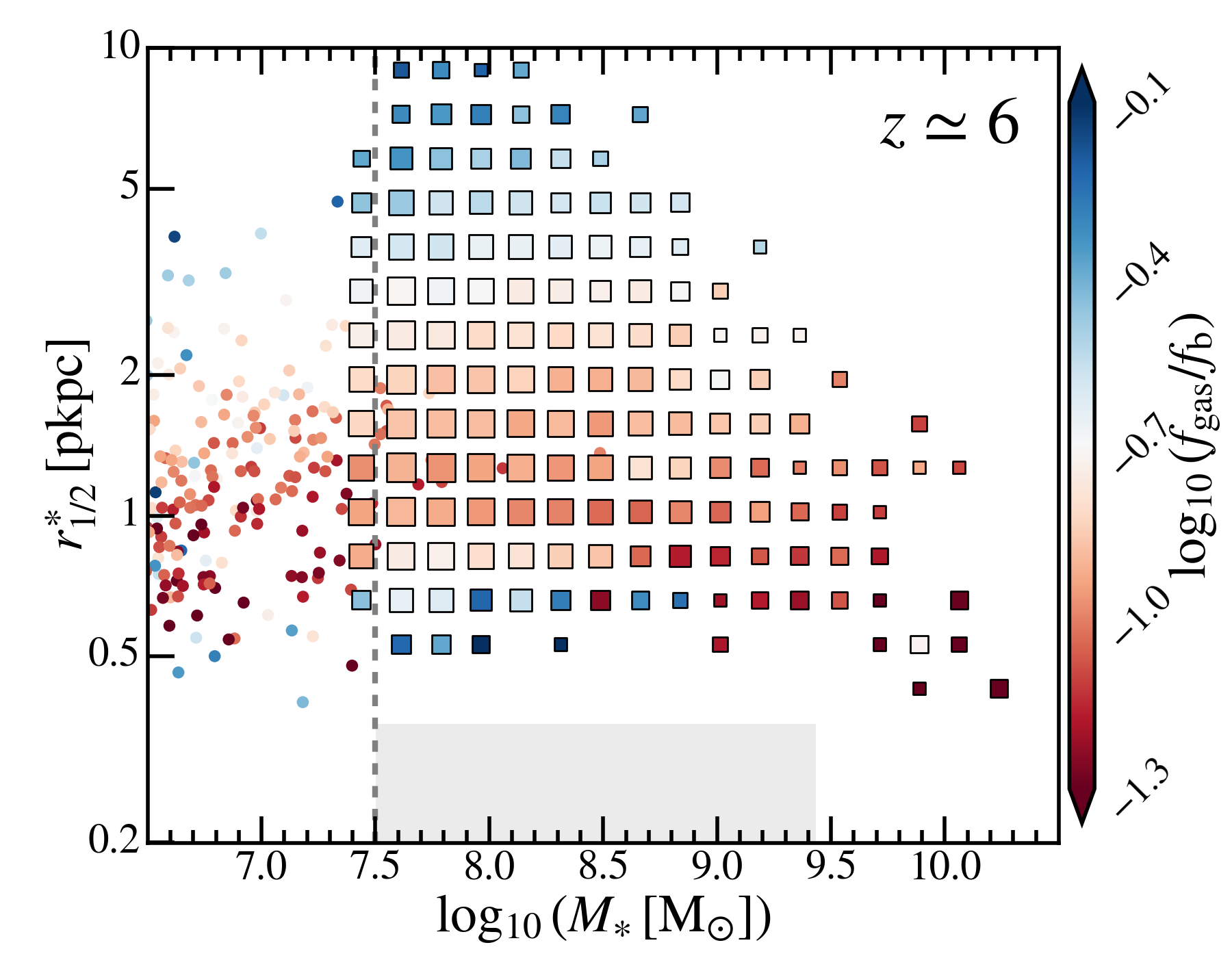}
    \includegraphics[width=0.49\linewidth]{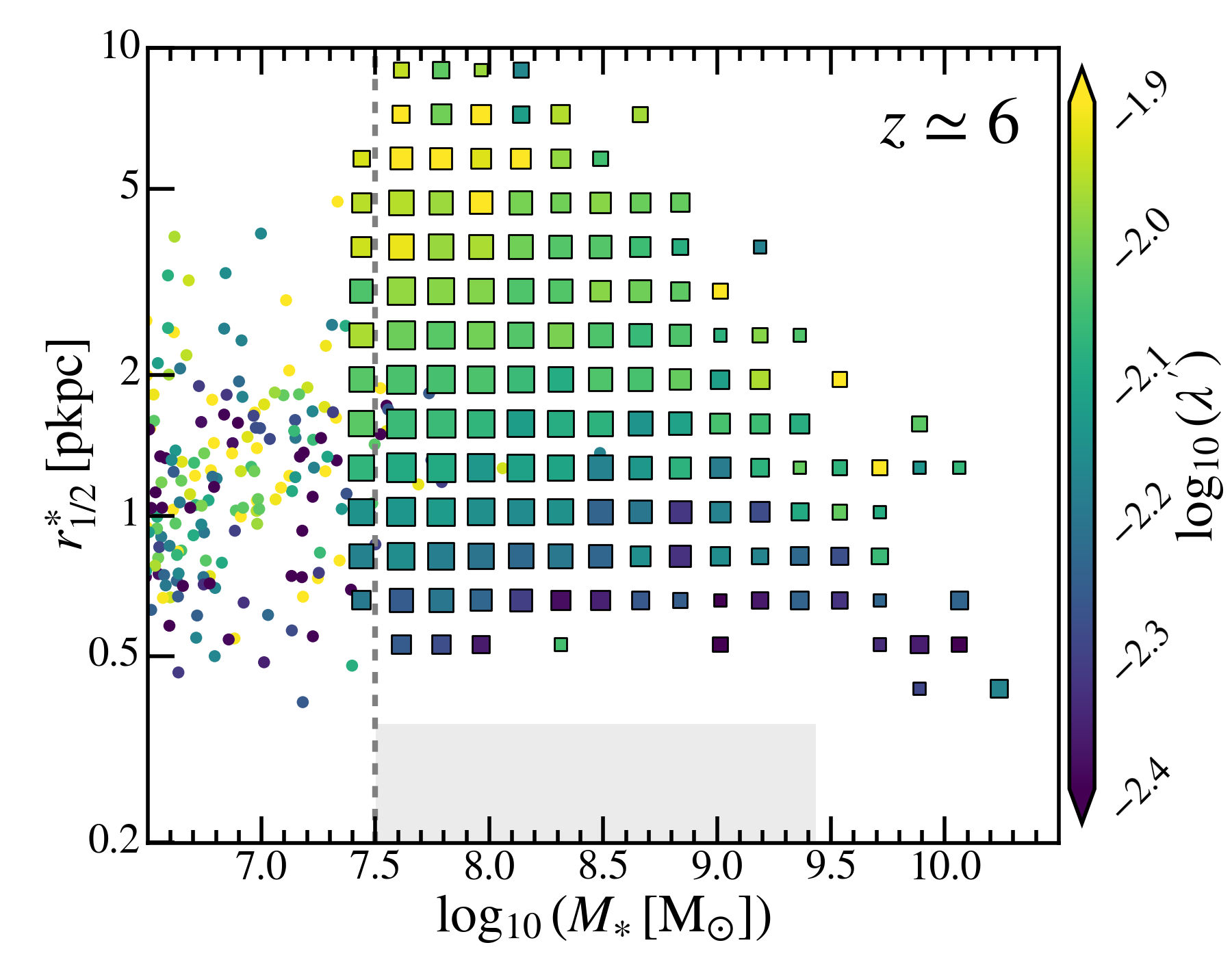}
    \includegraphics[width=0.49\linewidth]{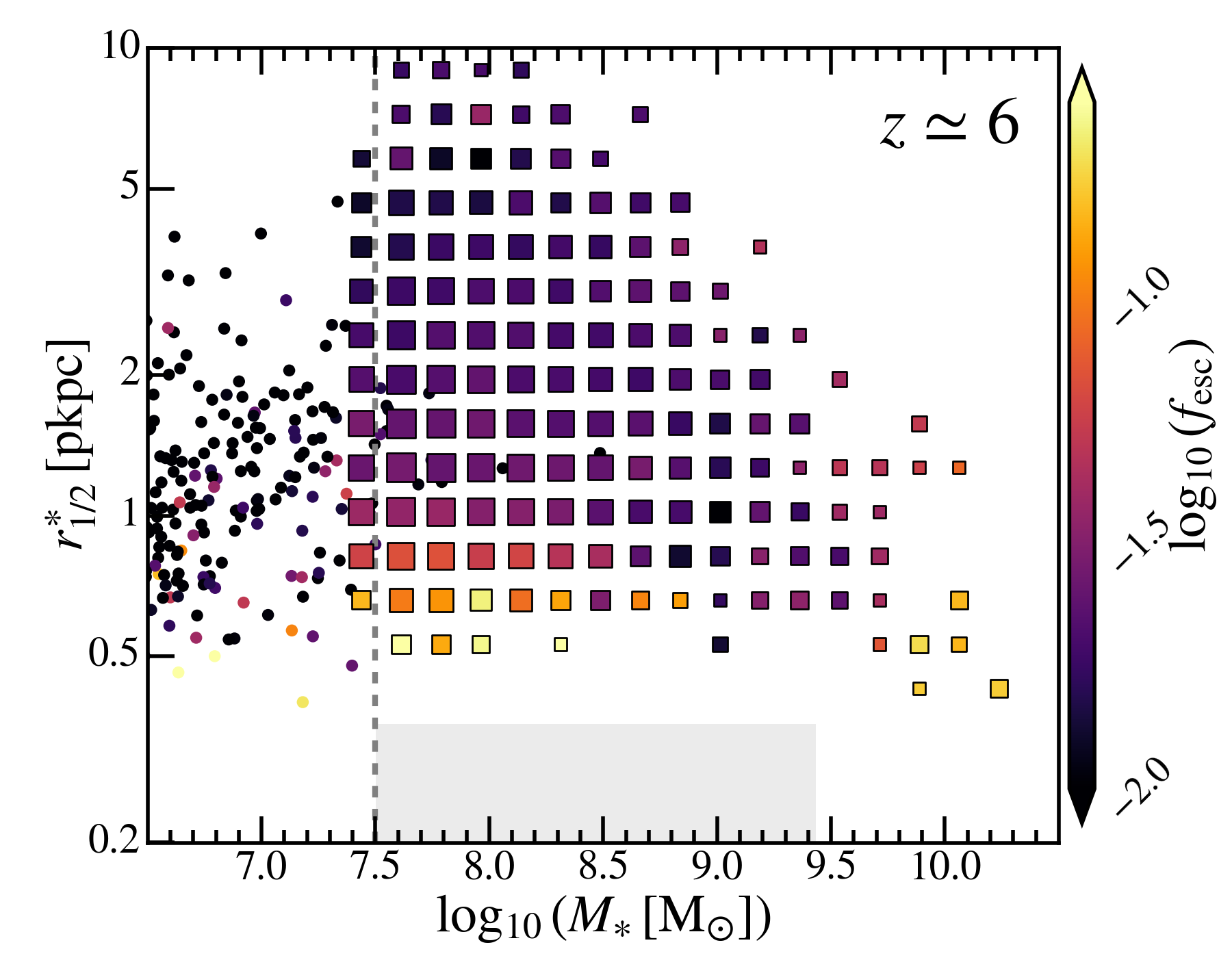}
    \includegraphics[width=0.49\linewidth]{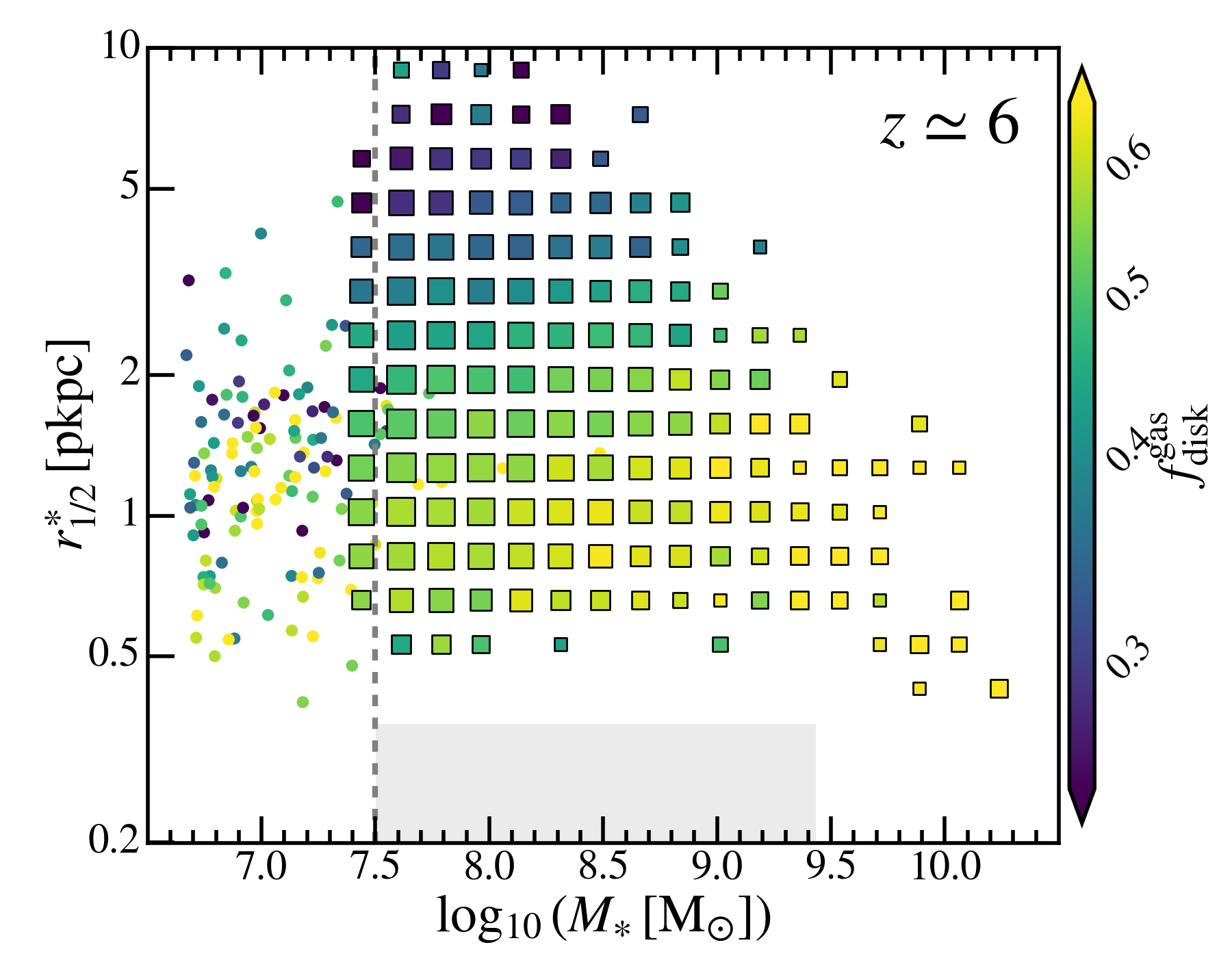}
    \caption{Galaxy intrinsic size--mass relation at $z\simeq 6$ color-coded by the distance to the SFMS (top left), SMBH accretion rate (top right), gas fraction (middle left), halo spin (middle right), Lyman continuum escape fraction (bottom left), and gas disk fraction (bottom right). The galaxies from \thesanone are shown by squares and the square sizes are proportional to the logarithm of galaxy density in the size--mass plane. The \thesanhr galaxies are directly shown with circles. The vertical dashed line indicates the stellar mass limit we choose for \thesanone. The shade regions indicate where numerical effects could prevent the formation of most compact galaxies (see Figure~\ref{fig:size-mass}). Galaxies below the main sequence tend to be more compact and the extreme cases are the compact quenched galaxies at the massive end. At fixed stellar mass, more compact galaxies tend to have more intense SMBH accretion, lower gas abundance, lower halo spin, higher ionizing photon escape fractions, and higher gas disk fraction.}
    \label{fig:size-mass-correlation}
\end{figure*}

\begin{figure}
    \centering
    \includegraphics[width=1\linewidth]{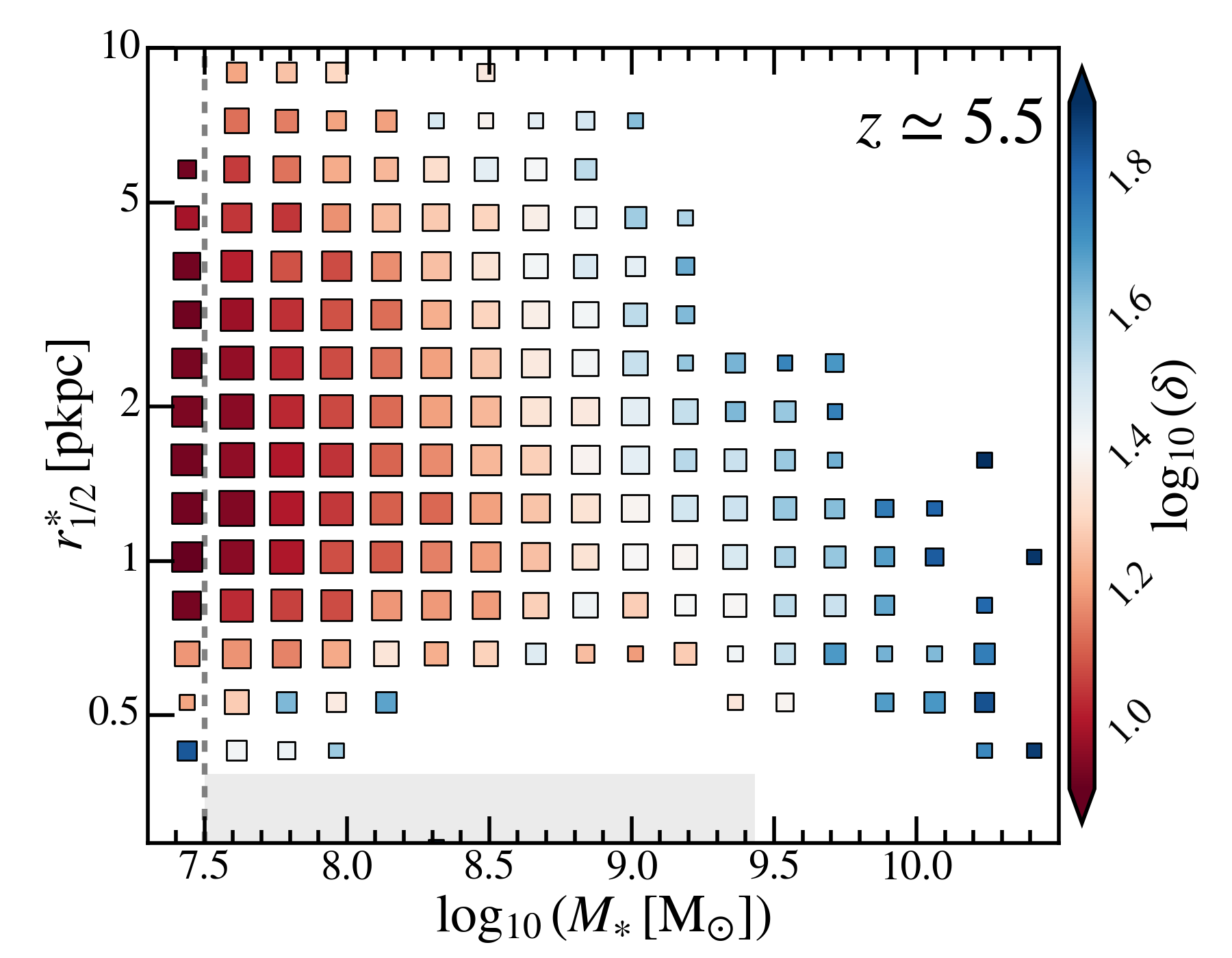}
    \includegraphics[width=1\linewidth]{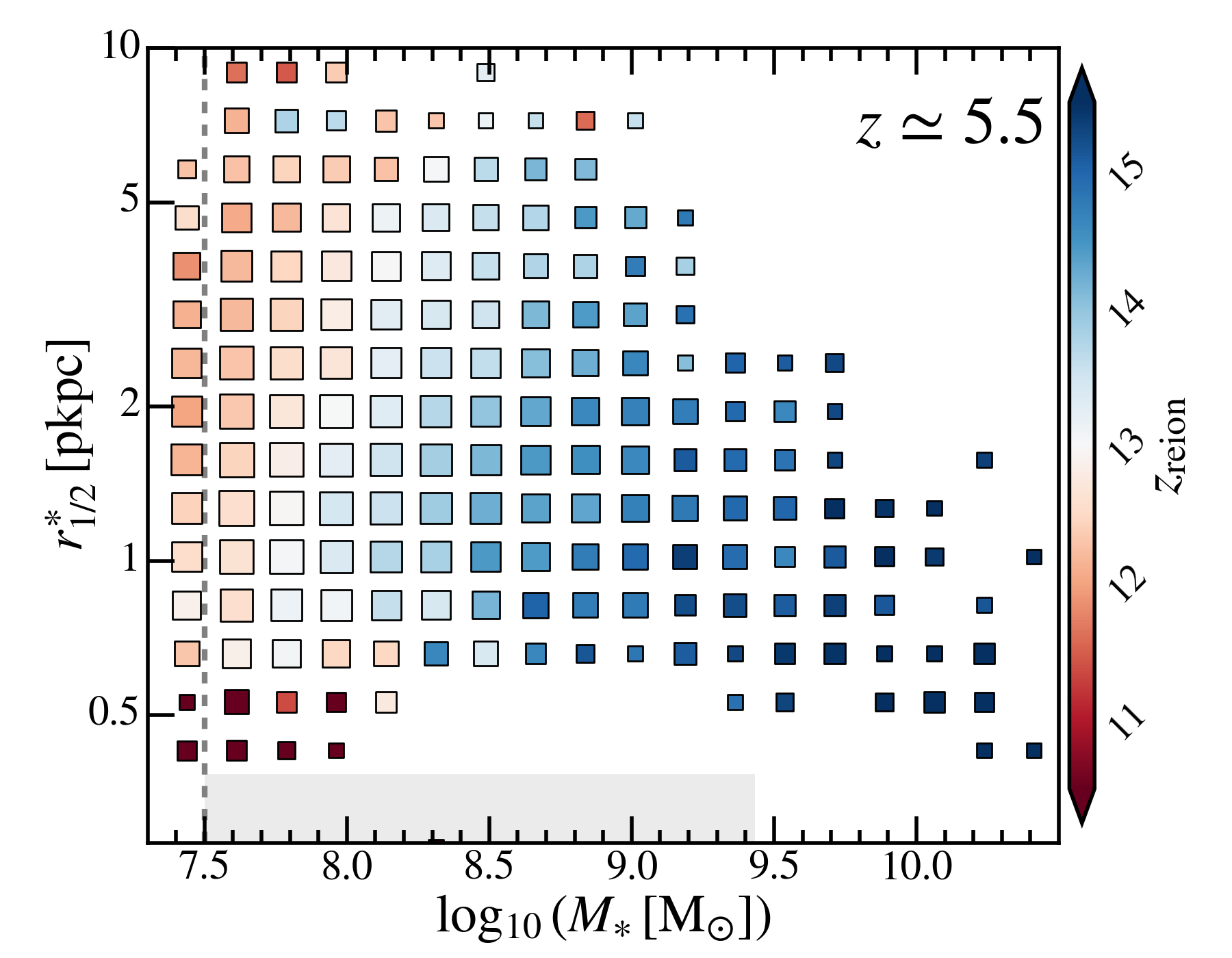}
    \caption{Galaxy intrinsic size--mass relation at $z\simeq 5.5$ color-coded by the matter overdensity (top), and redshift of reionization (bottom) in the nearby environment of galaxies. The notations and plotting styles are the same as Figure~\ref{fig:size-mass-correlation}. To the first order, galaxies with larger stellar masses tend to reside in overdense regions, and their local environments are reionized earlier. Nevertheless, at fixed stellar mass, galaxies in regions that are underdense or have earlier reionization tend to be more compact. However, a population of compact low-mass galaxies resides in extremely dense regions, and their sizes could be significantly affected by environmental effects.}
    \label{fig:size-mass-correlation-env}
\end{figure}

\subsection{Correlation of size with galaxy properties}
\label{sec:correlation}

To understand the physical driver of galaxy size evolution found above, we will first check the correlation of $r^{\ast}_{1/2}$ with different galaxy properties. In Figure~\ref{fig:size-mass-correlation}, we show the intrinsic size--mass relation of galaxies at $z\simeq 6$ in the \thesan simulations. The sizes of the square markers are proportional to the logarithm of galaxy number densities in the size--mass plane. The colors map various galaxy properties, including (1) the relative positions of galaxies to the SFMS (top left), (2) the accretion rates of SMBHs normalized by galaxy SFR (top right), (3) the halo gas fraction normalized by the universal baryon fraction ($f_{\rm b}\equiv \Omega_{\rm b}/\Omega_{\rm m}$; middle left), (4) the halo spin (middle right), (5) the Lyman continuum escape fraction ($f_{\rm esc}$; bottom left), and (6) the gas disk fraction (bottom right). We remind the readers about definitions of relevant galaxy properties provided in Table~\ref{tab:prop}. In addition to them, $f_{\rm esc}$ is defined as the ratio between the number of ionizing photons that escape and are intrinsically emitted from all sources within the virial radius. This is calculated using the method outlined in \citet{Yeh2023}. We will focus on \textbf{at fixed stellar mass}, how these secondary parameters affect galaxy sizes. We note that typically in the massive end, the results are affected by small statistics of massive galaxies in \thesanone and should therefore be interpreted with caution.

In the top left panel, we find that galaxies at fixed stellar mass below the SFMS tend to be more compact than the ones above the SFMS. The extreme version of this is the compact quenched galaxies at the massive end, which appear smoothly connected to the compact star-forming population. Similar trends are found at redshift up to $\simeq 10$. This indicates a potential causal correlation between compaction and quenching of galaxies. This is supported by the trends in the middle left panel, where compact galaxies show lower halo gas mass fractions. In one way, the compact sizes of galaxies could be driven by concentrated starbursts from rapid gas inflow and depletion. Meanwhile, the stellar and AGN feedback associated with the central starburst can drive gas outflows that further decrease the gas abundance and eventually lead to the quenching of galaxies. The causal connection of these signatures will be explored in Section~\ref{sec:physical-driver} below.

In the top right panel, we find that compact galaxies at $M_{\ast}\gtrsim 10^{8}\msun$ tend to have larger SMBH accretion rates with respect to the galaxy SFR. The SMBH accretion and the concentrated starburst are likely sustained by the same gas inflow towards the centre of the galaxies. We will show a clear relationship between compaction and the ignition of AGN activity in Section~\ref{sec:physical-driver:evo} below. 

In the middle right panel, we find that galaxy sizes show a positive correlation with halo spin, which is expected from the classical disk formation theory~\citep[e.g.][]{Mo1998}. The role of halo spin in determining galaxy size will be investigated in detail in Section~\ref{sec:spin} below. As supporting evidence, in the bottom right panel, we find massive and compact galaxies display prominent disk morphology with more than half of their gas content in a co-rotating component (in terms of angular momentum alignment and within the $2\,r^{\ast}_{1/2}$ aperture as defined in Table~\ref{tab:prop}). More massive and more compact galaxies tend to have higher gas disk fractions. A similar trend exists also for the disk fraction of the stellar component. These findings are qualitatively consistent with the picture of disk instability-driven compaction, as will be discussed in Section~\ref{sec:discussion:instability}.

In the bottom left panel, we find that the most compact galaxies tend to have orders of magnitude higher Lyman continuum escape fractions than extended galaxies at the same stellar mass. This is consistent with earlier findings in simulations that the gas outflow driven by concentrated starbursts can leave young massive stars (temporarily) transparent to neutral gas absorption and act as the main channel of ionizing photon escape~\citep[e.g.][]{Smith2019,Ma2020,Barrow2020,Rosdahl2022}. 

\subsection{Correlation of size with environments}
\label{sec:correlation-env}

With the on-the-fly radiative transfer method, the \thesan simulations can self-consistently track the ionization states of the intergalactic medium around galaxies and enable detailed studies of the relationship between galaxy formation and large-scale environments~\citep[e.g.][]{Borrow2022, Garaldi2022}. We utilize the so-called Cartesian outputs of the \thesan simulations~\citep{Kannan2022,Garaldi2023} at $z\simeq 5.5$, which save a subset of gas/dark matter properties on a 3D Cartesian grid. This includes the total matter overdensity ($\delta$) and the redshift of reionization ($z_{\rm reion}$), calculated following the method described in \citet{Garaldi2022,Neyer2023} and Zhao et al. 2024 in prep.. For each halo, we define $z_{\rm reion}$ as the latest time at which the (mass-weighted) ionized fraction of the gas in the closest grid cell exceeds the threshold value $0.5$. For matter overdensity, we take the grid resolution of $512^3$ (cell size $\sim 200\,{\rm ckpc}$). For $z_{\rm reion}$, we take the grid resolution of $1024^{3}$ (cell size $\sim 100\,{\rm ckpc}$). The results have no significant dependence on the choice of grid resolution. For the results presented below, we adopt a smoothing length of $250\,{\rm ckpc}$ for relevant fields. Results smoothed at protocluster scale ($1\,{\rm cMpc}$) are presented in Appendix~\ref{appfig:env} for comparison. 

In Figure~\ref{fig:size-mass-correlation-env}, we show the size--mass relation at $z\simeq 5.5$ color-coded by variables related to the environment in which galaxies reside. In the top panel, we show the impact of local matter overdensity. There is a clear leading-order dependence of $\delta$ on galaxy stellar mass. More massive galaxies hosted by more massive haloes tend to reside in denser regions. Nevertheless, \textbf{at fixed stellar mass}, more compact galaxies are more likely to be found in underdense regions while diffuse galaxies preferentially stay in overdense regions. Mergers and gas accretion (associated with denser environments) could play an important role in building extended stellar content of galaxies. This is supported by the larger gas fractions and perturbed morphology of these galaxies as seen in Figure~\ref{fig:size-mass-correlation}. On the other hand, internal physical processes are likely responsible for the formation of compact galaxies. The exception is the population of extremely compact galaxies at $M_{\ast}\gtrsim 10^{7.5}-10^{8.5}\msun$, which resides in extremely overdense regions. As shown in Figure~\ref{appfig:central}, these galaxies are mostly satellite galaxies and are likely affected by the tidal fields and ram pressure of their hosts. We also note that the impact of matter overdensity shown here is insensitive to the choice of smoothing length. Similar results are found when using a larger $1{\rm cMpc}$ smoothing length as shown in Appendix~\ref{appfig:env}.

In the bottom panel, we show the dependence on the redshift of reionization of gas in the environment. Smoothed over the fiducial $250\,{\rm ckpc}$ scales, there is a strong leading-order dependence that gas around more massive galaxies is reionized earlier in cosmic time. On top of this, \textbf{at fixed stellar mass}, the gas around more compact galaxies tends to experience earlier reionization. The physical interpretation is that, at $250\,{\rm ckpc}$ scales, local radiation sources and gas budget start to be more relevant for reionization. Massive galaxies that produce more ionizing photons reionize their nearby environment earlier. And \textbf{at fixed stellar mass}, extended galaxies tend to be richer in gas (as seen in Figure~\ref{fig:size-mass-correlation}) and it will take longer time to reionize their local environment given roughly the same amount of ionizing photon budget. We note that, unlike the case of matter overdensity, the dependence of $z_{\rm reion}$ does change when using a larger smoothing length as shown in Appendix~\ref{sec:app-env}. At cMpc scales, $z_{\rm reion}$ exhibits a tight positive correlation with matter overdensity. Denser regions reionize earlier due to larger amount of ionizing radiation sources~\citep{Garaldi2022}. Therefore, the $z_{\rm reion}$ in the size-mass plane closely follows the trend of overdensity as shown in Appendix~\ref{appfig:env}. As a consequence, at fixed stellar mass, $z_{\rm reion}$ at cMpc scales is lower around more compact galaxies since they reside in underdense environments. The trend is opposite to what has been found for $z_{\rm reion}$ on 250 ckpc scales.

As a summary of the size correlations, in Table~\ref{tab:correlation_coeff}, we present the Pearson and Spearman rank-order correlation coefficient between $r^{\ast}_{1/2}$ and the galaxy or environmental properties discussed above. We perform the measurements around two stellar masses $10^{8}$ and $10^{9}\msun$. In both mass bins, at least $300$ galaxies are included for measuring the correlation coefficients. Out of all the variables considered, the gas fraction has the strongest correlation with galaxy size, followed by halo spin, sSFR, and gas disk fraction. The overdensity shows a strong correlation only at the massive end because of the tip of compact satellite galaxies at the low-mass end. The correlation for overdensity is not sensitive to the choice of the smoothing length. However, for $z_{\rm reion}$, opposite correlations are found when using a smoothing length of 250 ckpc versus 1 cMpc. As discussed above, this reflects a transition from large-scale reionization, which is more affected by overdensity and the amount of external radiation sources, to small-scale reionization, which is more sensitive to local sources and gas abundance. These correlations suggest that (1) a causal connection could exist between compaction, gas depletion, and quenching, (2) external perturbations are likely not the driver for compact but instead generally make galaxies puffier, (3) internal processes related to the disk morphology are likely the trigger of compaction. These aspects will be discussed in more detail in Section~\ref{sec:physical-driver}.

\begin{table}
    \addtolength{\tabcolsep}{-1pt}
    \renewcommand{\arraystretch}{1.2}
    \centering
    \caption{Summary of the correlation of intrinsic size ($r^{\ast}_{1/2}$) with various galaxy/environment properties (in log-log space). The Pearson ($\alpha_{\rm p}$) and Spearman rank-order ($\alpha_{\rm s}$) correlation coefficients are evaluated based on galaxies with stellar mass $\log_{10}{(M_{\ast}/\msun)}=9\pm 0.3$ or $\log_{10}{(M_{\ast}/\msun)}=8\pm 0.3$. We sort the properties based on the amplitudes of the Spearman correlation coefficients at $M_{\ast} \sim 10^{9}\msun$. For overdensity and redshift of reionization, we show results when the smoothing length of fields is chosen to be $250$ ckpc (fiducial) and 1 cMpc, respectively. Overall, the strongest correlation shows for the gas abundance, followed by halo spin, local matter overdensity, and sSFR with respect to the main sequence.}
    \begin{tabular}{lccccr}
        \hline
        Properties & $\alpha^{9}_{\rm p}$ & $\alpha^{9}_{\rm s}$ & $\alpha^{8}_{\rm p}$ & $\alpha^{8}_{\rm s}$ & Ref.  \\
        \hline 
        \hline
        \textbf{Galaxy:} & \\
        $f_{\rm gas}/f_{\rm b}$ & 0.689 & 0.756 & 0.451 & 0.597 & Figure~\ref{fig:size-mass-correlation}\\
        $\lambda^{\prime}$ & 0.347 & 0.393 & 0.289& 0.298 & Figure~\ref{fig:size-mass-correlation}\\
        ${\rm sSFR}/{\rm sSFR}_{\rm MS}$ & 0.366 & 0.388 & 0.269 & 0.285 & Figure~\ref{fig:size-mass-correlation}\\
        $f^{\rm gas}_{\rm disk}$ & $-0.363$ & $-0.322$ & $-0.417$ & $-0.425$ & Figure~\ref{fig:size-mass-correlation}\\
        $\dot{M}_{\rm BH}/{\rm SFR}$ & $-0.304$ & $-0.319$ & $-0.293$ & $-0.317$ & Figure~\ref{fig:size-mass-correlation}\\
        $f_{\rm esc}$ & 0.238 & 0.265 & $-0.289$ & $-0.251$ & Figure~\ref{fig:size-mass-correlation}\\
        \hline 
        \textbf{Environment:} & \\
        $\delta$ (250 ckpc) & 0.297 & 0.388 & 0.026 & 0.209 & Figure~\ref{fig:size-mass-correlation-env}\\
        $\delta$ (1 cMpc) & 0.283 & 0.325 & 0.092 & 0.114 & Figure~\ref{appfig:env}\\
        $z_{\rm reion}$ (250 ckpc) & -0.256 & -0.255 & -0.105 & -0.210 & Figure~\ref{fig:size-mass-correlation-env}\\
        $z_{\rm reion}$ (1 cMpc) & 0.164 & 0.208 & 0.140 & 0.106 & Figure~\ref{appfig:env}\\
        \hline
    \end{tabular}
    \label{tab:correlation_coeff}
    \renewcommand{\arraystretch}{0.9090909090909090909}
\end{table}

\subsection{Comparison to analytical model predictions}
\label{sec:spin}

\begin{figure}
    \centering 
    \includegraphics[width=0.95\linewidth, clip, trim={0.5cm 0 0 0}]{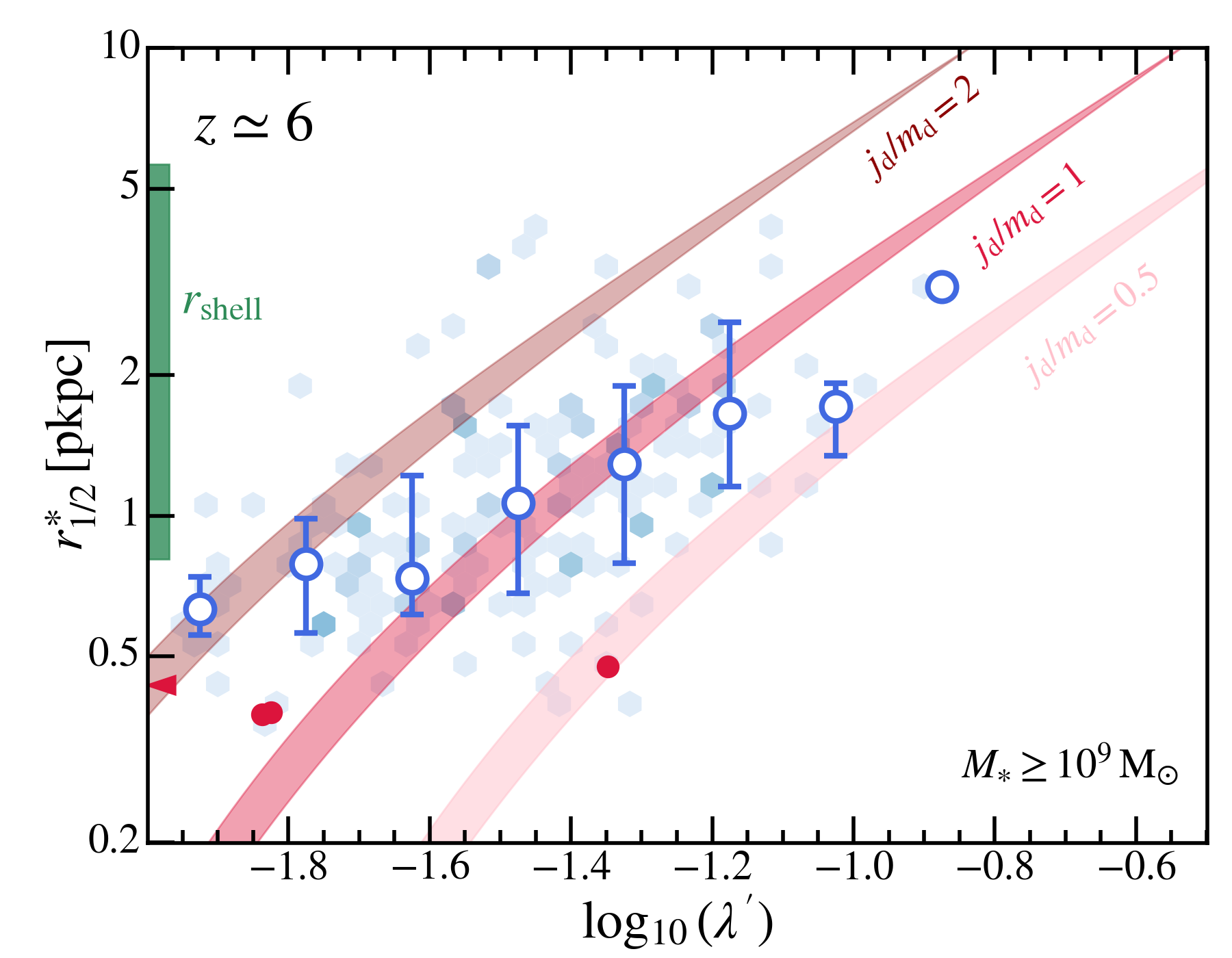}
    \includegraphics[width=0.95\linewidth, clip, trim={0.5cm 0 0 0}]{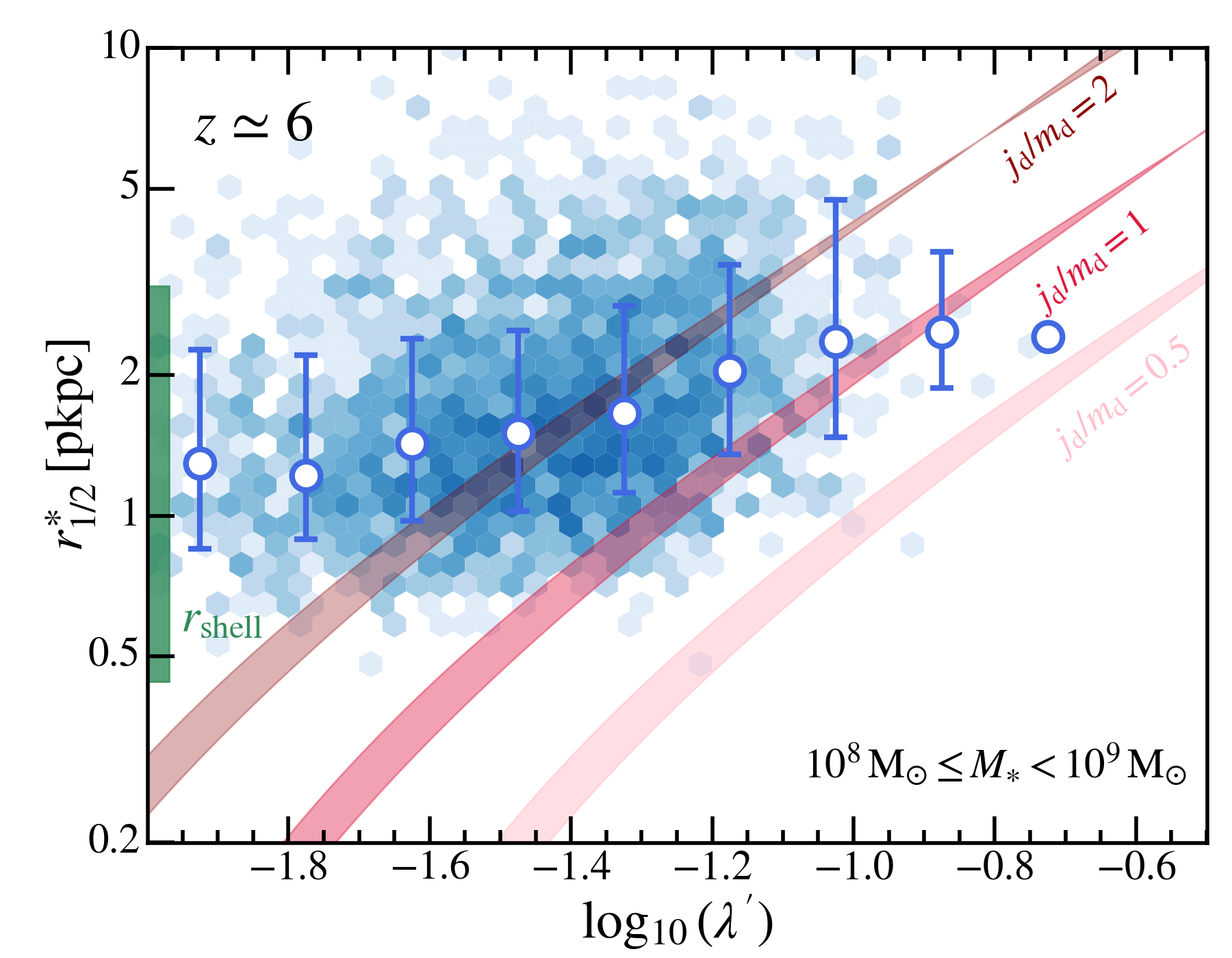}
    \includegraphics[width=0.95\linewidth, clip, trim={0.5cm 0 0 0}]{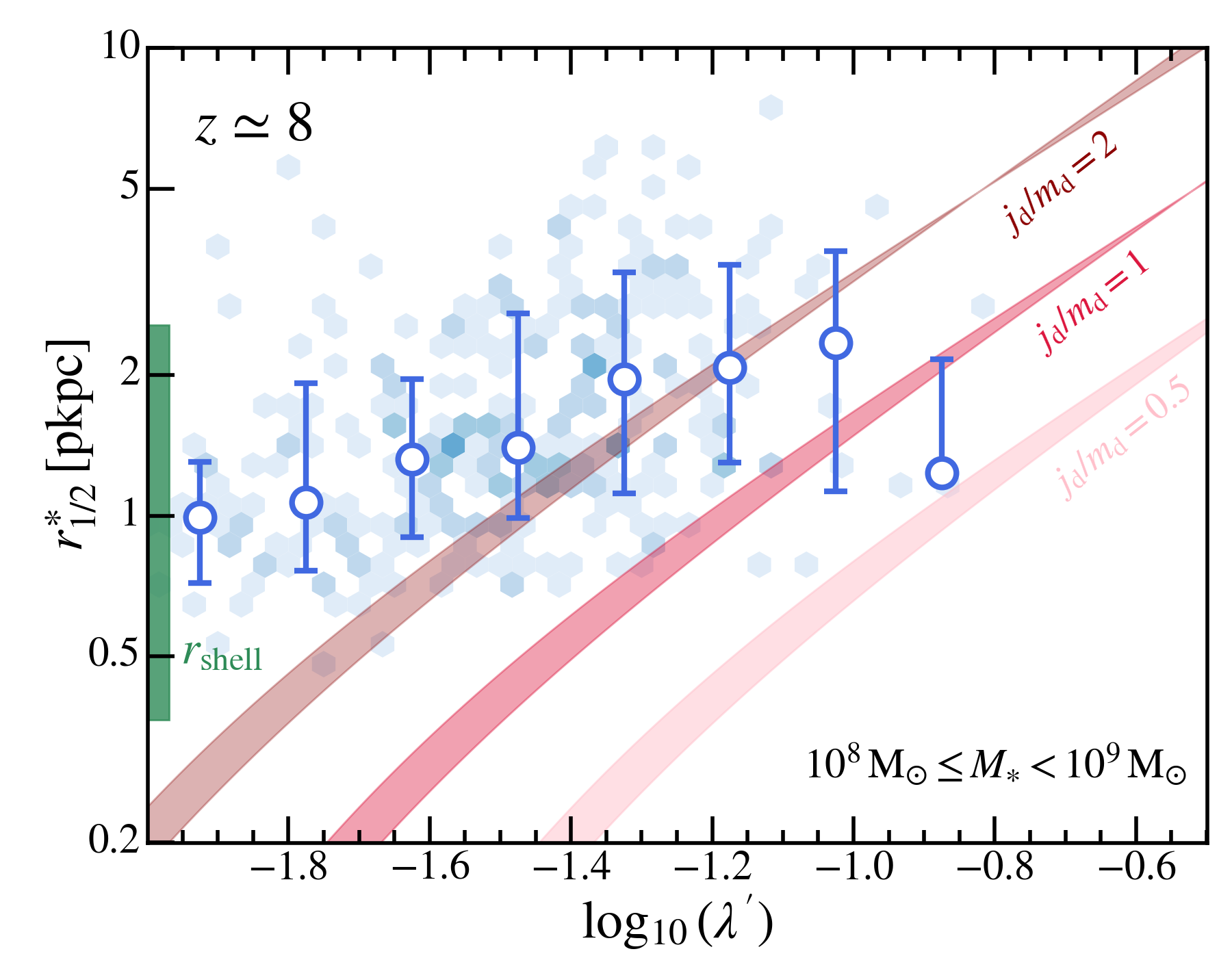}
    \caption{Galaxy intrinsic size versus halo spin at $z=$ 6 and 8 in \thesanone. In each panel, the simulation results are shown by the blue contour with circles and error bars indicating the median relation and 1-$\sigma$ scatter. The colored bands are predictions of an analytical disk formation theory, assuming $m_{\rm d}=0.01-0.05$ and $j_{\rm d}/m_{\rm d}$ values as labelled. The vertical green band shows the prediction from a spherical shell model. In the top two panels, we show the results in mass bin $M_{\ast}\geq 10^{9}\msun$ and $10^{8}\msun\leq M_{\ast} \leq 10^{9}\msun$ at $z=6$. In the bottom panel, we show the results in the mass bin $10^{8}\msun\leq M_{\ast} \leq 10^{9}\msun$ at $z=8$ (the number of galaxies in the high-mass bin at this redshift is not sufficient enough to form reliable results).}
    \label{fig:size-spin}
\end{figure}

\subsubsection{Disk formation theory}

The correlation of galaxy size with halo spin we see in Figure~\ref{fig:size-mass-correlation} is expected from the classical theory of disk formation~\citep[e.g.][]{Mo1998,Kravtsov2013}. In this theory, the half-mass radius of the stellar disk is predicted to be
\begin{equation}
    \dfrac{r^{\ast}_{1/2}}{R_{\rm vir}} = 1.678\,\dfrac{R_{\rm d}}{R_{\rm vir}} =\dfrac{1.678}{\sqrt{2}}\,\left(\dfrac{j_{\rm d}}{m_{\rm d}}\lambda\right)\,\dfrac{f_{\rm R}(\lambda, j_{\rm d}, m_{\rm d}, c)}{\sqrt{f_{\rm c}(c)}} \, ,
    \label{eq:rdisk}
\end{equation}
where $R_{\rm d}$ is the scale length of the exponential disk, $j_{\rm d}$ ($m_{\rm d}$) is the angular momentum (mass) ratio of the stellar disk with respect to the host halo, $f_{\rm c}$ and $f_{\rm R}$ are functions defined in \citet{Mo1998}, and the virial radius $R_{\rm vir}$ of a halo has been defined in Table~\ref{tab:prop}. We also note the difference between the two definitions of spin ($\lambda$ and $\lambda^{\prime}$) as discussed in Table~\ref{tab:prop}. If we assumed that the baryons and dark matter initially share the same distribution of specific angular momentum~\citep{Fall1980} and the disk formation process conserves the angular momentum~\citep{Mestel1963} (i.e. $j_{\rm d}/m_{\rm d}\sim 1$), the disk size will be mainly determined by the DM halo properties. In practice, the disk mass fraction $m_{\rm d}$, which enters $f_{\rm R}$, has little impact on the disk size. The halo concentration also has a mild influence on galaxy size. According to the canonical halo mass-concentration relations~\citep[e.g.][]{Wechsler2002, Dutton2014, Diemer2019}, high-redshift haloes have $c\simeq 4$ at $z\gtrsim 6$ with a weak dependence on halo mass. On the other hand, the halo spin parameter $\lambda$ was thought to be the dominant factor for disk sizes~\citep[e.g.][]{Somerville2008, Guo2011, Benson2012, Somerville2018}. %In Figure~\ref{fig:size-mass-correlation}, we have found the same qualitative trend that at fixed stellar mass, haloes with higher spin tend to host larger galaxies.

To quantitatively compare galaxy sizes in the \thesan simulations with the disk formation theory predictions, in Figure~\ref{fig:size-spin}, we show $r^{\ast}_{1/2}$ versus halo spin $\lambda^{\prime}$ (as defined in Table~\ref{tab:prop}) in different stellar mass bins in \thesanone. First, the spin parameter distribution from the simulation is consistent with previous studies using N-body simulations that $\lambda^{\prime}$ follows a log-normal distribution with median value $\sim 0.035$ and 1-$\sigma$ scatter $\sim 0.5$ dex~\citep[e.g.][]{Bullock2001,vandenBosch2002,Maccio2007}. We also show the predicted $r^{\ast}_{1/2}$ from disk formation theory, given the median halo mass of galaxies in the stellar mass bin and assuming $m_{\rm d}$ in the range $0.01-0.05$, $j_{\rm d}/m_{\rm d}=0.5, 1$ or $2$.

For galaxies with $M_{\ast}\geq 10^{9}\msun$, they on the bulk agree with the disk formation model with conserved specific angular momentum ($j_{\rm d}/m_{\rm d}=1$). Galaxies with lower halo spin have smaller sizes. However, we need to note that, within this mass range at fixed halo spin, more massive galaxies tend to have smaller sizes and agree better with model predictions with lower $j_{\rm d}/m_{\rm d}$. This could be due to many non-adiabatic processes that can affect the angular momentum of gas fueling star-formation, which will be discussed in Section~\ref{sec:physical-driver}. The four compact quenched galaxies are typically hosted by low-spin haloes. 

\subsubsection{Spherical shell model}

However, galaxies with lower stellar masses ($10^{8}\msun \leq M_{\ast}< 10^{9}\msun$) display almost no dependence on $\lambda^{\prime}$ in the low-spin end and only a mild positive dependence in the high-spin end. In the low-mass, low-spin haloes, the geometrical assumption of a stellar disk may break down in the first place. The morphology of the galaxy can be severely disturbed by more frequent mergers and stronger impact of feedback given the shallower potential well of early low-mass haloes~\citep[e.g.][]{elBadry2018, Jiang2019}. 

An alternative proxy for galaxy size is the radius where feedback-driven outflow encounters the cold gas inflow from CGM and creates a shock. This will be referred to as the spherical shell model. The balance of ram pressure at the shock radius $R_{\rm sh}$ gives~\citep{Dekel2023,Li2023}
\begin{equation}
    \dfrac{R_{\rm sh}}{R_{\rm str}} = \sqrt{\dfrac{V_{\rm w}\,\dot{M}_{\rm out}}{4\,V_{\rm in}\,\dot{M}_{\rm in}}} \, ,
    \label{eq:rshell}
\end{equation}
where $V_{\rm in}$ is the gas inflow velocity approximated as the halo virial velocity $V_{\rm vir}$, $\dot{M}_{\rm in}$ is the inflow rate estimated to be ${\rm SFR}/{\rm SFE}$ (SFE is the galaxy-averaged star-formation efficiency, for which we take the empirical parameterization SFE($M_{\rm halo}$) outlined in \citealt{Shen2023,Shen2024}). $R_{\rm str}$ is the effective radius of accreting gas streams, which can be estimated as~\citep{Mandelker2018,Dekel2023}
\begin{equation}
    R_{\rm str} \simeq 1.66\,\lambda_{\rm s}\,\left(\dfrac{1+z}{10}\right)^{1/2}\,R_{\rm vir} \, ,
\end{equation} 
where $\lambda_{\rm s}$ is the contraction factor between the gas stream and dark matter filament and we vary it between $0.02 - 0.14$~\citep{Mandelker2018}. $V_{\rm w}$ is the wind velocity, which for the IllustrisTNG galaxy formation model adopted by the simulations, has an empirical dependence on halo mass as~\citep{Pillepich2018}
\begin{equation}
    V_{\rm w} = {\rm MIN}\left(V^{\rm min}_{\rm w}, \kappa_{\rm w} \times 110 \kms \left(\dfrac{M_{\rm halo}}{10^{12}\msun}\right)^{1/3}\,\left(\dfrac{H_{\rm 0}}{H(z)}\right)^{1/3}\right)\, ,
\end{equation}
where $\kappa = 7.4$ and $V^{\rm min}_{\rm w}=350 \kms$. $\dot{M}_{\rm out}$ is the outflow rate and is related to SFR via the mass loading factor $\eta$. In the IllustrisTNG model, at low-metallicities ($Z \ll Z_{\rm w,ref} = 0.002$),
\begin{equation}
    \label{eq:mass-loading}
    \eta \equiv \dot{M}_{\rm out}/{\rm SFR} = \dfrac{2}{V^2_{\rm w}}\,\bar{e}_{\rm w}\,N_{\rm SNII}\,E_{\rm SNII}\,(1-\tau_{\rm w}) \, ,
\end{equation}
where $\bar{e}_{\rm w} = 3.6$ and $\tau_{\rm w}=0.1$ are constants as defined in \citet{Pillepich2018}, $N_{\rm SNII}=0.0118\,\msun^{-1}$ is the number of core-collapse supernovae (SNII) per formed stellar mass, $E_{\rm SNII}=10^{51}\erg$ is the available energy per SNII. 

In Figure~\ref{fig:size-spin}, we show $R_{\rm sh}$ with the green band and the value agrees well with the size plateau reached by low-spin galaxies in the stellar mass bin $10^{8}\msun \leq M_{\ast}< 10^{9}\msun$ at both $z\simeq 6$ and $z\simeq 8$. At fixed halo mass, $R_{\rm sh}$ has no dependence on halo spin and only a mild dependence on redshift. This is consistent with the slow redshift evolution of galaxy sizes at the low-mass end we found in Section~\ref{sec:size-mass}. Meanwhile, Equation~\ref{eq:rshell}-\ref{eq:mass-loading} imply that $R_{\rm sh} \propto (V_{\rm w}\,\eta\,{\rm SFE})^{1/2}$ at fixed halo mass. This indicates the potentially strong influence of the feedback prescription on the sizes of low-mass galaxies. Since the wind prescription in the IllustrisTNG model adopted is tightly correlated with halo mass as well, this reduces to $R_{\rm sh} \propto {\rm SFE}^{1/2}$ is qualitatively consistent with our findings that galaxy size correlates with their position relative to the main sequence. 

\section{Physical drivers of compact galaxy formation}
\label{sec:physical-driver}

\subsection{Evolution of massive compact galaxies}
\label{sec:physical-driver:evo}

\begin{figure}
    \centering
    \includegraphics[width=1\linewidth, trim={0.5cm 1cm 0 0}, clip]{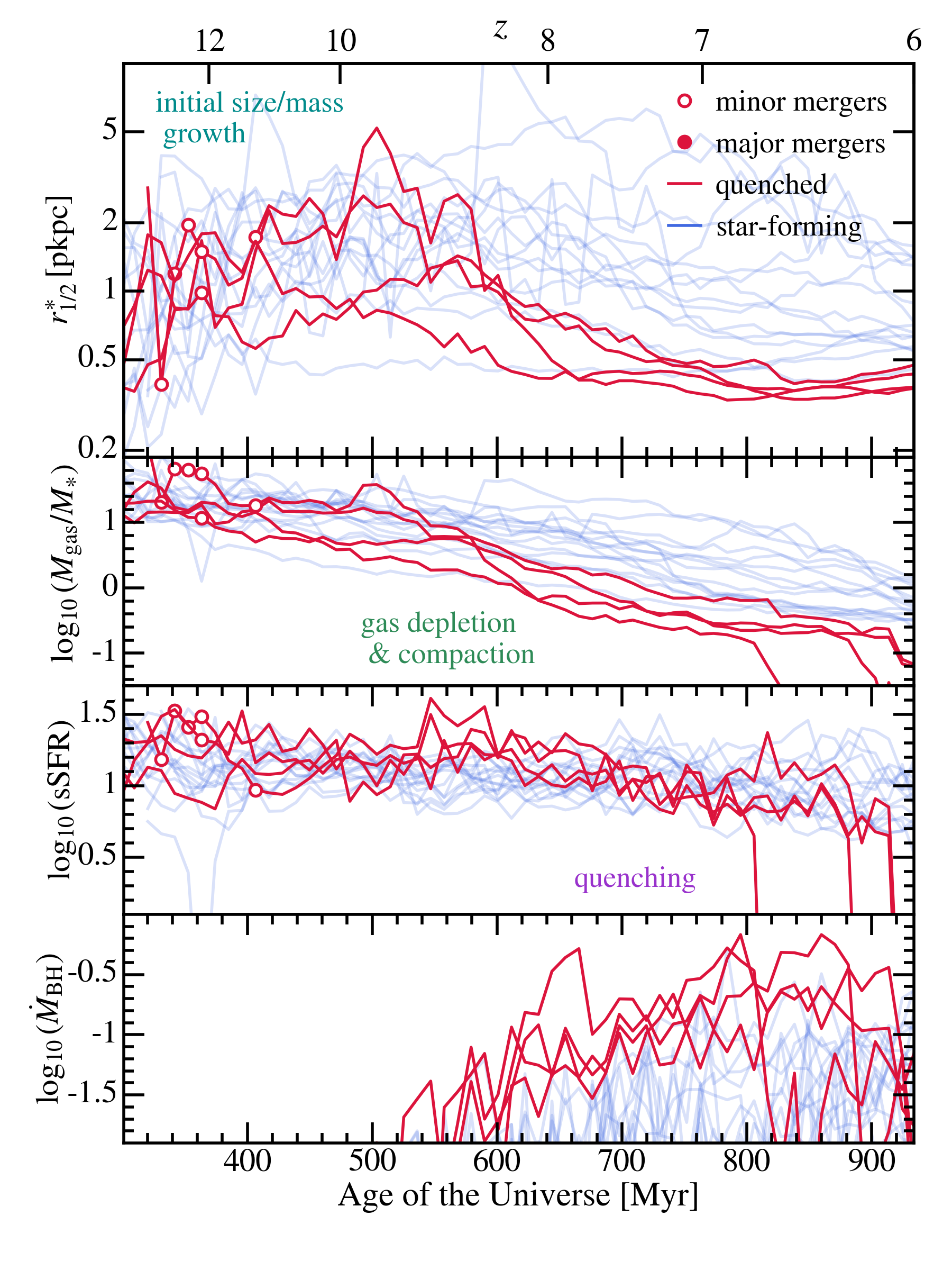}
    \caption{From top to bottom, we show the evolution of size, gas relative abundance ($M_{\rm gas}/M_{\ast}$), sSFR (in unit of $\Gyr^{-1}$), and SMBH accretion rate (in unit of $\msun\,{\rm yr}^{-1}$) of galaxies from $z\simeq 13$ to $z\simeq 6$. Galaxies are selected at $z\simeq 6$ with $M_{\ast}>10^{9.5}\msun$ and are divided into star-forming galaxies and quenched galaxies (see details in the main text). For each galaxy, we follow the main progenitors and document the merger history. The open and solid circles show the times when minor and major mergers happen for the selected quenched galaxies. The evolution of compact quenched galaxies can be divided into three phases, the initial growth in mass and size driven by (minor) mergers, the rapid compaction phase with gas depletion and concentrated star-formation, and the quenching phase.}
    \label{fig:trace-quenched}
\end{figure}

\begin{figure}
    \centering
    \includegraphics[width=1\linewidth, trim={1cm 0.5cm 0 0}, clip]{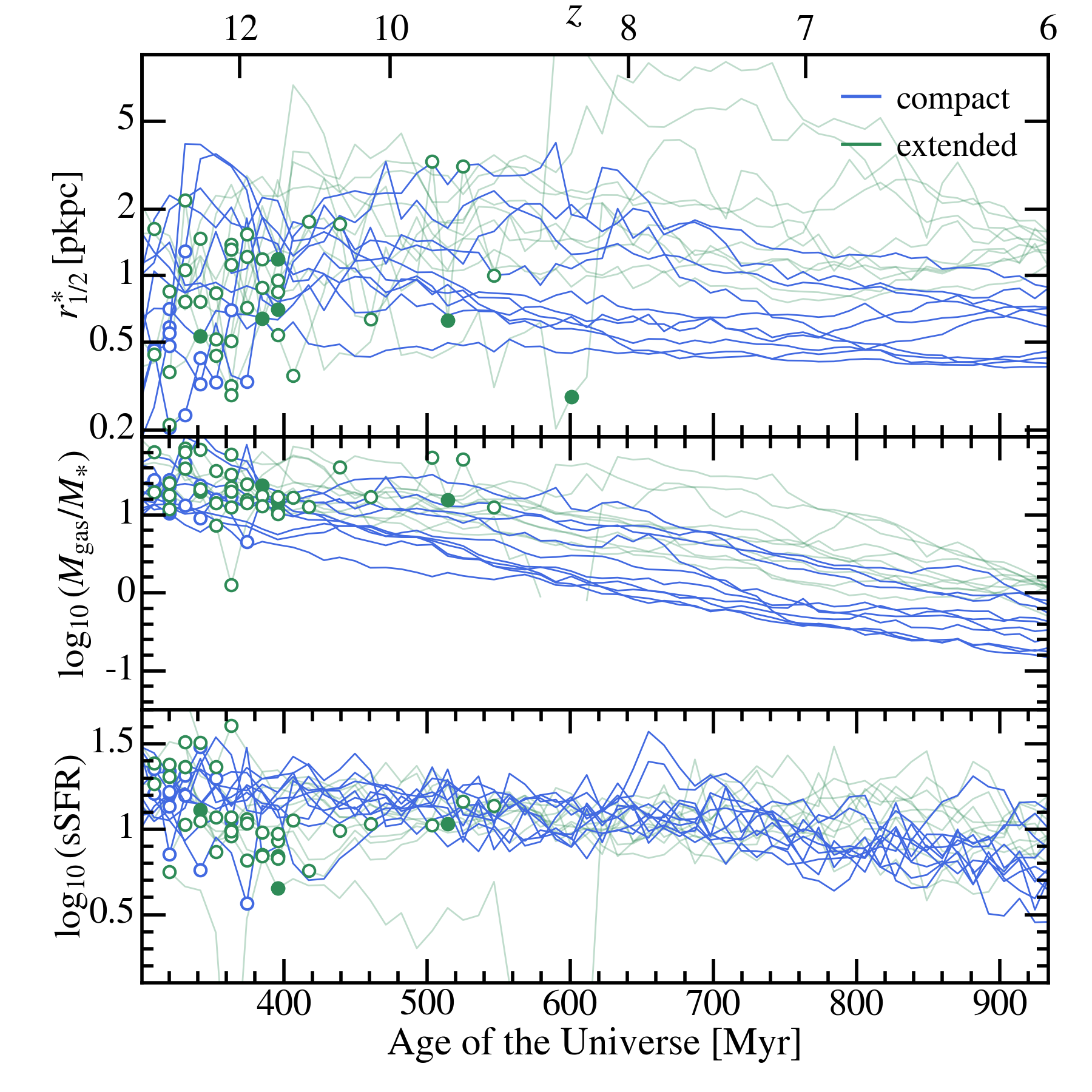}
    \caption{Similar to Figure~\ref{fig:trace-quenched}, here we show the evolution of star-forming galaxies divided into the compact  ($r_{\ast} < 1\pkpc$) and extended ($r_{\ast}\geq 1\pkpc$) populations at $z\simeq 6$. The notations are the same as Figure~\ref{fig:trace-quenched}. We find the compact group tends to have quieter merger histories early on. They later experience stronger gas depletion and have lower sSFR in the end compared to their extended counterparts.}
    \label{fig:trace-starforming}
\end{figure}

In this section, we explore the causal connections between galaxy sizes and various galaxy/environment properties found in previous sections. We will first investigate the most compact galaxies at the massive end and follow their evolution history. To achieve this, we subtract the main progenitor branches from the merger trees of galaxies constructed using the \textsc{LHaloTree} algorithm~\citep[described fully in the supplementary information of][]{Springel2005}. We track various galaxy properties and identify major (minor) mergers with a merging halo mass ratio threshold of $0.1$ ($0.01$). The analysis here focuses on massive well-resolved galaxies ($M_{\ast}>10^{9.5}\msun$) with unambiguous merger histories. We classify these galaxies into three categories: extended ($r_{\ast}\geq 1\pkpc$) star-forming galaxies, compact ($r_{\ast} < 1\pkpc$) star-forming galaxies, and quenched galaxies (all turn out to be compact). 

In the top panel of Figure~\ref{fig:trace-quenched}, we show the time evolution of the sizes of four quenched galaxies selected at $z\simeq 6$, compared to all the star-forming galaxies at a similar mass scale. In the bottom three panels, we show the time evolution of the galaxy gas abundance (normalized by stellar mass, $M_{\rm gas}/M_{\ast}$), sSFR, and SMBH accretion rates of the same galaxies, respectively. The major and minor mergers of the quenched galaxies are shown by solid and open circles. We identify three phases of the evolution of a compact quenched galaxy in the first Gyr of the universe. The later quenched population is initially indifferent from the star-forming galaxies with stellar mass and size growth dominated by minor mergers and smooth accretion. During this epoch, the baryon content of the galaxies is dominated by gas and galaxies grow in a self-similar fashion maintaining a stable sSFR. In the second phase beginning around $z\sim 10$, rapid gas depletion and compaction of the stellar content happen at the same time in these later quenched galaxies. Right after compaction starts, the sSFRs of these galaxies rise temporarily compared to the normal star-forming counterparts followed by a mild decline. During this phase, SMBH accretion gets activated at the same time as concentrated star-formation happens. None of these phenomena are associated with major or minor merger events, suggesting internal processes dominating the size evolution. In the final phase, the sSFR drops in quenched galaxies in an extremely short time scale ($\lesssim 10 \Myr$) and so does the SMBH accretion. The gas is completely depleted and the sizes of these quenched galaxies stabilize at $\lesssim 0.5\pkpc$ later on. 

In Figure~\ref{fig:trace-starforming}, we show the same type of evolution history for the star-forming galaxies at $z\simeq 6$, divided into compact and extended populations. Within the star-forming galaxies, we find a distinction between the compact and the extended populations and the trend is similar to what we found for quenched and star-forming galaxies above. Starting around $z\sim 10$, the compact star-forming galaxies have comparably lower gas abundance as their sizes deviate from the extended population, followed by declined sSFRs later on. These findings are broadly consistent with the trends we found in Figure~\ref{fig:size-mass-correlation} that compact galaxies tend to have lower sSFR and gas abundance. These compact star-forming galaxies might have gone through a similar gas depletion and compaction process to the quenched population but to a milder extent or at shifted timings due to lower masses or higher spins. Compared to the extended ones, the compact star-forming galaxies also have quieter merger histories, which is consistent with the underdense environments they reside in as found in Figure~\ref{fig:size-mass-correlation-env}. This again suggests that internal processes are responsible for the compaction of galaxies rather than external perturbations.

The rapid quenching of the few compact galaxies is likely driven by the rapid removal of cold gas reservoirs by stellar and AGN feedback. Evidences are shown in the gas images in Figure~\ref{fig:gas-image}, where the central gas distribution of the quenched galaxy is disturbed and displays shock-like features. However, we are not able to disentangle the relative contribution of stellar and AGN feedback in quenching. In addition, the development of a compact bulge can help stabilize the disk instability and drive morphological quenching~\citep[e.g.][]{Martig2009}. However, this may not completely shut down star-formation in the entire galaxy but only prevents mass inflow and central starburst. Other external mechanisms for quenching will operate on longer time scales than we observe in these simulations. For example, secular halo quenching can happen due to the development of a hot virialized inner CGM, which suppresses cold gas fueling the galaxy. These could be relevant for the mild decrease in sSFR in compact star-forming galaxies.

\begin{figure}
    \centering
    \includegraphics[width=1\linewidth]{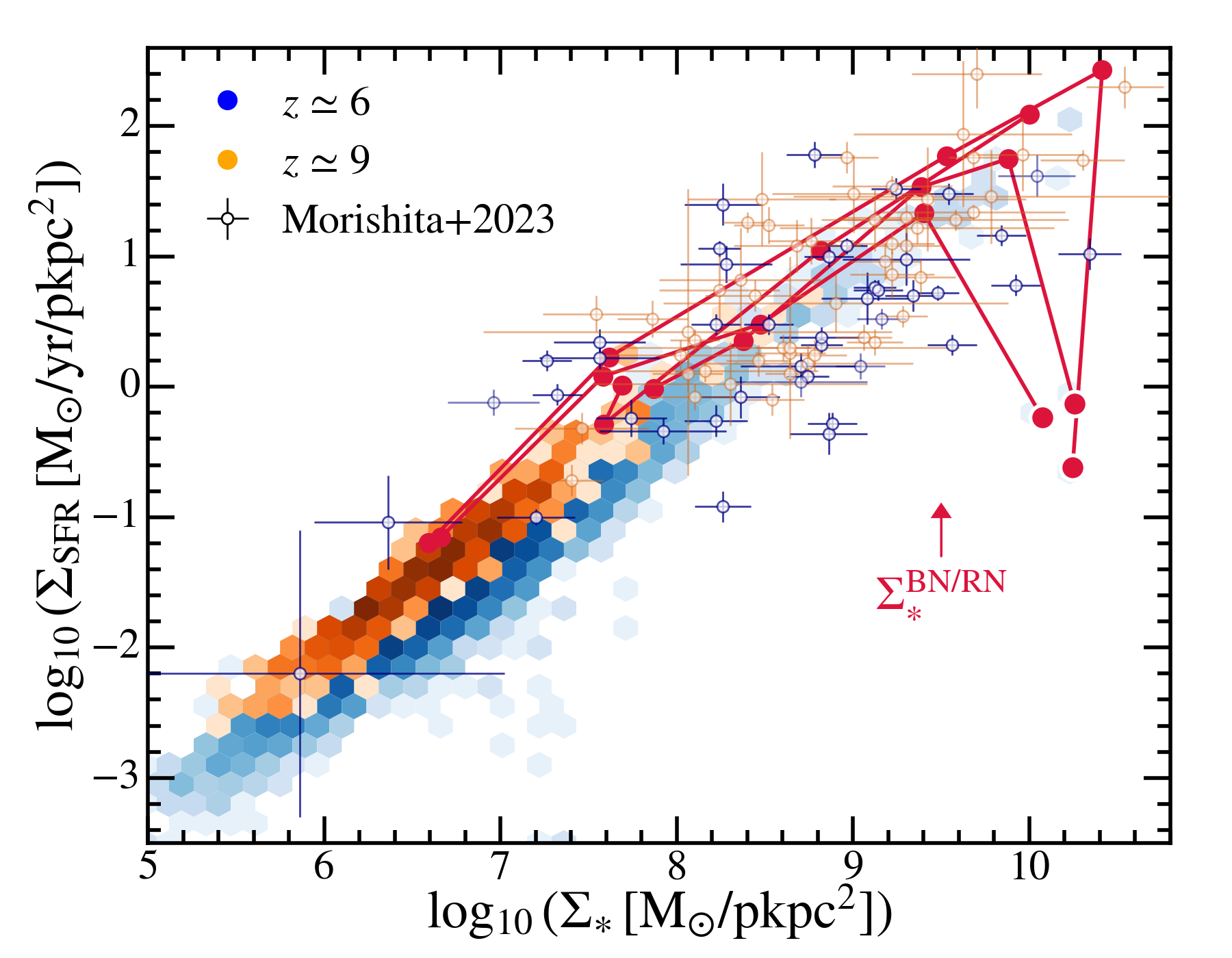}
    \caption{SFR surface density versus stellar mass density of galaxies in the \thesanone simulation compared to observational samples~\citep{Morishita2023}. We highlight the evolution trajectories of compact quenched galaxies from $z\simeq 12$ to $z\simeq 6$. Each marker is separated by $\Delta z = 1$. The arrow indicates the critical surface mass density threshold motivated by previous theoretical studies of massive compact galaxies (known as ``blue/red nuggets'', BN/RN) at cosmic noon.}
    \label{fig:trace-surface-density}
\end{figure}

\begin{figure*}
    \raggedright
    \rotatebox{90}{\hspace{0.5cm}\textbf{a quenched compact galaxy}}
    \includegraphics[width=0.24\linewidth]{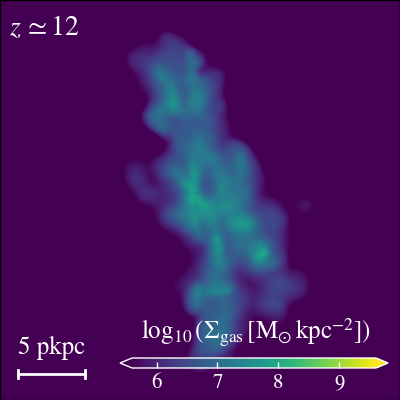}
    \includegraphics[width=0.24\linewidth]{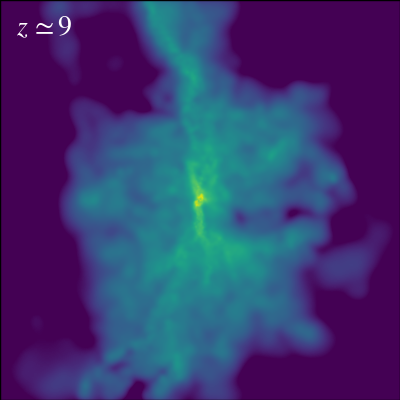}
    \includegraphics[width=0.24\linewidth]{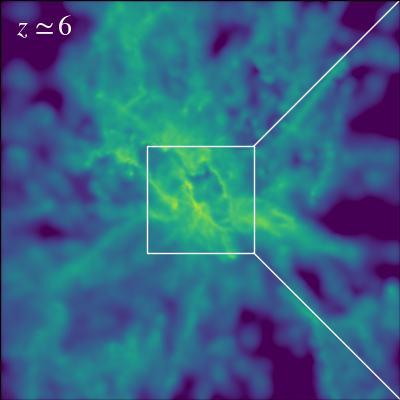}
    \hspace{-0.14cm}
    \includegraphics[width=0.24\linewidth]{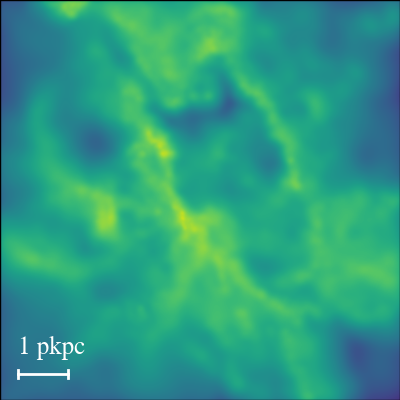} \\
    \rotatebox{90}{\hspace{0.3cm}\textbf{a star-forming extended galaxy}}
    \includegraphics[width=0.24\linewidth]
    {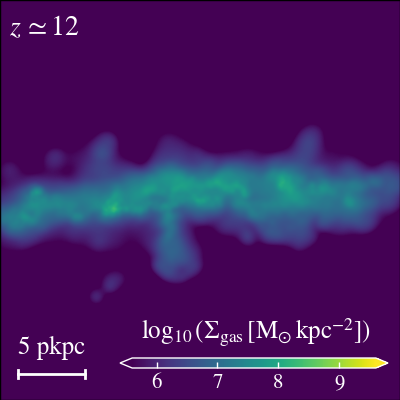}
    \includegraphics[width=0.24\linewidth]{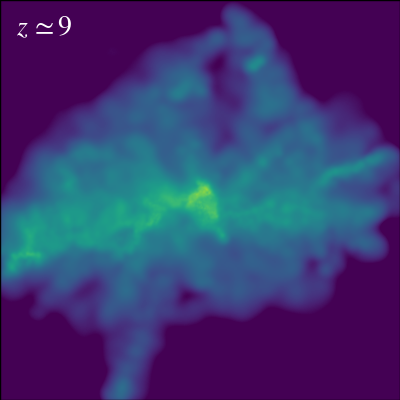}
    \includegraphics[width=0.24\linewidth]{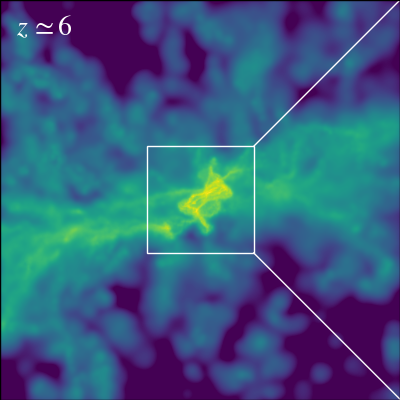}
    \hspace{-0.14cm}
    \includegraphics[width=0.24\linewidth]{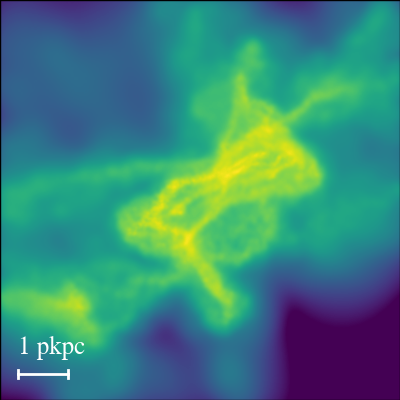}
    \caption{Surface density maps of the moderately cold ($T<3\times 10^{5}\,{\rm K}$) gas distribution around two simulated galaxies selected at $z\simeq 6$ and their main progenitors out to $z\simeq 12$. The first galaxy is a compact quenched galaxy while the second is an extended star-forming galaxy. The cold gas distributions in the CGM are dominated by filament-like gas streams. The star-forming galaxy has a central peak in the cold gas distribution while the cold gas distribution of the quenched one is disturbed by feedback at the centre.}
    \label{fig:gas-image}
\end{figure*}

In Figure~\ref{fig:trace-surface-density}, we show the SFR surface density ($\Sigma_{\rm SFR}\equiv {\rm SFR}/(2\,\pi\,r^{\ast\,\,\,2}_{1/2})$) versus stellar mass surface density ($\Sigma_{\ast}\equiv M_{\ast}/(2\,\pi\,r^{\ast\,\,\,2}_{1/2})$) of galaxies in the \thesanone simulation. They are compared to the observational constraints from \textit{JWST}~\citep{Morishita2023}. We highlight the evolution trajectories of the quenched galaxies from $z\simeq 12$ to $z\simeq 6$. $\Sigma_{\rm SFR}$ follows a tight correlation with $\Sigma_{\ast}$ with little redshift evolution and shows good agreement with the observed sample despite lower scatter. The later quenched galaxies grow their stellar mass following this correlation until reaching a stellar mass surface density threshold and become quenched afterward. This threshold is about $\Sigma_{\ast}\sim 10^{9} - 10^{10} \msun\,\pkpc^{-2}$, corresponding to $\Sigma_{\rm SFR}\sim 10^{1} - 10^{2} \msun\,{\rm yr}^{-1}\,{\pkpc}^{-2}$. This threshold happens to be consistent with the compaction and quenching phases discussed extensively in literature for galaxies at cosmic noon~\citep[e.g.][]{Dekel2009, Dekel2014, Zolotov2015, Tacchella2016, Lapiner2023} as well as the early proto-bulge formation found in observations~\citep[e.g.][]{Baggen2023,Baker2023}.

\begin{figure*}
    \raggedright
    \includegraphics[width=0.49\linewidth, clip, trim={0 0 0 0}]{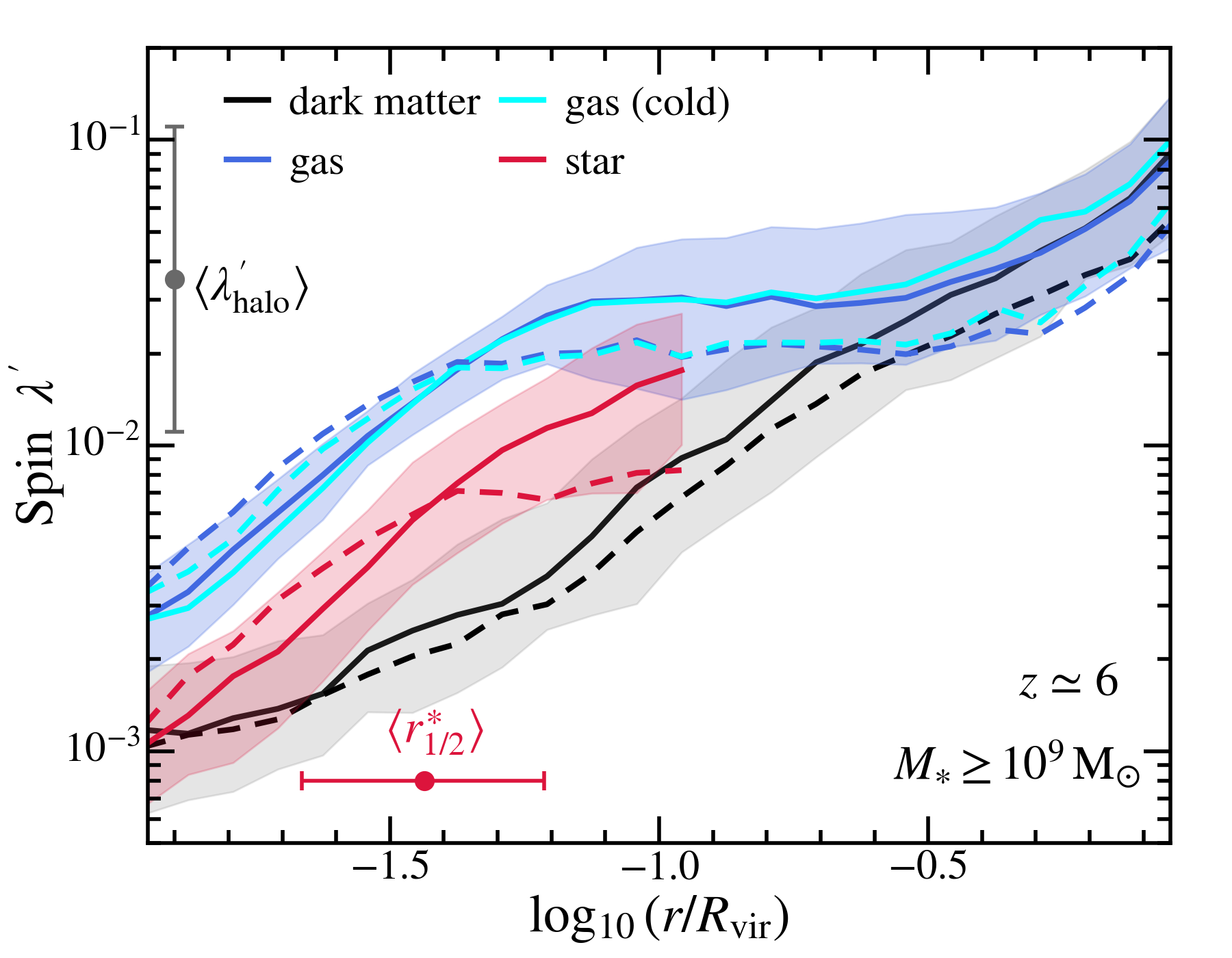}
    \includegraphics[width=0.49\linewidth, clip, trim={0 0 0 0}]{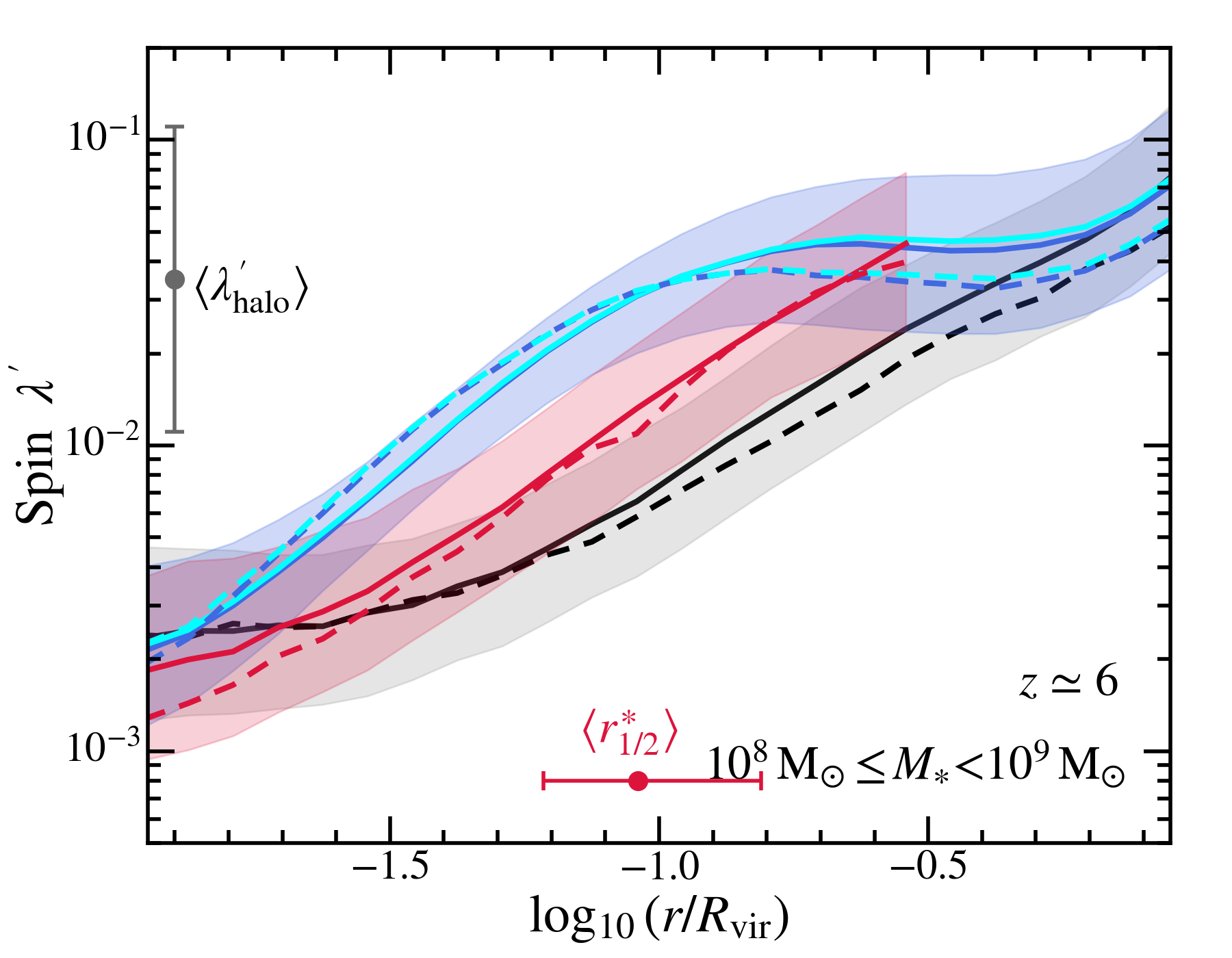}
    \includegraphics[width=0.49\linewidth, clip, trim={0 0 0 0}]{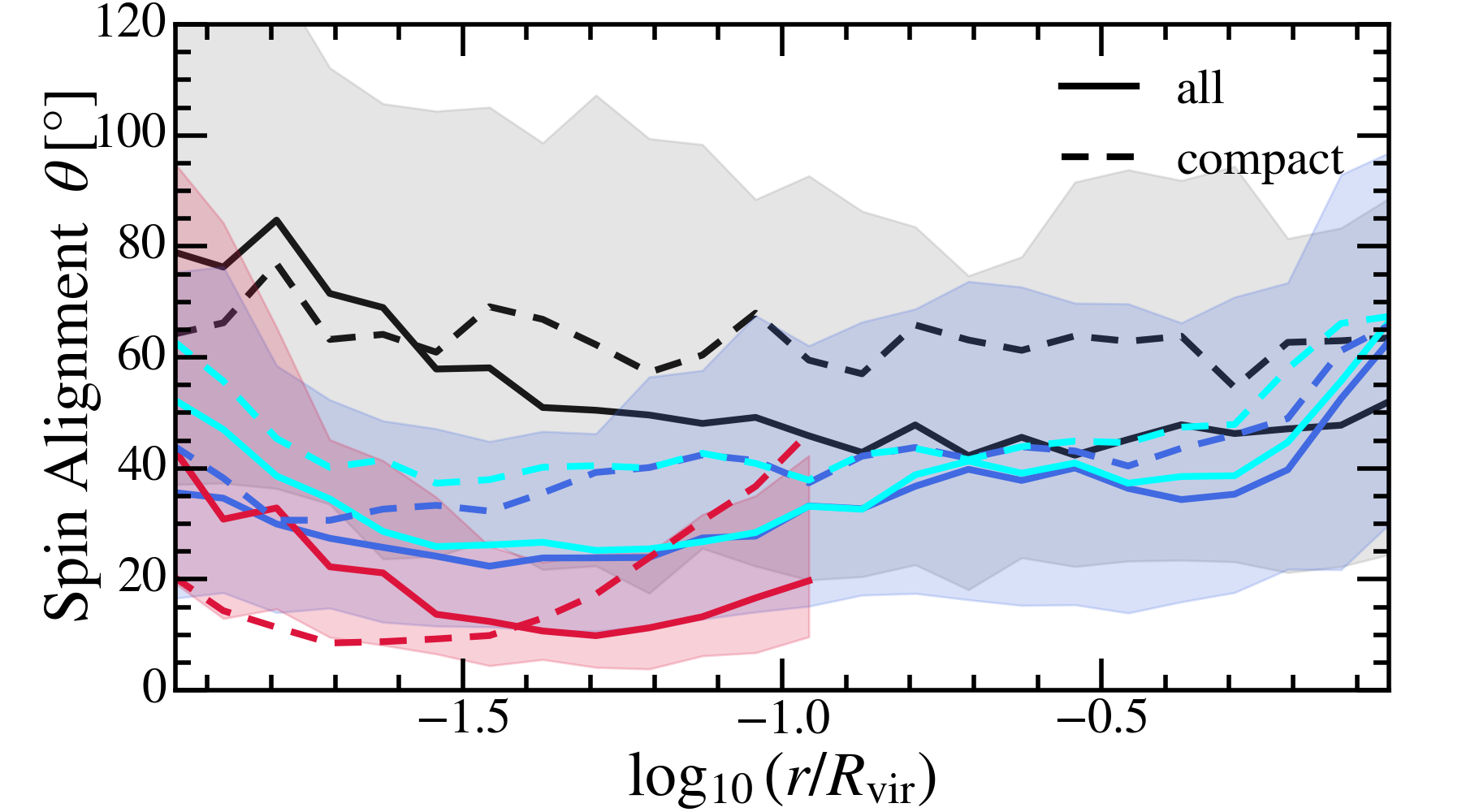}
    \includegraphics[width=0.49\linewidth, clip, trim={0 0 0 0}]{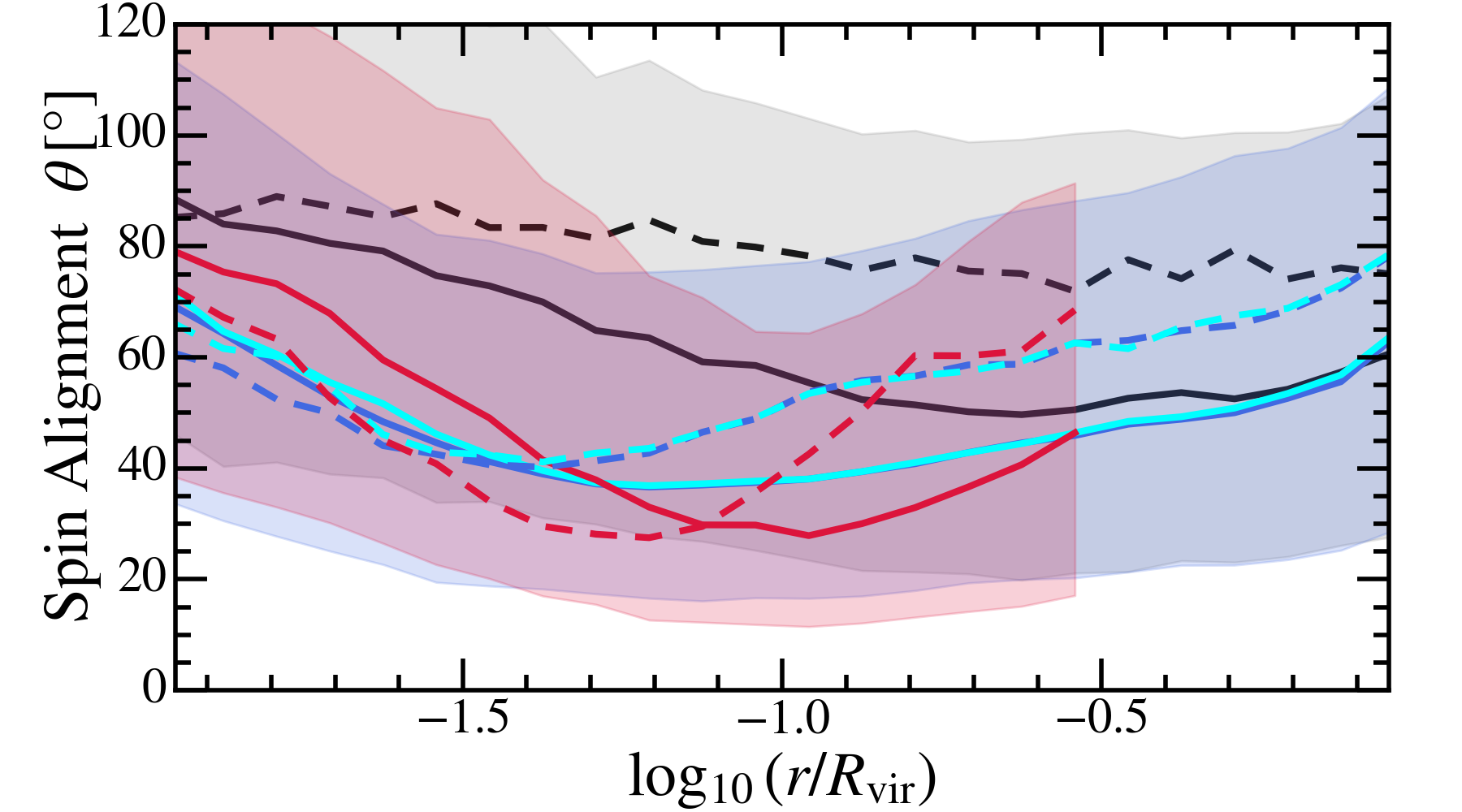}
    \caption{{\it Top}: Spin profile of dark matter, gas, and star at $z=6$ in \thesanone. We show the stacked results of galaxies in the mass bin of \textbf{$M_{\ast}\geq 10^{9}\msun$ (left)} and \textbf{$10^{8}\msun\leq M_{\ast}\leq 10^{9}\msun$ (right)}. We show the results of all galaxies in the mass range in solid and the compact ($r^{\ast}_{1/2}<1\pkpc$) subset of galaxies in dashed lines. The gray circle with the error bar indicates the median spin of dark matter haloes $\langle \lambda^{\prime}_{\rm halo}\rangle \simeq 0.035$. The red circle with the error bar indicates the median and 1-$\sigma$ scatter of $r^{\ast}_{1/2}$ of stacked galaxies. {\it Bottom}: Spin alignment of dark matter, gas, and star to the galaxy face-on direction defined based on stellar content (as defined in Table~\ref{tab:prop}). $\theta$ is the angle between the spin vector of each component with the galaxy face-on direction.}
    \label{fig:spin_profile}
\end{figure*}

\subsection{Gas angular momentum}
\label{sec:physical-driver:jgas}

To understand the physical reason for compaction, we need to start from the beginning when the gas fueling star-formation is accreted into the halo. The angular momentum of the gas has long been considered one of the most important factors for regulating galaxy sizes. We can gain insights into the time evolution of gas angular momentum by examining the specific angular momentum (or equivalently spin) profile of the halo. 

In Figure~\ref{fig:spin_profile}, we show the spin profile of galaxies in \thesanone at $z\simeq 6$ grouped in two stellar mass bins. The spin of a particle species $x$ is defined as $j_x/(\sqrt{2}\,V_{\rm vir}\,R_{\rm vir})$, where $j_x$ is the specific angular momentum of the corresponding particle species. This definition is chosen such that, for dark matter, it becomes equivalent to the halo spin defined in Table~\ref{tab:prop}. Here we consider the spin profile of dark matter, stars (restricted to $r< 2\,r^{\ast}_{1/2}$), all gas, and specifically cold gas ($T<10^{5}\,{\rm K}$). The spin of dark matter increases nearly linearly with radius and the central angular momentum has little correlation with the outskirts. Similar spin or specific angular momentum profiles have been found in previous cosmological $N$-body simulations~\citep[e.g.][]{Bullock2001,Bett2010} and is consistent with a toy model for halo buildup by tidal stripping of satellites as they spiral inward via dynamical friction~\citep{Bullock2001, Maller2002}. %Material in subhaloes that sinks deeper continuously transfers orbital angular momentum to the dark matter by dynamical friction and therefore has lower specific angular momentum.

In the classical theory, haloes acquire angular momentum from the gravitational tidal torques exerted by neighbouring density perturbations~\citep[e.g.][]{Hoyle1951,Peebles1969,Doroshkevich1970,White1984,Barnes1987} and through mergers~\citep[e.g.][]{Vitvitska2002, Maller2002}. Baryons were expected to initially follow the dark matter mass distribution and share the same specific angular momentum distribution. This motivates the classical disk formation theory described in Section~\ref{sec:spin}. In this picture, halo gas goes through a joint semi-spherical infall with dark matter from outside the halo before being heated to $T_{\rm vir}$ by a virial shock and becoming pressure-supported. The halo gas then cools down, contracts to the bottom of the halo potential while conserving angular momentum, and eventually forms a rotation-supported disk~\citep[e.g.][]{Fall1980,Mo1998,Bullock2001}. 

However, at high redshifts, halo gas accretion is dominated by the ``cold mode''~\citep[e.g.][]{Keres2005, Dekel2006, Keres2009, Dekel2009nature, Wetzel2015} driven by supersonic cold gas streams penetrating the halo directly to the central disk (often spatially associated with dark matter filaments) as well as gas clumps in merging subhaloes. The cold gas distribution shown in Figure~\ref{fig:gas-image} clearly demonstrates these features. Therefore, deviations of the specific angular momentum of gas and dark matter are expected~\citep[e.g.][]{Keres2009, Stewart2013, Nelson2015}. At the outskirt of the halo ($r\sim R_{\rm vir}$), the specific angular momentum of initially accreted cold gas can be larger than the dark matter due to stronger quadruple torques experienced by the gas~\citep{Danovich2015}, but this difference will be weak at high redshifts. In the \thesan simulations, we do find a small enhancement of the spin of cold gas compared to dark matter at the halo outskirts. Meanwhile, we also find that the gas spin is generally higher than that of the dark matter component at $r \lesssim 0.5\,R_{\rm vir}$ and often features a plateau-like feature until reaching $r \sim 0.1\,R_{\rm vir}$. This is consistent with many earlier studies~\citep[e.g.][]{Sharma2005,Kimm2011,Danovich2015,Teklu2015}. This is because the newly accreted cold gas has not yet mixed with the earlier accreted ones that have lower angular momentum and cluster in the central region due to cooling. The inflowing gas streams roughly conserve their angular momentum~\citep{Stewart2013, Danovich2015}. The plateau of gas spin happens to be similar to the halo average spin $\lambda^{\prime}$ even for the relatively compact population. This suggests that there is enough angular momentum and rotation support of accreted gas before reaching the vicinity of the galaxy, and it should be the physical processes within the galactic radius that drive the compaction of the stellar content.

In the messy region below the impact parameter of the streams, $\lesssim 0.1\,R_{\rm vir}$, tidal torques from the inner galactic structure as well as complex interactions of streams with other streams and feedback-driven outflows can substantially decrease the angular momentum of gas~\citep[e.g.][]{Ceverino2010,Danovich2015,Nelson2015}. Closer to the star-forming region, the spin of gas starts be aligned with the stellar disk. Further loss of gas angular momentum at smaller radii is driven by the disk instability~\citep[e.g.][discussed later in Section~\ref{sec:discussion:instability}]{Dekel2009,Dekel2014}. These processes together result in the decline of gas spin profile at small galactocentric radii, which can be substantially smaller than the overall averaged halo spin. Disk instability is more likely to be the driver of the compaction of massive galaxies where feedback-driven outflows are less effective and disk morphologies show up.

At similar radial ranges, we find the spin alignment of gas and stars becomes poorer below $r^{\ast}_{1/2}$, which is a consequence of the proto-bulge formation from concentrated star-formation. We also find that the spin of gas is about $2-3$ times larger than that of the stellar disk. Similar gaps have been found in observational studies as well~\citep{Obreschkow2014,Cortese2016}. As a more direct tracer of the recently arrived mass, gas is expected to come in with a higher specific angular momentum, while stars are made of baryons that arrived earlier with lower specific angular momentum~\citep[e.g.][]{Danovich2015, Teklu2015}. In addition, feedback from young massive stars and potentially AGN can drive galactic outflows, which preferentially remove the low-angular momentum gas in the dense central regions of the galaxy and effectively increase the specific angular momentum of the remaining gas~\citep[e.g.][]{Maller2002, Governato2010, Brook2011, Brook2012a, Guedes2011, Romanowsky2012, Agertz2016, Lagos2017}. Such effects are likely more prominent at high redshifts, where star-formation is more clustered and bursty~\citep[e.g.][]{Smit2016, Sparre2017, Emami2019, Flores2021, Shen2023, SunJ2023}.

Compared to galaxies in the low-mass bin, massive galaxies have better alignments of gas and stellar angular momentum at $r\lesssim 0.1\,R_{\rm vir}$ and clear signatures of a co-rotating stellar disk at $r^{\ast}_{1/2}\lesssim r \lesssim 2\,r^{\ast}_{1/2}$. This is consistent with the fact that massive galaxies are better described by the classical disk formation theory while low-mass galaxies are better described by the spherical shell model shown in the section above. This change in geometry could be due to the more frequent mergers and stronger impact of feedback-driven outflows of low-mass galaxies. Compared to the overall galaxy population, compact galaxies tend to have lower spins of both dark matter and gas upon accretion but similar spin profiles in the inner region of the halo. The initial lower spin makes these haloes more prone to disk instability~\citep{Dekel2014} and enters the compaction phase earlier.

\begin{figure*}
    \raggedright
    \includegraphics[width=0.247\linewidth]{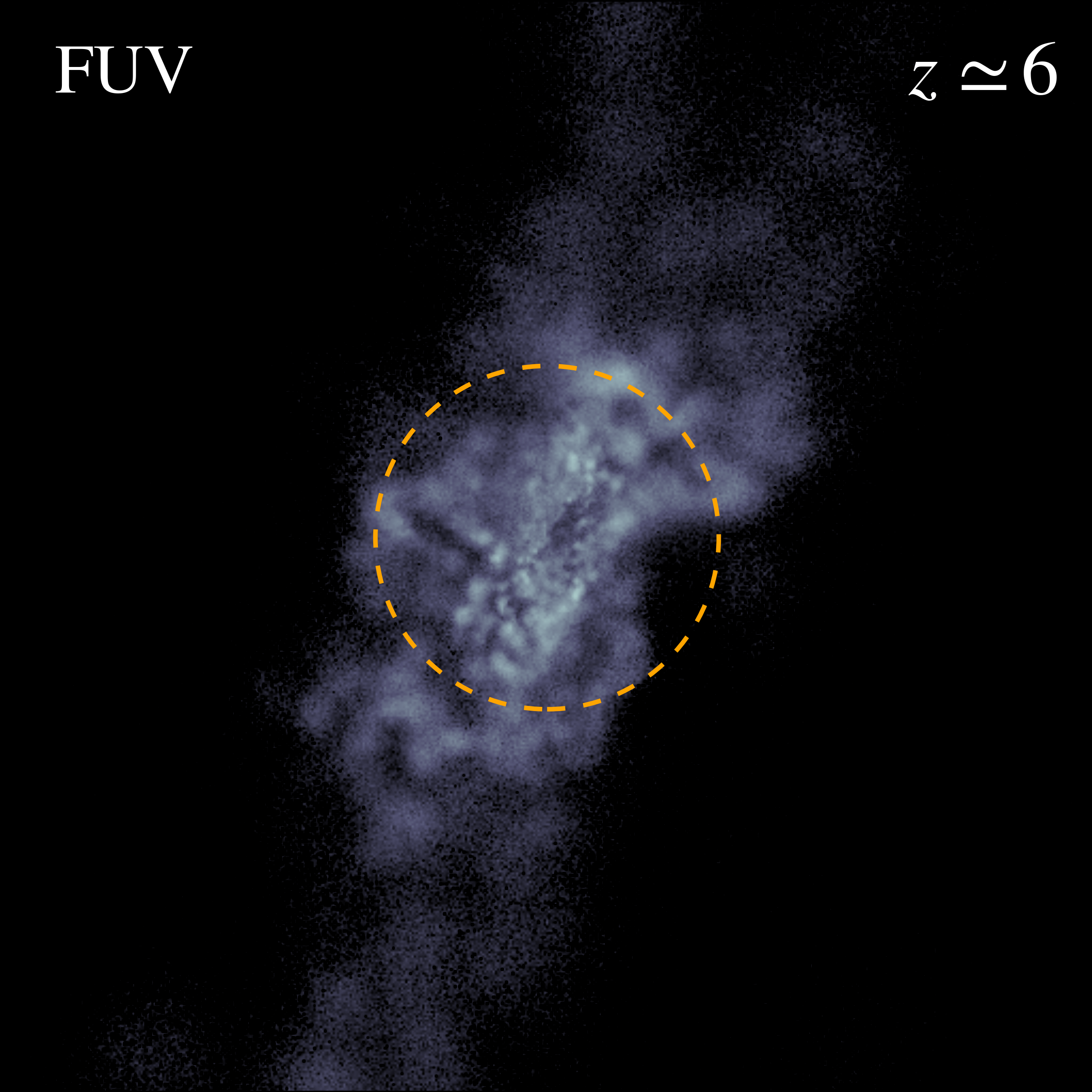}
    \hspace{-0.13cm}
    \includegraphics[width=0.247\linewidth]{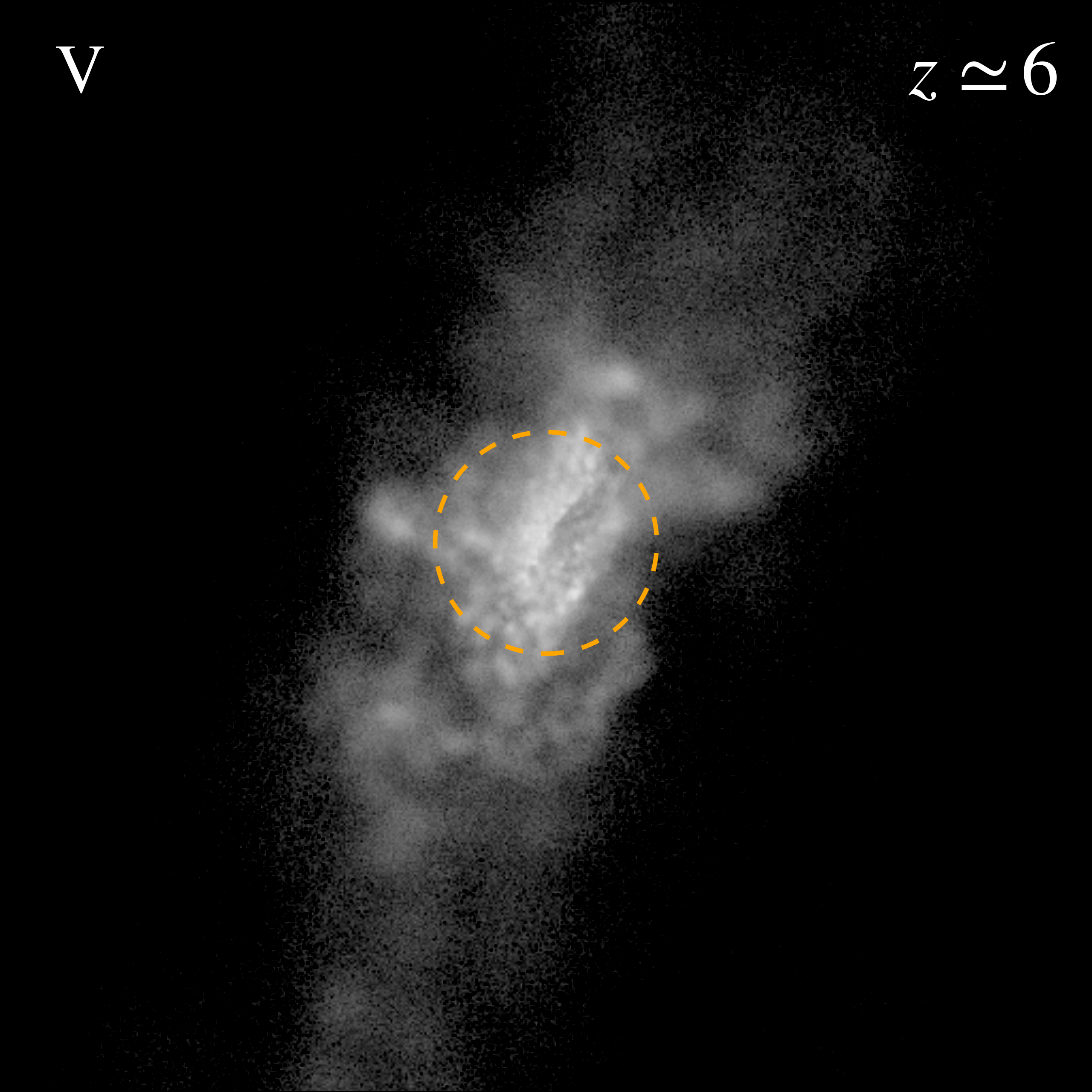}
    %\hspace{0.13cm}
    \includegraphics[width=0.247\linewidth]{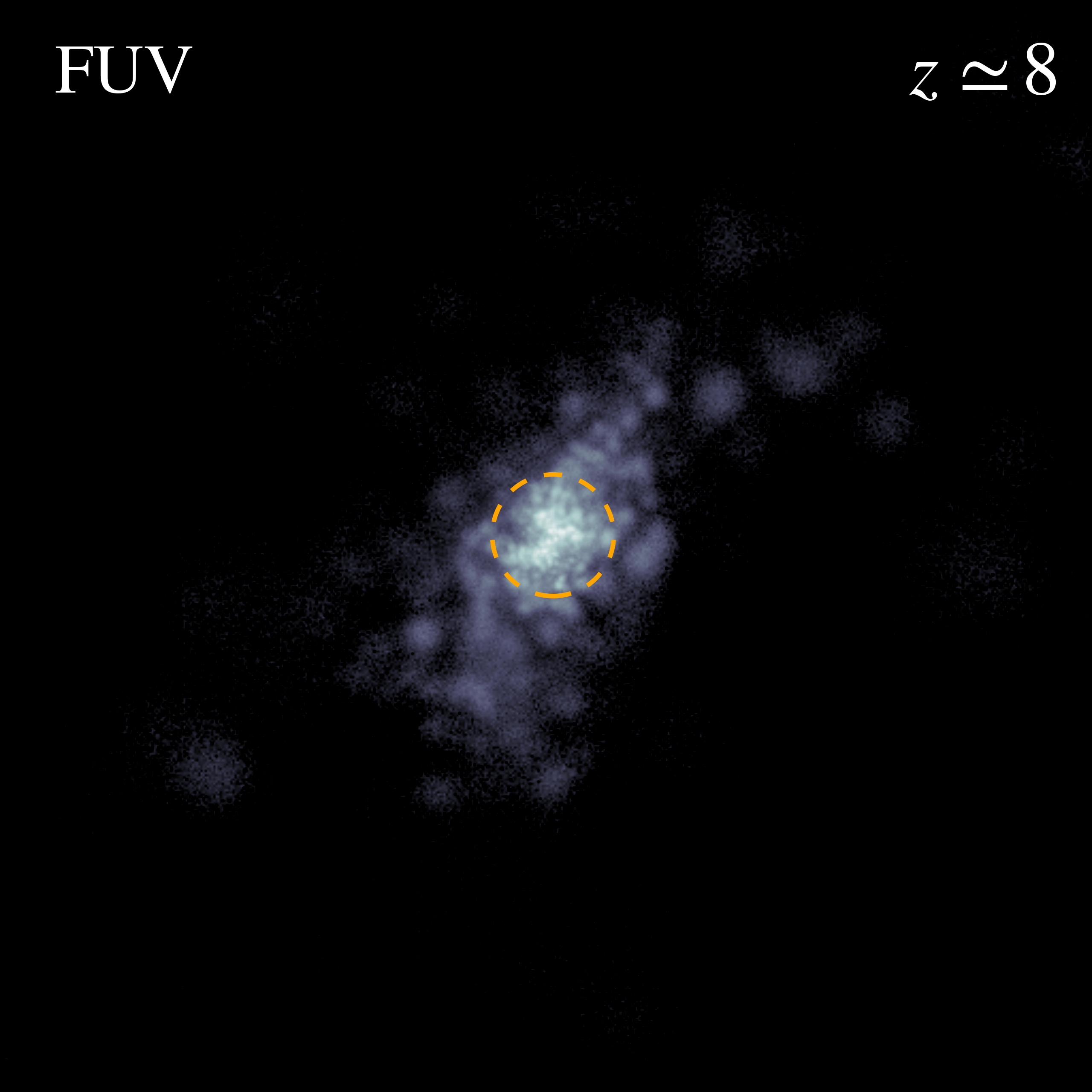}
    \hspace{-0.13cm}
    \includegraphics[width=0.247\linewidth]{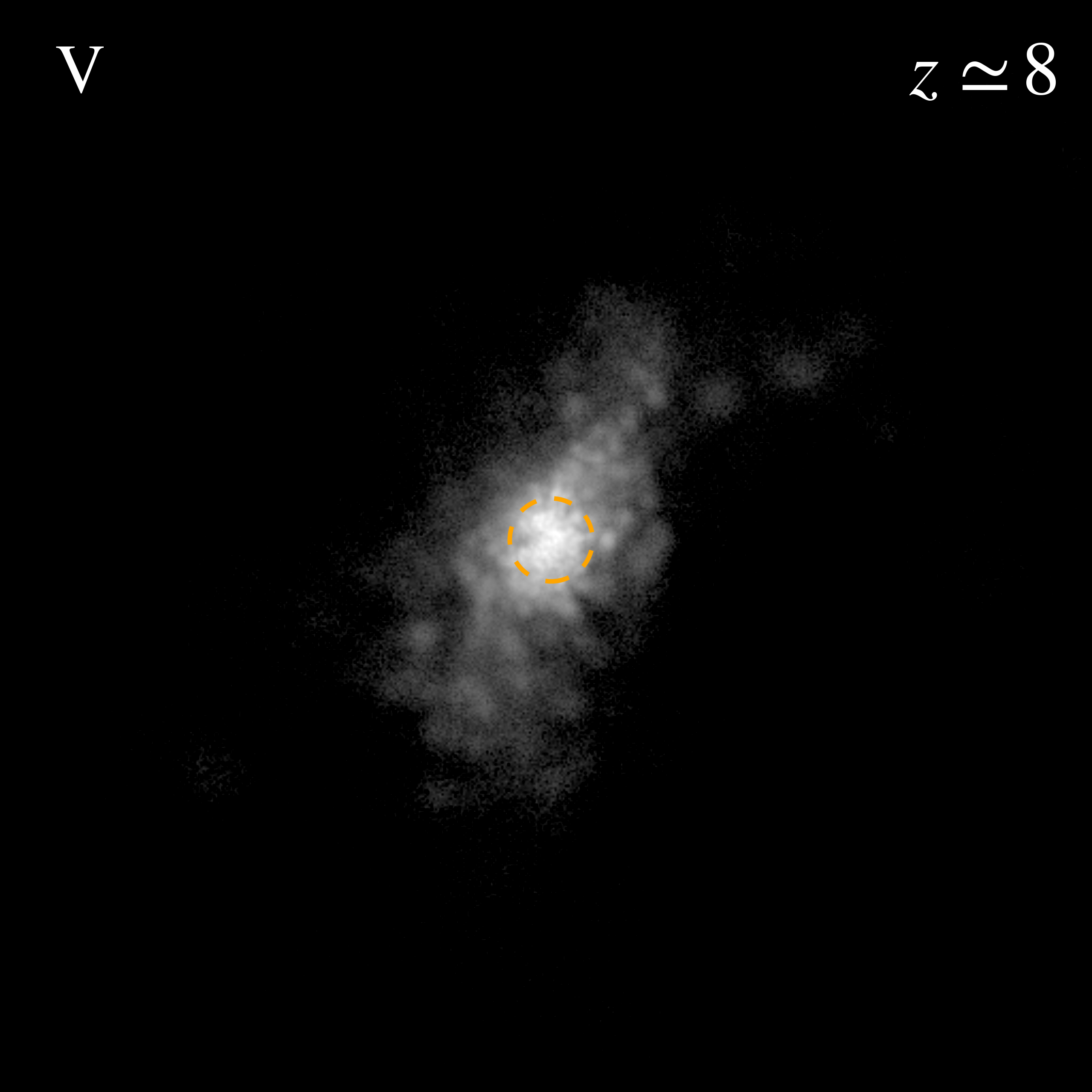}
    \vspace{-0.02cm}
    \includegraphics[width=0.247\linewidth]{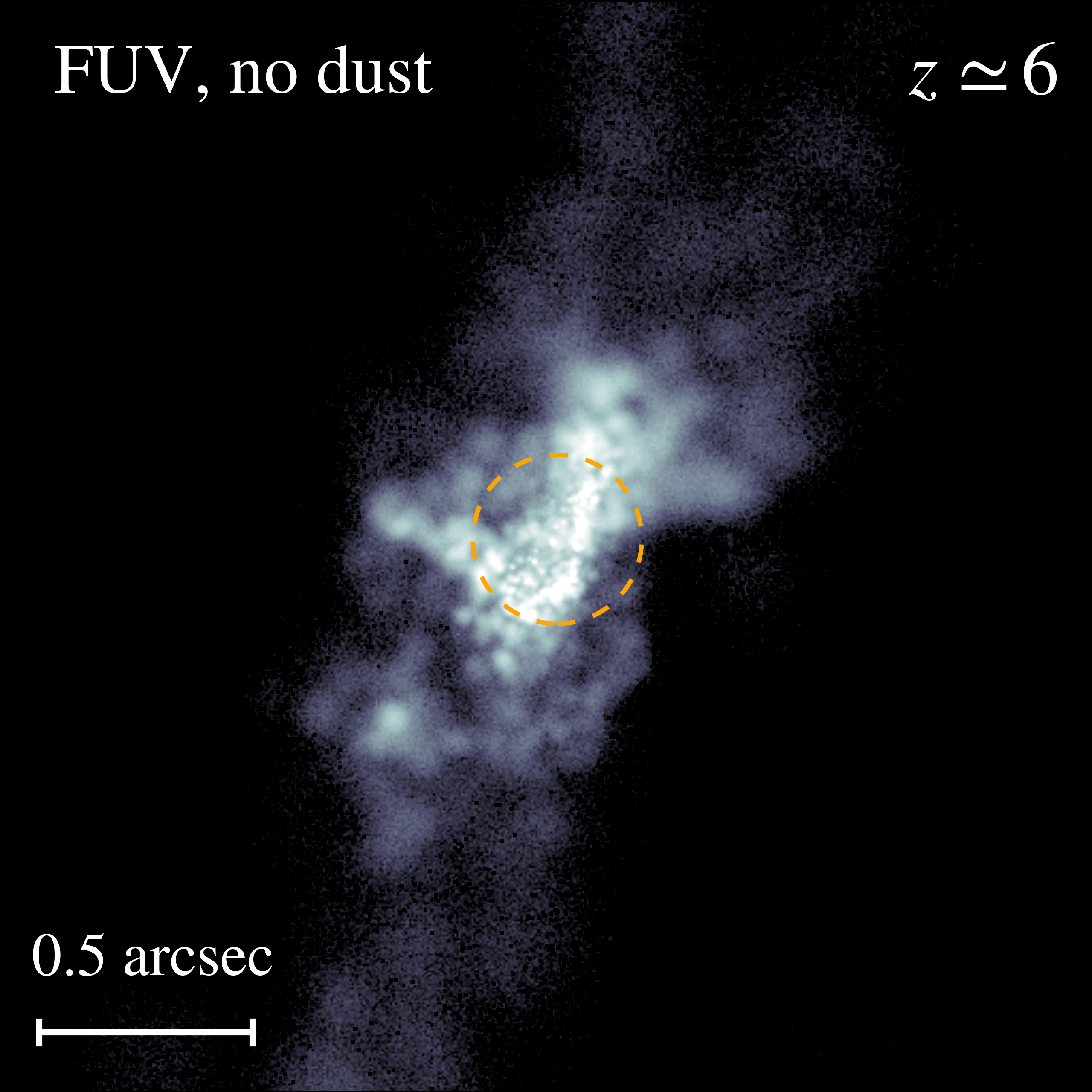}
    \hspace{-0.13cm}
    \includegraphics[width=0.247\linewidth]{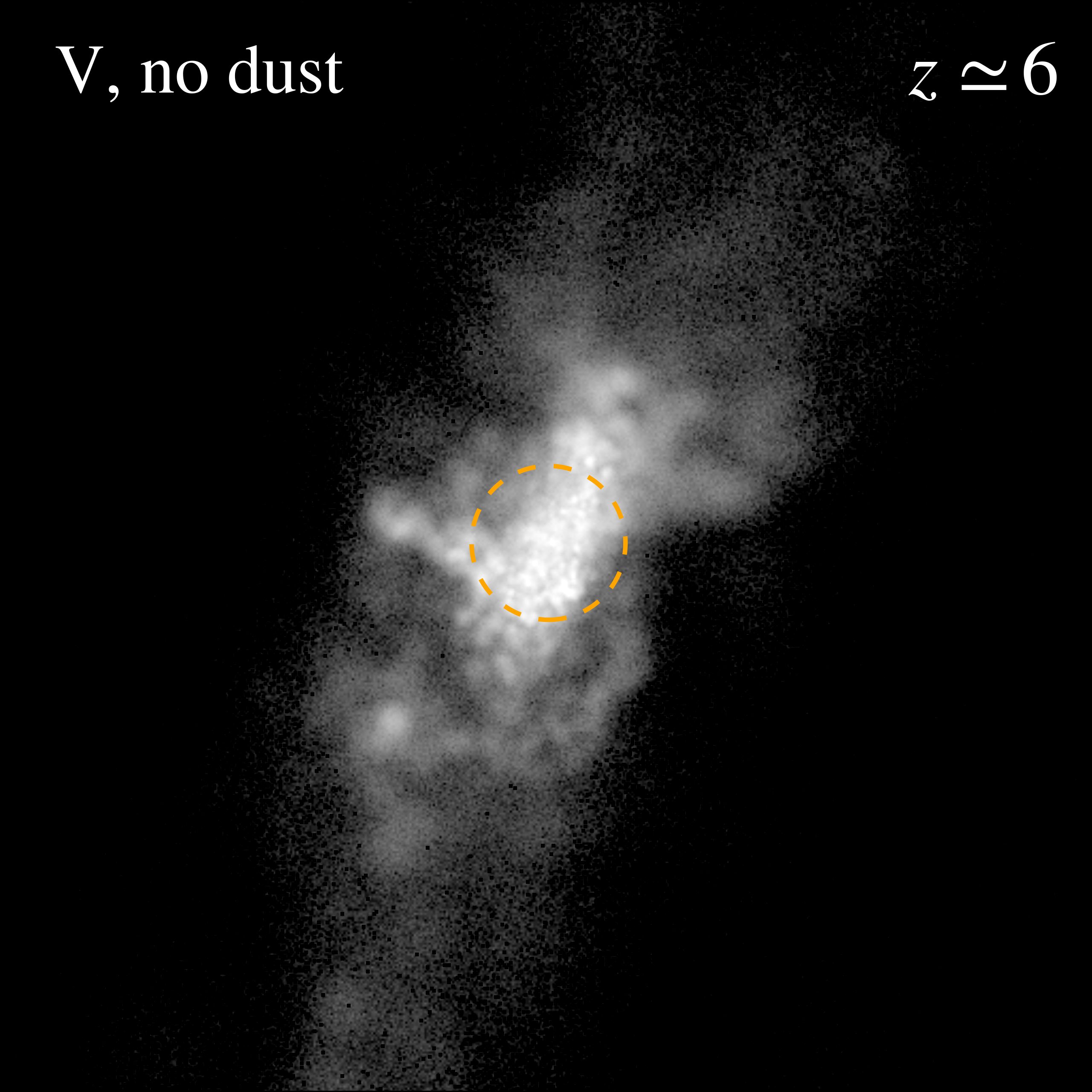}
    %\hspace{0.13cm}
    \includegraphics[width=0.247\linewidth]{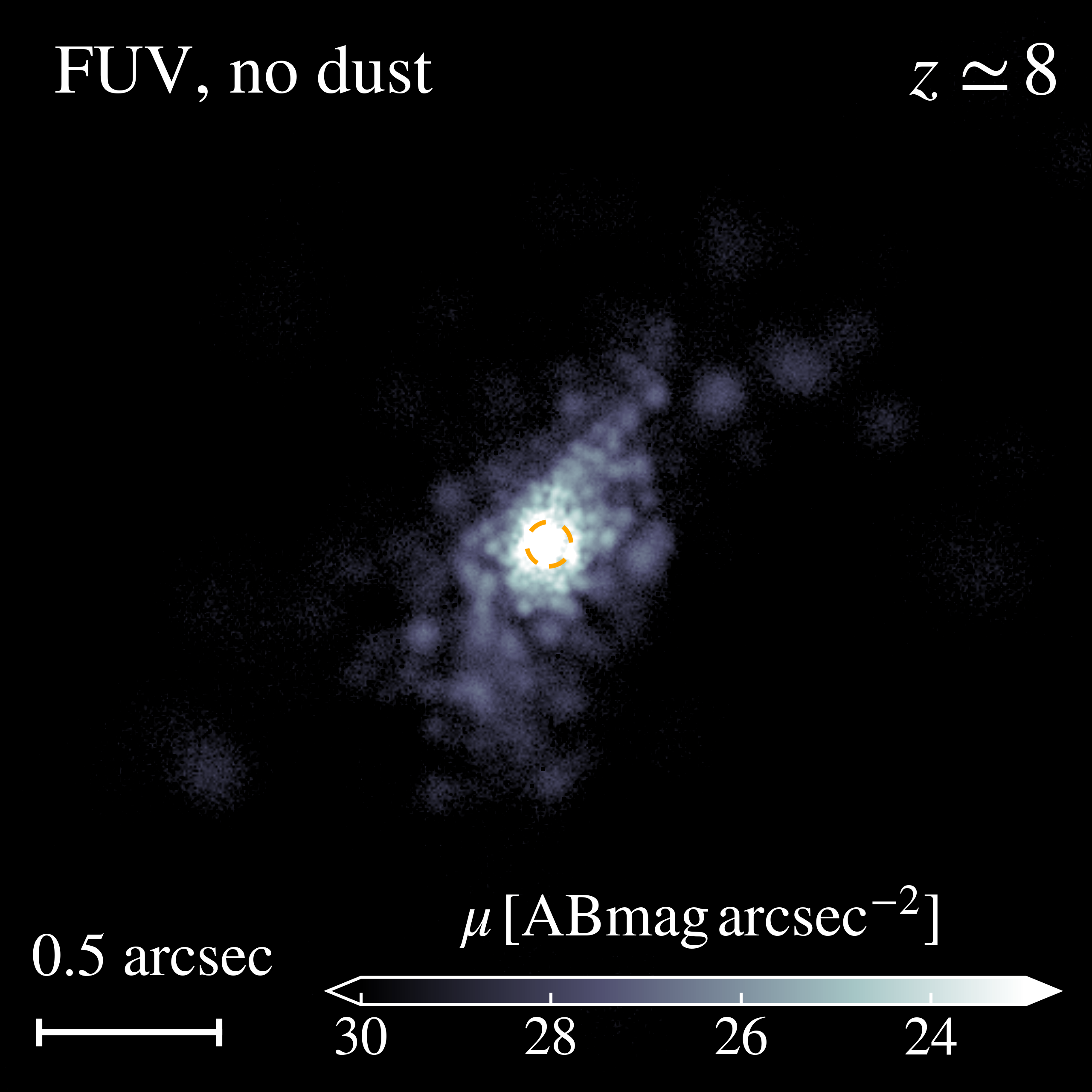}
    \hspace{-0.13cm}
    \includegraphics[width=0.247\linewidth]{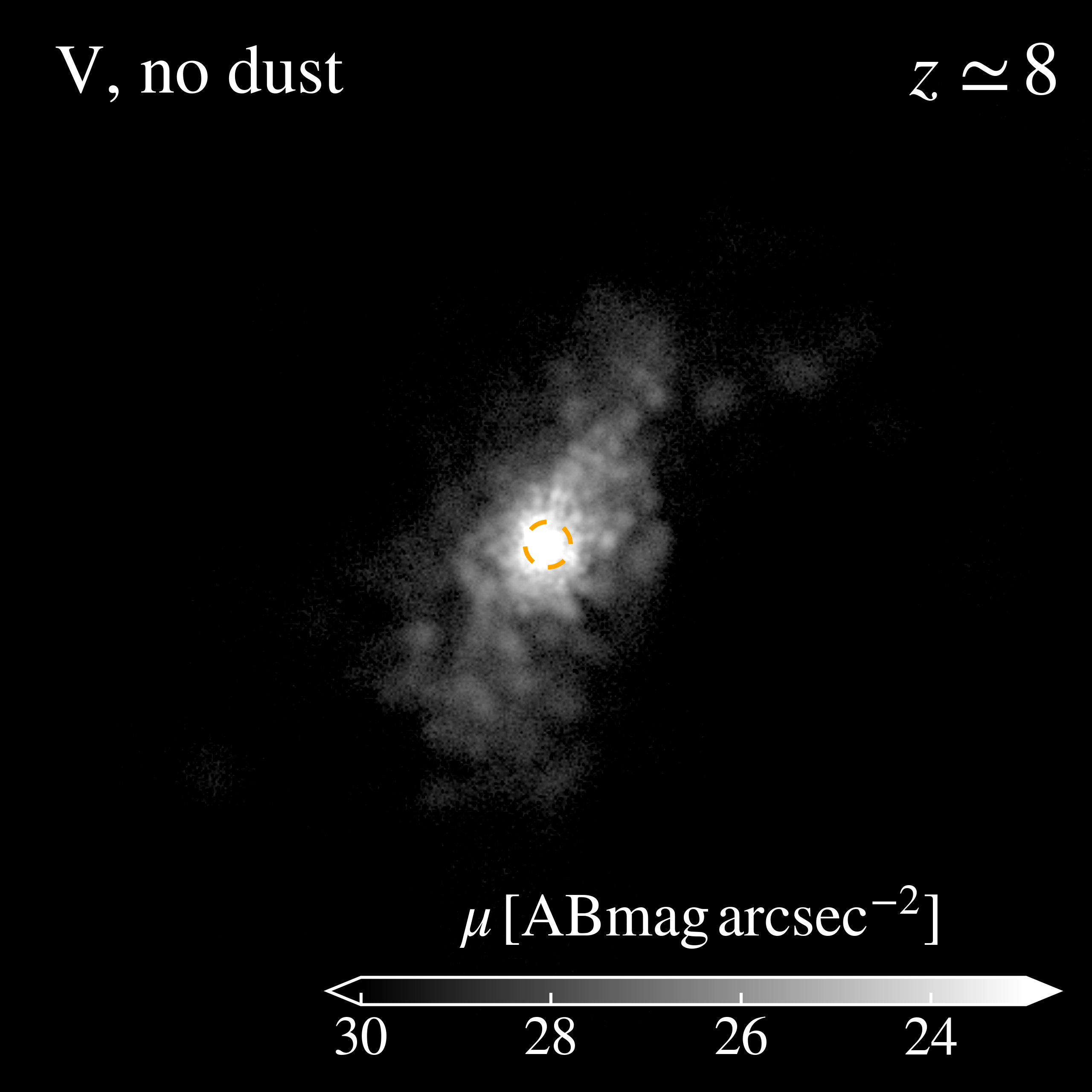}
    \caption{Images of two galaxies at $z\simeq 6$ and $z\simeq 8$ in the \thesanone simulation in rest-frame UV and optical (V band) with and without dust attenuation. The images are generated following the procedure described in Section~\ref{sec:method:image}. The orange circles indicate the effective radius in the corresponding band centered on the surface brightness-weighted centre of the galaxy.}
    \label{fig:mock-images}
\end{figure*}

\subsection{Disk instability}
\label{sec:discussion:instability}

One important driver for the compaction of massive galaxies at $z\gtrsim 6$ is disk instability. As shown in Figure~\ref{fig:size-mass-correlation} and Figure~\ref{fig:spin_profile}, we find more prevalent signatures of disk morphology for both the stellar and gas components of these galaxies. This is expected from disk formation theory as massive galaxies tend to have deeper and more concentrated gravitational potentials~\citep[e.g.][]{Hopkins2023} and increased pressured confinement from the hot virialized CGM~\citep[e.g.][]{Stern2021}. However, the gaseous disks in high-redshift star-forming galaxies are likely not stable or in a marginally stable state. 

The large cold gas abundance and rapid gas cooling can trigger global instability in the disk. This can be understood through a modified turbulent Toomre instability~\footnote{Here we focus on the gas-rich early phase and neglect the stellar content~\citep[e.g.][]{Romeo2011}.} criterion of a differentially rotating disk~\citep{Toomre1964}
\begin{equation}
    Q = \dfrac{\sqrt{c^2_{\rm s} + \sigma_{\rm r}^2}\,\kappa}{\pi\,G \,\Sigma_{\rm tot}} > Q_{\rm c} \, ,
\end{equation}
where $\kappa$ is the epicyclic frequency (e.g. $\sqrt{2}\,\Omega$ for a flat rotation curve, where $\Omega$ is the angular circular velocity), $c_{\rm s}$ is the thermal sound speed, $\sigma_{\rm r}$ is the radial velocity dispersion, $\Sigma_{\rm tot}$ is the total matter surface density, and $Q_{\rm c}$ is an order-unity constant. Disk instability happens if the LHS drops below $Q_{\rm c}$, which can be a consequence of either cooling or increased $\Sigma_{\rm tot}$ from gas accretion. Note that even when turbulence dominates over the thermal support, turbulence cascade to the viscous scale can still turn turbulent energy into thermal energy, which is eventually carried away by cooling radiation. 

In high-redshift galaxies, both scenarios are highly possible. For example, in the standard $\Lambda$CDM cosmology, the typical gas accretion time scale can be estimated as $t_{\rm acc} = M_{\rm halo}/\dot{M}_{\rm halo} \simeq 316\Myr\, (1+z/7)^{-2.25}\,(M_{\rm halo}/10^{12}\msun)^{-0.15}$~\citep{Dekel2009}, where the gas accretion rate of a halo has been assumed to scale with the total mass growth rate by a constant factor. The ratio of $t_{\rm acc}$ to the Hubble time $t_{\rm H}\equiv 1/H(z)$ will become lower at higher redshift and can drop significantly below unity at $z\gtrsim 6$ to allow instabilities to develop in disks. A similar thing applies to the typical cooling time in the star-forming disk~\citep[e.g.][]{Dekel2023} which is almost always below $\sim 100\Myr$. 

\begin{figure}
\centering
\includegraphics[width=1\linewidth]{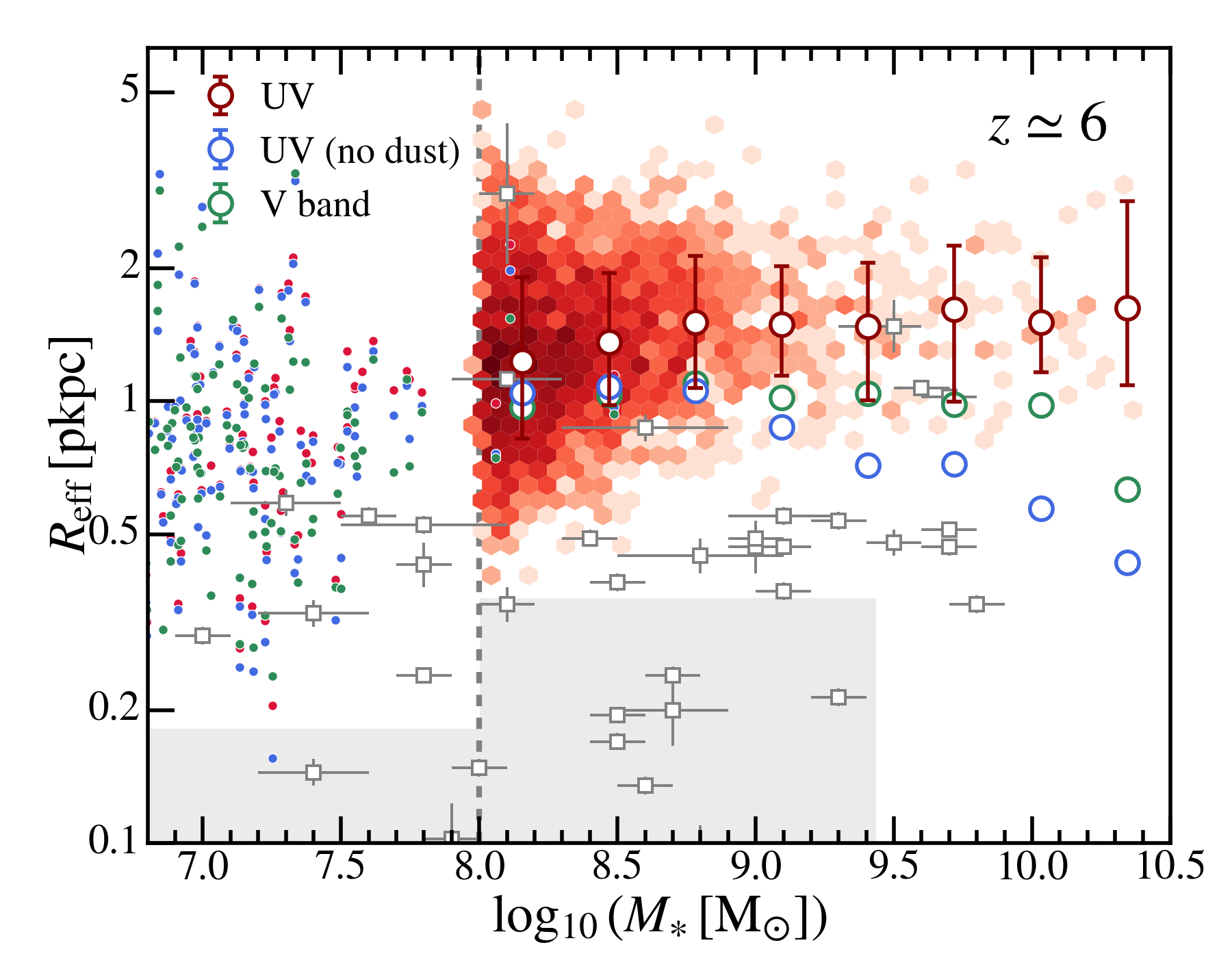}
\includegraphics[width=1\linewidth]{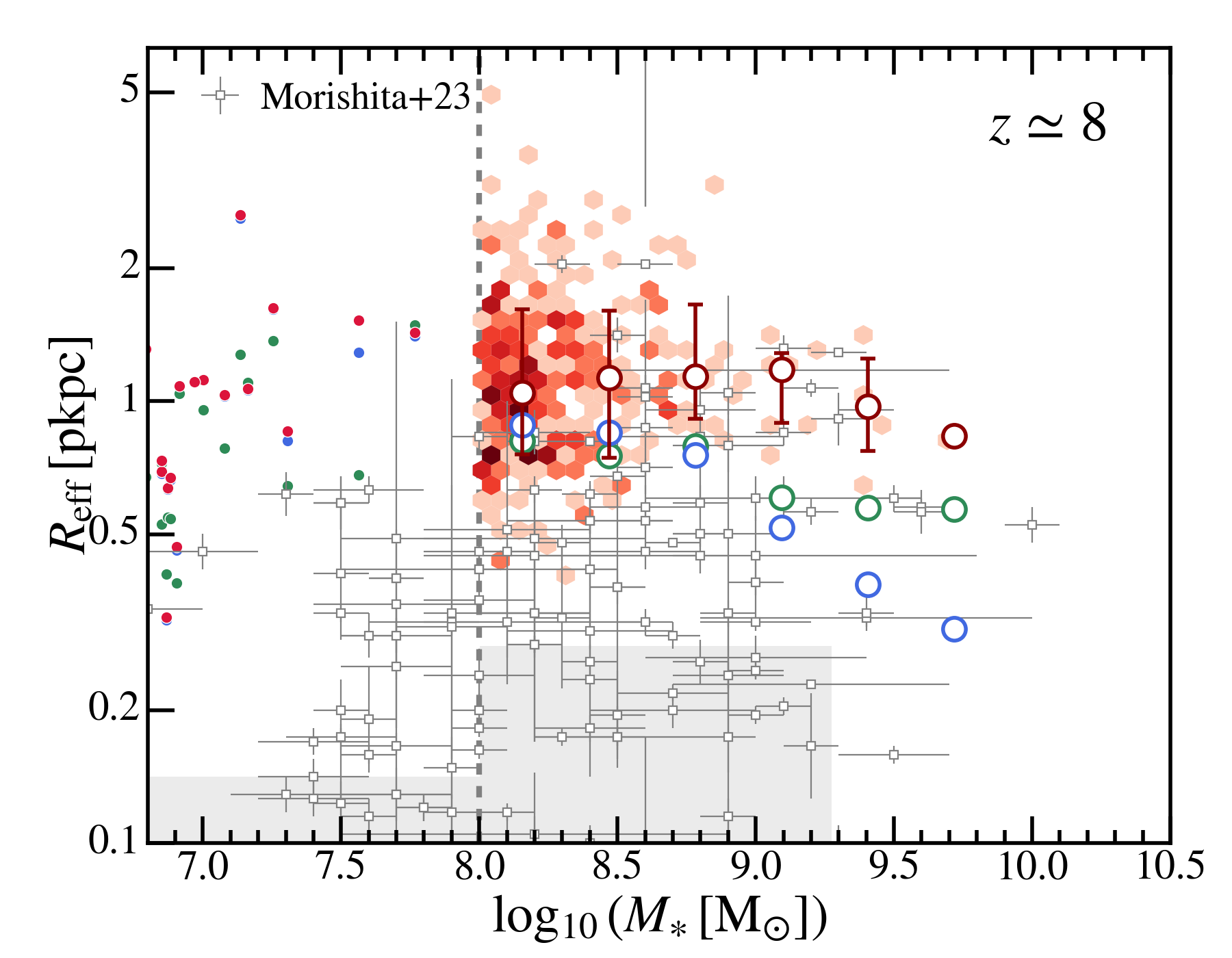}
\includegraphics[width=1\linewidth]{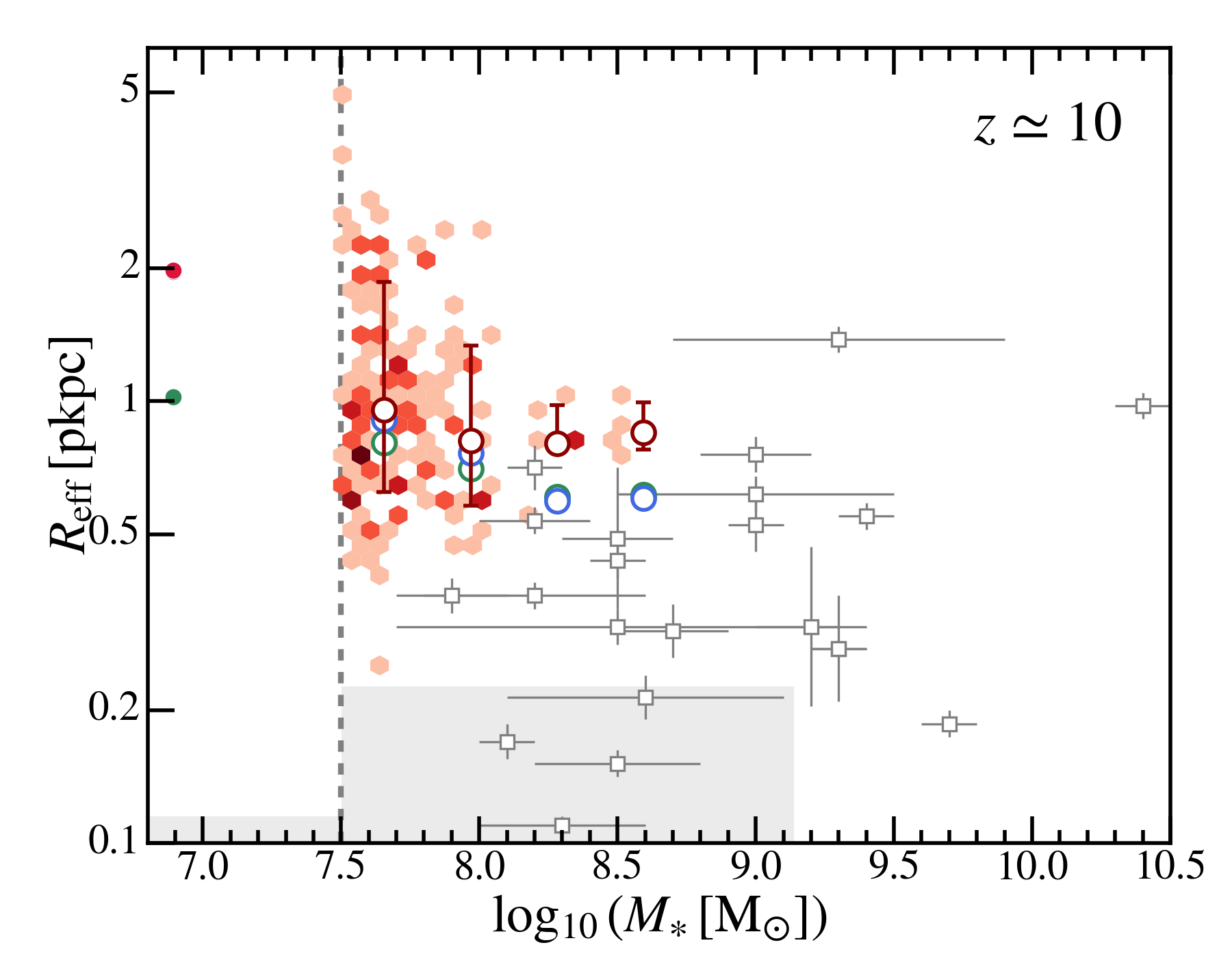}
\caption{Apparent size--mass relation at $z\simeq$ 6, 8, 10. The open circles and error bars show the median and 1-$\sigma$ scatter of galaxy effective radii ($R_{\rm eff}$) in \thesanone. Different colors correspond to the size in UV, V band, and UV without dust attenuation. The background colormap shows the full distribution of $R_{\rm eff}$ in UV versus stellar mass. The solid circles show individual galaxies in \thesanhr on this plane. The vertical dashed line indicates the stellar mass limit we choose for \thesanone mock images. We compare simulation results with observational constraints from \citet{Morishita2023} as shown by open gray squares with error bars.}
\label{fig:obs-size-mass}
\end{figure}

Once instabilities get triggered, clumps with sizes comparable to the Jeans length will form over a dynamical crossing time scale 
\begin{equation}
    t_{\rm d} \equiv \dfrac{1}{\Omega} = \dfrac{R_{\rm d}}{V_{\rm c}} \simeq 10\Myr\,\left(\dfrac{R_{\rm d}}{1\pkpc} \right)\,\left(\dfrac{V_{\rm c}}{100\kms}\right)^{-1} \, ,
\end{equation}
where $R_{\rm d}\sim r^{\ast}_{1/2}$ is the characteristic size of the disk and $V_{\rm c}$ is the circular velocity at $R_{\rm d}$. These clumps will migrate toward the centre of the galaxy through clump-clump interactions and meanwhile drive general mass inflows~\footnote{At the same time, there exists some feedback mechanism to keep the disk in a marginally stable state, including the mass inflow that reduces $\Sigma_{\rm tot}$, clump-clump encounters stirring up the disk and increasing turbulence, and thermal or mechanical feedback from star-formation. In this self-regulated state, mass inflow and clump migration can be sustained.}. In a self-regulated state, \citet{Dekel2009} showed that the evacuation time scale of the disk due to clump migration is
\begin{equation}
    t_{\rm evac} \simeq 20\,t_{\rm d}\,\left(\dfrac{\alpha}{0.2}\right)^{-1}\,\left(\dfrac{Q_{\rm c}}{0.68}\right)^{2}\,\left(\dfrac{\delta}{0.5}\right)^{-2} \, ,
\end{equation}
where $\delta$ is the disk mass to total mass ratio within $R_{\rm d}$, $\alpha$ is the disk mass locked in giant gas clumps as defined in \citet{Dekel2009}, and $Q_{\rm c}=0.68$ for an isothermal thick disk~\citep{Goldreich1965a}. The compaction happens faster in gas-rich environments with high $\delta$. Since baryons dominate the mass budget in the central halo and gas dominates over stars in the early phase of galaxy growth (see Figure~\ref{fig:trace-quenched}), $\delta$ is comparable to the $f^{\rm gas}_{\rm disk}$ measured in Figure~\ref{fig:size-mass-correlation} and can reach $\gtrsim 0.5$ in massive galaxies on the track of compaction. The time scale for instability-driven compaction will be of order a few hundred Myr, which is quite consistent with the evolution track shown in Figure~\ref{fig:trace-quenched}. As argued in \citet{Dekel2014}, once disk instability gets triggered, $\delta$ will monotonically increase during the gas contraction phase and this will be a runaway process. This could create a bimodal track of galaxy size evolution that depends on halo spin, which has been discussed extensively in the context of compact galaxies at $z\sim 2-3$~\citep[e.g.][]{Dekel2014,Zolotov2015,Lapiner2023}. However, at higher redshifts, the compaction track will quickly become much more likely to happen~\citep{Dekel2014}.

The gas inflow is also likely to be ``wet''~\citep{Dekel2014,Krumholz2018}. In the self-regulated marginally stable disk, star-formation happens at the time scale $t_{\rm sf} = t_{\rm ff}/{\rm SFE}_{\rm ff} \simeq 50\, t_{\rm d}\,({\rm SFE}_{\rm ff}/0.01)^{-1}$,
where ${\rm SFE}_{\rm ff}$ is the star-formation efficiency in molecular phase per free-fall times and we take the value $\sim 0.01$ suggested by observations~\citep[e.g.][needs to be distinguished from the galaxy-scale SFE defined above]{Sun2023}, $t_{\rm ff}\simeq 0.5 t_{\rm d}$ is the free-fall time assuming star-formation happens in regions that are denser than the mean density of the disk by a factor of few~\citep{Dekel2014}. $t_{\rm sf}$ is usually longer than $t_{\rm evac}$, especially when $\delta$ is large, cold gas inflow can be sustained against local star-formation in the disk, and therefore, fuels the starburst at galaxy centres and helps form compact galaxies.

\begin{figure}
\centering
\includegraphics[width=1\linewidth]{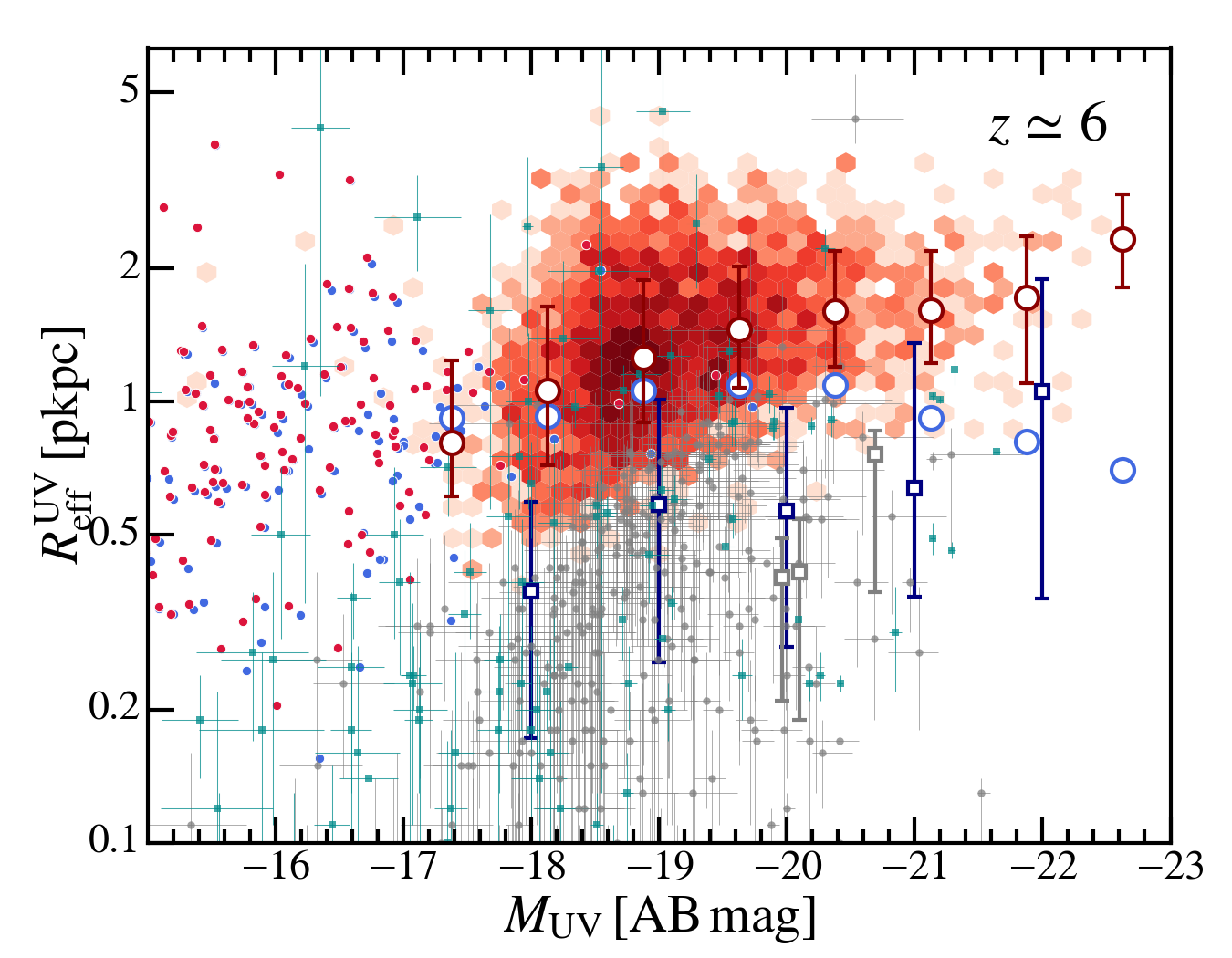}
\includegraphics[width=1\linewidth]{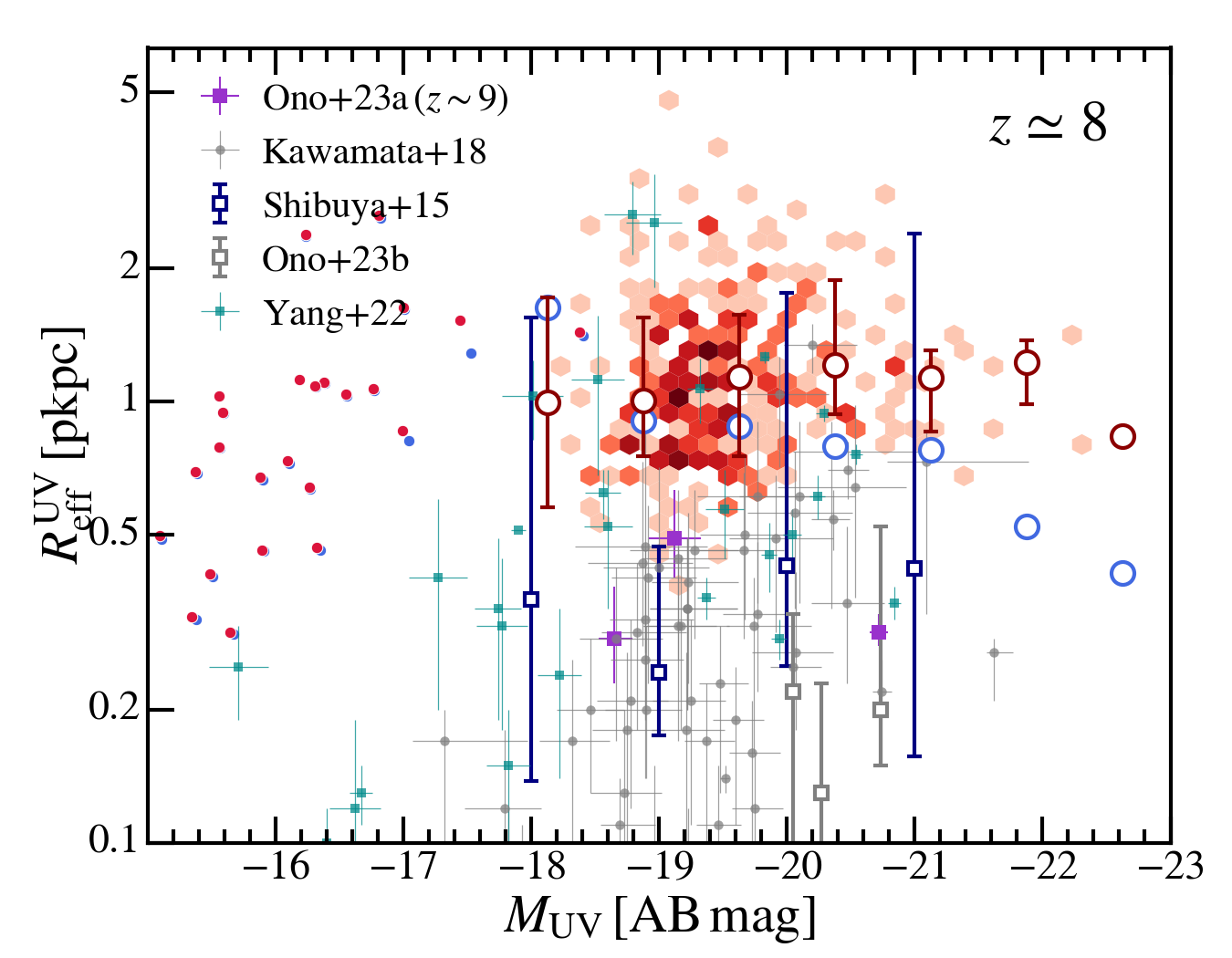}
\includegraphics[width=1\linewidth]{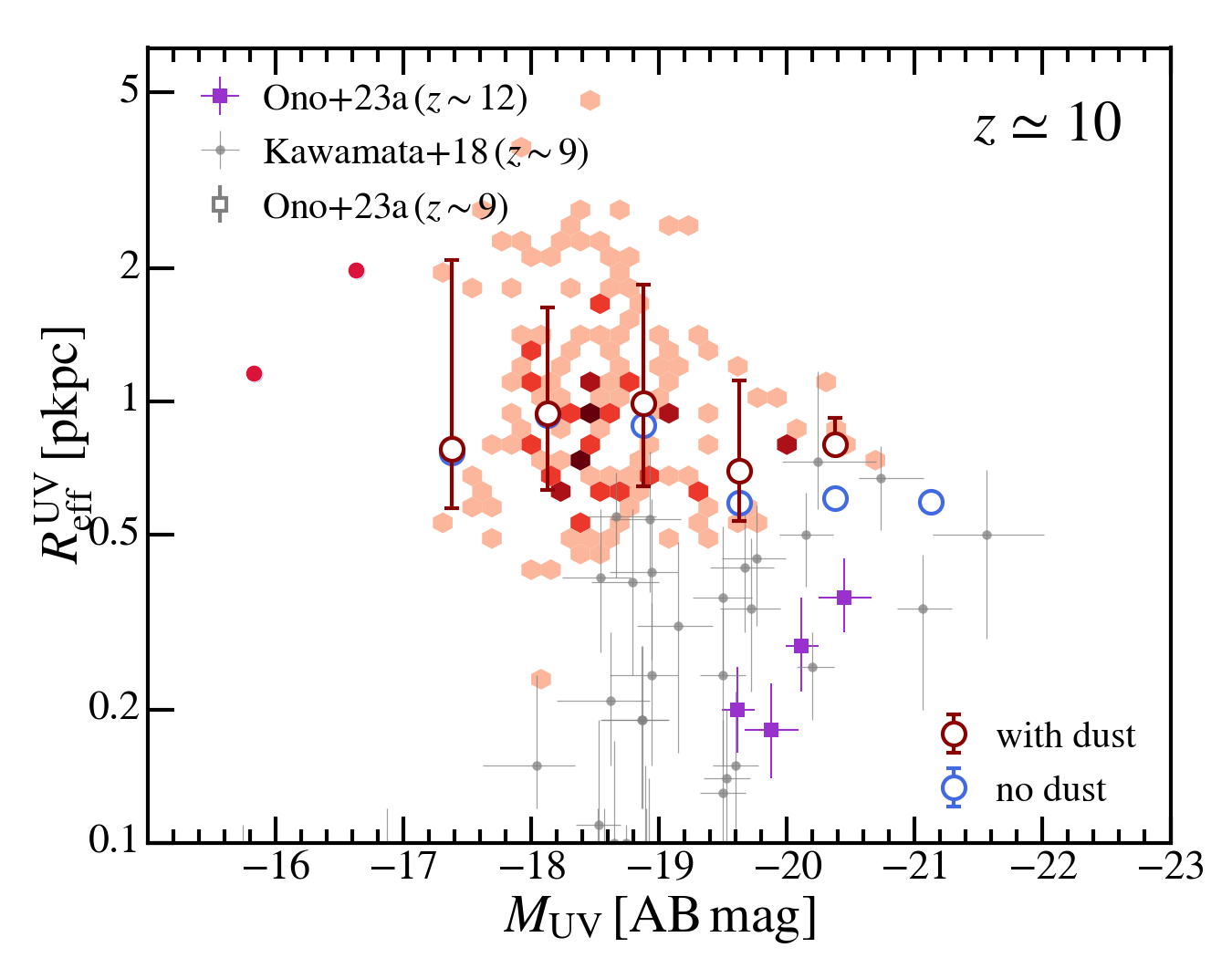}
\caption{Apparent size-luminosity relation at $z\simeq$ 6, 8, 10. The open circles and error bars show the median and 1-$\sigma$ scatter of galaxy UV effective radii ($R^{\rm UV}_{\rm eff}$) in \thesanone with and without dust attenuation. The background colormap shows the full distribution of $R^{\rm UV}_{\rm eff}$ versus $M_{\rm UV}$. The solid circles show individual galaxies in \thesanhr on this plane. We compare simulation results with observational constraints from \citet{Shibuya2015-size,Kawamata2018,Yang2022,Ono2023a,Ono2023b}.}
\label{fig:obs-size-lum}
\end{figure}

\section{Comparison with observations}
\label{sec:observation}

\subsection{Apparent size-mass(luminosity) relations}
\label{sec:apparent-size}

The analyses above focus on the intrinsic physical sizes of galaxies. However, in the observations, galaxy sizes are often measured by fitting the surface brightness profile/distribution of galaxies in certain bands and quoted in terms of the effective radius ($R_{\rm eff}$, 2D ``half-light'' radius). To enable a fair comparison of simulation results with observations, we utilize the mock galaxy images described in Section~\ref{sec:method:image}. We find $R_{\rm eff}$ as the circular aperture centered on the pixel flux-averaged centre\footnote{The median offset between the flux-average centre and the galaxy centre defined using potential minimum in Section~\ref{sec:method:galprop} is $\sim 500$ physical pc, and, in $\gtrsim 80\%$ of the cases, the offset is $\lesssim 1 \pkpc$. This is not a negligible effect given the small apparent size of some of these galaxies.} of the image that contains half of the total light of all image pixels. In Figure~\ref{fig:mock-images}, we show examples of these mock galaxy images in rest-frame UV and optical V band with and without dust attenuation. The galaxy $R_{\rm eff}$ in UV and V bands are fairly close to each other before dust attenuation. However, dust attenuation increases the apparent size of galaxies preferentially in UV, which will be discussed in detail below.

In Figure~\ref{fig:obs-size-mass}, we show the galaxy $R_{\rm eff}$ versus stellar mass relations in rest-frame UV and V band in the \thesan simulations. We also show the UV $R_{\rm eff}$ when dust attenuation is not taken into account. The discrepancies of size from band choice or dust attenuation mainly show up in massive galaxies. Unlike the intrinsic $r^{\ast}_{1/2}$, the $R_{\rm eff}$ in UV display a mild positive correlation with galaxy stellar mass. The simulation results are compared to the measurements in \citet{Morishita2023}, which were based on 341 galaxies at $5<z<14$ in a suite of nine public \textit{JWST} extragalactic fields\footnote{The fields included in \citet{Morishita2023} analysis are PAR1199~\citep[PID 1199;]{Stiavelli2023}, J1235~(PID 1063;
PI: Sunnquist), NEP~\citep[PID
2738;]{Windhorst2023}, PRIMER-UDS and COSMOS~(PID 1837; PI: Dunlop), NGDEEP~\citep[PID 2079; PI: Finkelstein;]{Bagley2023b}, CEERS~\citep[(PID 1345;]{Bagley2023a, Finkelstein2023}, GLASS-JWST/UNCOVER~\citep[PID 1324;]{Treu2022}, and JADESGDS~\citep{Eisenstein2023, Robertson2023, Tacchella2023}.} in Cycle 1. We find that the observed galaxies are systematically more compact than the simulated ones. A large population of compact galaxies exists at $M_{\ast}\lesssim 10^{9}\msun$ with $R_{\rm eff}\lesssim 200\,{\rm pc}$. We note that they are still larger than the full width at half maximum (FWHM) of the PSF of short-wavelength NIRCam filters (e.g. for F070W, F090W, F115W, FWHM$\sim 0.03$ arcsec $\sim 150\,{\rm pc}$ at $z\sim 8$; \citealt{Rigby2023}). 

In Figure~\ref{fig:obs-size-lum}, we show the galaxy $R_{\rm eff}-M_{\rm UV}$ relation from the \thesan simulations. The absolute UV magnitude $M_{\rm UV}$ is derived by integrating the fluxes of galaxy images in rest-frame UV. Similar to the size--mass relation, the size-luminosity relations after dust attenuation show a mildly positive slope at $z\simeq 6$ and are almost flat at $z\gtrsim 8$. We compare the simulation results with observational data based on HST from \citet{Shibuya2015-size} and \citet{Kawamata2018} as well as the ones based on \textit{JWST} from \citet{Yang2022,Ono2023a,Ono2023b}. Similar discrepancies between simulations and observations are found at the faint end ($M_{\rm UV}\gtrsim -20$) at about 1-$\sigma$ level. 

Due to the numerical limitation, the \thesan simulations could miss a population of compact galaxies below the convergence scale and the gravitational softening length (as discussed in Section~\ref{sec:size-mass}), which is indicated by the gray shaded region in the figure. However, the offset between the median relations is robust against numerical effects when taking the \thesanhr results into account. On the other hand, the discrepancy in size could indicate limitations of the feedback model employed by the \thesan simulations (and also IllustrisTNG). We note this is not restricted to \thesan, but essentially many other simulations shown in Figure~\ref{fig:size-mass-comparision} fail to produce compact ($R_{\rm eff}\lesssim 300\,{\rm pc}$) galaxies at the stellar mass $\lesssim 10^{9}\msun$ at $6 \lesssim z \lesssim 10$. 

In \citet{Ono2023b}, a population of extremely compact ($R_{\rm eff}\lesssim 100\,{\rm pc}$) galaxies was compared to the FOREVER22~\citep{Yajima2022} and Renaissance simulations~\citep{Xu2016,Barrow2017}, which produced a few analogs at higher redshifts ($z\gtrsim 10$) in overdense regions. These galaxies were argued to be formed under the impact of galaxy mergers or interactions. This scenario is similar to the subpopulation of compact galaxies at $M_{\ast}\lesssim 10^{8}\msun$ we find in \thesan in extremely overdense regions (see Figure~\ref{fig:size-mass-correlation-env}). However, they are not representative of the general galaxy population. In contrast, we find that the sizes of the majority of the low-mass galaxies increase during the phase of merger-driven growth and are positively correlated with matter overdensity. 

In the SERRA simulation~\citep{Pallottini2022}, compact galaxies were consistently produced at $z\simeq 7.7$ with $R^{\rm UV}_{\rm eff}\sim 100\,{\rm pc}$ at $M_{\ast}\sim 10^{9}\msun$ and were on the other hand slightly below the observational constraints. In the high-resolution variants of the \textsc{Simba} simulations~\citep{Romeel2019,Wu2020}, galaxy sizes were found to decrease down to $\sim 200\,{\rm pc}$ at $M_{\rm UV}\sim -18$ at $z\simeq 6$. These dramatic and qualitative differences between size predictions of different simulations may indicate incompleteness in many current feedback implementations. The prescriptions configured to prevent excessive star-formation at a similar level might result in rather different kinematic behavior of the gas that fuels star-formation~\citep{Rosdahl2017,Hopkins2018-feedback} and different stellar morphology. This highlights the potential of using galaxy morphology to constrain galaxy formation models at high redshifts.

\begin{figure*}
    \centering
    \includegraphics[width = 1\linewidth, clip, trim={1cm 0 0 0}]{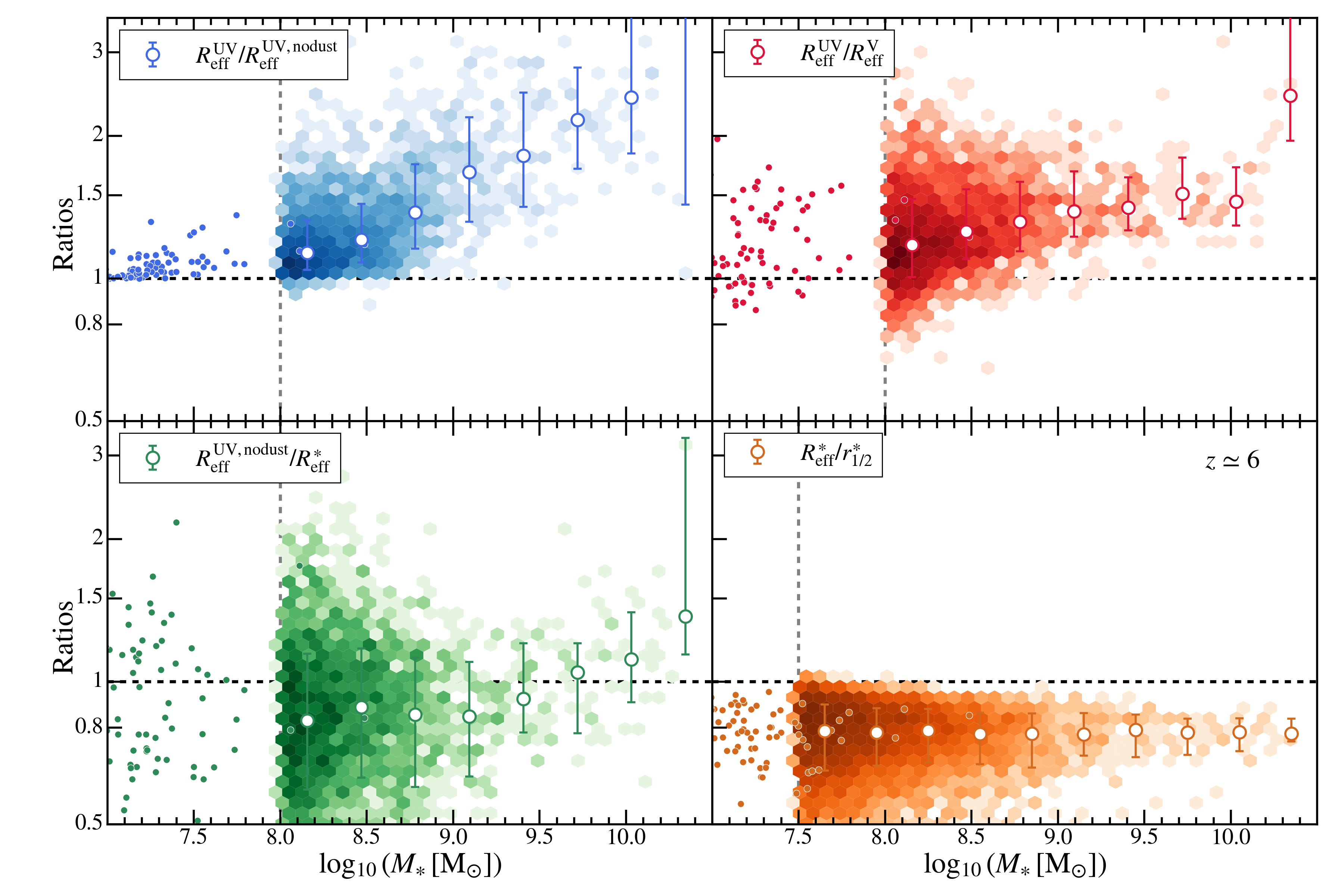}
    \caption{An illustration of key factors in bridging the apparent and intrinsic galaxy size at $z\simeq 6$. From top left to bottom right, we show the size ratios between (1) $R_{\rm eff}$ in UV after and before dust attenuation, (2) $R_{\rm eff}$ in UV and V band, (3) $R_{\rm eff}$ in UV and $R_{\rm eff}$ of the stellar mass, (4) $R_{\rm eff}$ of stellar mass and the intrinsic 3D $r^{\ast}_{1/2}$, respectively. The vertical dashed line indicates the stellar mass limit we choose for \thesanone mock images. Dust attenuation makes galaxies appear larger in UV and this effect is more prominent in more massive galaxies. The $R_{\rm eff}$ in UV is slightly larger than at longer wavelengths mainly due to the difference level of dust attenuation. The $R_{\rm eff}$ in UV is slightly smaller than the $R_{\rm eff}$ of underlying stellar mass distribution due to more young stars forming in the central part of the galaxy but this difference diminishes at the massive end. The 2D sizes of galaxies are consistently smaller than the 3D sizes by about $20$ percent due to projection effects.}
    \label{fig:obs-factors}
\end{figure*}

\begin{figure}
    \centering
    \includegraphics[width=1\linewidth]{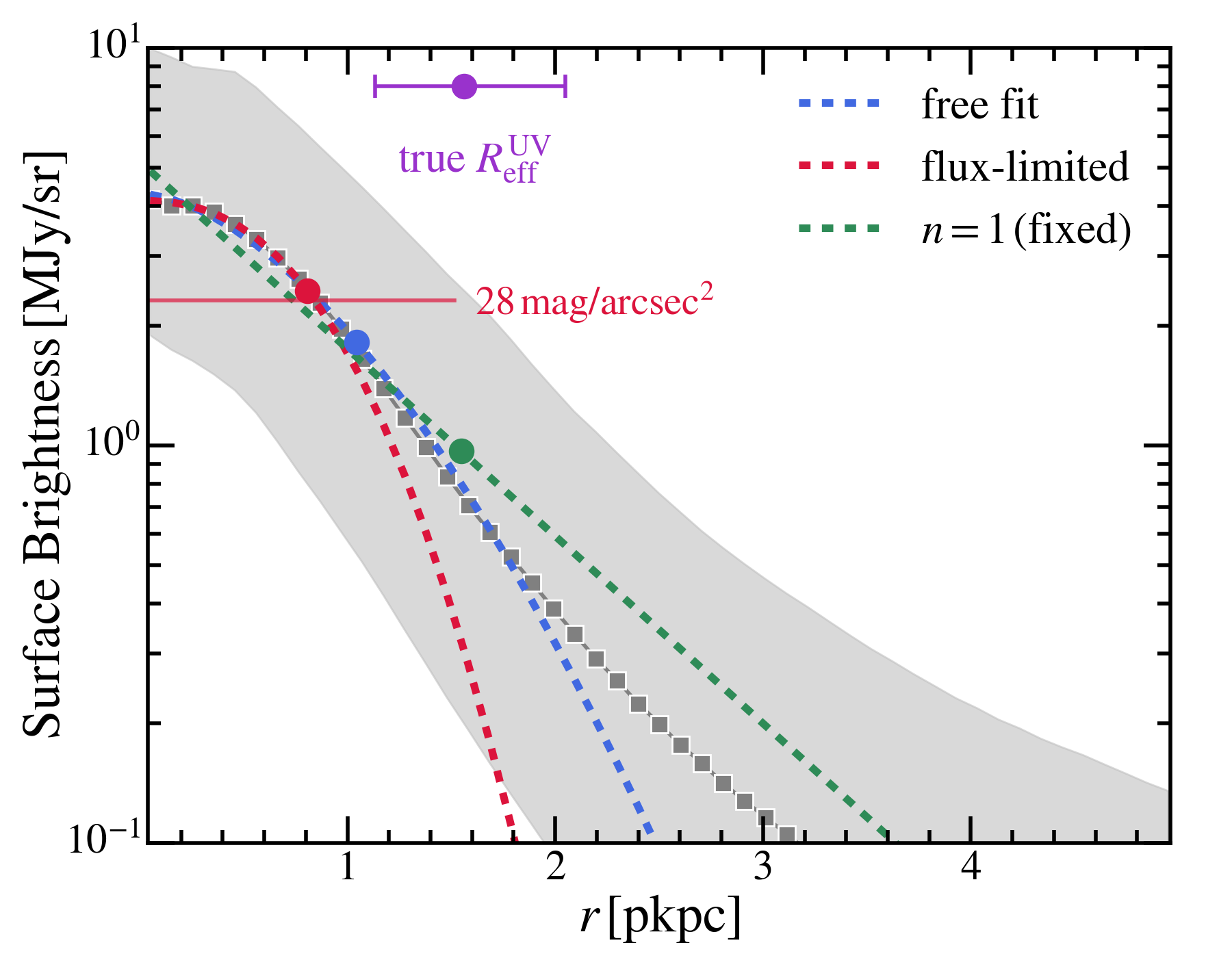}
    \includegraphics[width=1\linewidth]{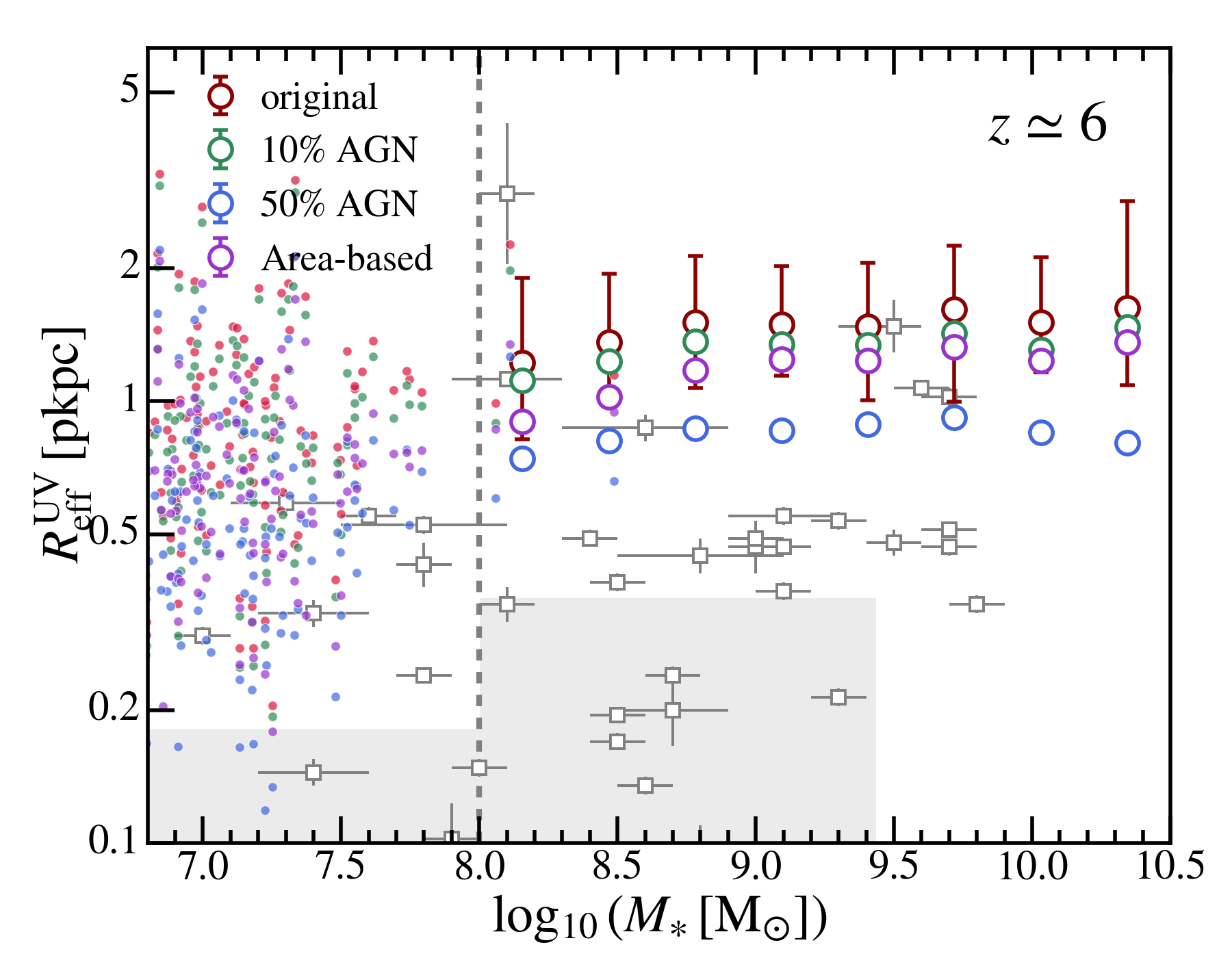}
    \caption{{\it Top}: Stacked UV surface brightness profile of $200$ galaxies with $M_{\ast}\sim 10^{9}\msun$ at $z\simeq 6$ in \thesanone. The gray squares and shaded region show the median surface brightness profile and 1-$\sigma$ scatter. We fit the profile with the S\'{e}rsic profile and test the impact of fixing $n$ to unity (an exponential profile) and applying a surface brightness limit of $28\mmag\,{\rm arcsec}^{-2}$. The dashed lines show the best-fit profiles and the colored circles indicate the corresponding $R^{\rm UV}_{\rm eff}$ derived. {\it Bottom}: A comparison of the fiducial $R^{\rm UV}_{\rm eff}$ with the one measured using a non-parametric area-based approach and the ones with different levels of AGN (point source) contamination. The area-based approach results in slightly smaller galaxy sizes due to the clumping of stellar distributions. An ambitious assumption of $50$ percent contamination from a point source can result in about a factor of two smaller galaxy sizes. But they are still larger than the observational constraints.}
    \label{fig:sb_profile}
\end{figure}

\subsection{Observational factors shaping apparent sizes}
\label{sec:apparent-size-factors}

In addition to the theoretical uncertainties discussed above, several observational effects could be relevant and cause order unity differences between the intrinsic and observed sizes. We will discuss them in this section. In Figure~\ref{fig:obs-factors}, we show the ratios of galaxy sizes with various definitions. The projection of the galaxy is taken to be the positive $z$ direction in the simulation coordinates and effectively represents a random viewing angle when analyzing an ensemble of galaxies. We summarize several key effects that set the difference between the observed galaxy size in certain band (usually rest-frame UV) and the intrinsic half-mass radius.

\vspace{0.1cm}

\noindent \textbf{Projection effects}: Since a spherical aperture always contains less mass or light than a cylindrical aperture, projection tends to make galaxies appear smaller. We find that the ratio between the 2D half-mass radius (with random projection) and the 3D one is insensitive to galaxy stellar mass and takes a median value of $\sim 0.78$. This roughly agrees with the estimates in \citet{Wolf2010} assuming spherical symmetry and a variety of surface brightness profiles. This also motivates the multiplication factor we used in comparing different simulations in Section~\ref{sec:size-mass-theory-compare}.

\vspace{0.1cm}

\noindent \textbf{Tracer effects}: We compare the $R_{\rm eff}$ in UV (before dust attenuation) with the 2D half-mass radius under the same viewing angle. We aim at understanding how well UV light acts as a tracer of the stellar mass in observations. We find that, at $M_{\ast}\lesssim 10^{9}\msun$, the UV effective radius is smaller than the stellar mass effective radius and differs by a constant factor of $\sim 0.8$. This is because recent starbursts take place in the central part of the galaxy while the old and potentially ex-situ stars reside in the outer part. At $M_{\ast}\gtrsim 10^{9}\msun$, the ratio gradually increases to unity due to the compaction of the galaxies and that most of the stellar mass is dominated by the recent central starburst.

\vspace{0.1cm}

\noindent \textbf{Dust attenuation}: We find that the UV effective radius after dust attenuation is larger than that before dust attenuation. The ratio between the two increases monotonically with stellar mass and can reach $\gtrsim 2$ at $M_{\ast}\gtrsim 10^{10}\msun$. This is mainly due to the relatively large abundance of dust in dense ISM and their spatial correlation with the young stellar populations. Therefore, dust attenuation is more concentrated in the central region of the galaxy and smooth out the surface brightness profiles of galaxies. As seen in Figure~\ref{fig:obs-size-mass}, it qualitatively changes the slope of the $R_{\rm eff}-M_{\ast}$ relation from intrinsically negative to positive, which is more consistent with observations. Similar effects have been discussed extensively in many earlier works~\citep[e.g.][]{Wu2020, Roper2022, Marshall2022, Popping2022, Cochrane2023}.

\vspace{0.1cm}

\noindent \textbf{Observing band}: We find a mild ($\sim 10 - 30$ percent) increase in galaxy $R_{\rm eff}$ in UV compared to the optical V band. The tracer effect discussed above could have a minor impact on the sizes at different wavelengths. However, the trend here is primarily driven by stronger dust attenuation in rest-frame UV. At the mass range around $M_{\ast}=10^{8}\msun$, there is almost no difference between the two since the dust attenuation is minimal in low-mass galaxies. This is in good agreement with observational results at a similar mass scale~\citep{Morishita2023}.

\vspace{0.1cm}

\noindent \textbf{Surface brightness bias}: Given the background noise, observations could completely miss sources if their peak surface brightness is below the detection limit. We find some signatures of this in Figure~\ref{fig:trace-surface-density} as the observed sample cuts off at $\Sigma_{\rm SFR}\sim 1 \msun\,{\rm yr}^{-1}\,\pkpc^{-2}$. This can be translated to the UV surface brightness (using the same \citealt{Kennicutt1998} relation adopted in \citealt{Morishita2023}) $\mu_{\rm lim} \sim 28 \mmag\,{\rm arcsec}^{-2}$, which is close to the detection limit of current \textit{JWST} surveys~\footnote{We perform a simple estimate by taking the 5-$\sigma$ point source limiting magnitudes $\sim 29\mmag$ in e.g. \citet{Morishita2023,Ormerod2023}, dividing the corresponding flux limit by the area of the assumed aperture $r_{0}=0.16\,{\rm arcsec}$, and converting it to 1-$\sigma$ limit. We obtain $\mu_{\rm lim}\simeq 29 + 2.5\,\log_{10}{(\pi\,r^{2}_{0})} + 2.5\,\log_{10}{5} \sim 28 \mmag\,{\rm arcsec}^{-2}$.}. The impact of this incompleteness has been discussed in \citet{Kawamata2018} and can lead to a $\sim 0.2-0.3$ dex increase in median galaxy sizes at $z \sim 6-9$ as well as increasing the steepness of the size-luminosity relation. 

Even for identified sources, the surface brightness profile can only be probed out to a certain radius beyond which extrapolations of the S\'{e}rsic profile are required. This can introduce a bias as the fits derived on the central region may fail dramatically at the wings and lead to artificially smaller apparent sizes for intrinsically diffuse galaxies~\citep{Roper2022}. In the top panel of Figure~\ref{fig:sb_profile}, we show the stacked surface brightness profile of 200 galaxies with $M_{\ast}\sim 10^{9}\msun$ at $z\simeq 6$ in \thesanone. After stacking, the surface brightness distribution is roughly axially symmetric and we measure the surface brightness profile in a set of linearly-spacing annuli. We then fit the profile with the S\'{e}rsic model. We find that the free fit is already biased by the central bright component and underpredicts the surface brightness at the wing, yielding a small S\'{e}rsic index $n\sim 0.6$ and $R_{\rm eff}$ smaller than the true value. Fixing $n=1$ results in larger fitted $R_{\rm eff}$ that is surprisingly more consistent with the true value, but the fitting overpredicts the surface brightness at the wing. If we limit the fitting to where the flux is above $28 \mmag\,{\rm arcsec}^{-2}$, the best-fit $R_{\rm eff}$ is further biased to smaller values and is about a factor of two smaller than the true $R_{\rm eff}$. Similar trends are found for galaxies at other stellar mass scales as well. This highlights the potential uncertainties in measuring galaxy sizes by fitting the surface brightness distribution. Optimistically speaking, the simulation predictions and observations could be reconciled by a combination of selection effects and fitting biases associated with the surface brightness limit.

\vspace{0.1cm}

\noindent \textbf{Clumpiness}: In addition, the clumpy nature of high-redshift galaxies can also bias the size measurements to the separation of individual star-forming complexes or young star clusters~\citep[e.g.][]{Vanzella2017,Vanzella2023,Chen2023,Claeyssens2023}. Individual clump sizes will be substantially smaller than the sizes reported by classical dark matter halo finders. As discussed in e.g. \citet{Curtis-Lake2016,Ma2018-size}, an alternative way to measure size is by calculating the effective area $S_{\rm eff}$ (not necessarily continuous) enclosing half of the total surface brightness. The effective radius is then $R_{\rm eff}\equiv \sqrt{S_{\rm eff}/\pi}$. This non-parametric approach will be more sensitive to the sizes of individual stellar clumps and less biased by their separations. We compare this with the fiducial method in the bottom panel of Figure~\ref{fig:sb_profile}. We find that the size measured with this approach can be slightly smaller than our fiducial approach, but not enough to explain the discrepancy between simulations and observations. However, we cannot rule out the possibility that \thesan underpredicts the clumpiness of the stellar distribution. The inability to resolve the multiphase ISM structure of galaxies in the IllustrisTNG model can result in different sub-kpc scale morphology of galaxies. Future studies using zoom-in simulations employing explicit ISM models are required to better understand the impact of clumpiness on galaxy size measurements.

\vspace{0.1cm}

\noindent \textbf{Unresolved sources}: AGN acts as a point source in galaxy photometry. In the observational sample, \citet{Morishita2023} used the [O{\small III}]--H$\beta$ ratios and the mass-extinction diagram to demonstrate that their compact sources are not AGN or have little light contribution from AGN. From simulations, we calculate the bolometric luminosities of the SMBH in simulations as
\begin{equation}
    L_{\rm bol} = (1-\epsilon_{\rm fb}) \, \epsilon_{\rm r} \, \dot{M}_{\rm BH} \, c^{2} \, ,
\end{equation}
where $\epsilon_{\rm fb}=0.1$ is the feedback efficiency in the high-accretion quasar mode~\citep{Weinberger2017} and $\epsilon_{\rm r}$ is the radiative efficiency, assumed to be $0.2$ in \thesan~\citep{Kannan2022}. We then convert this to the rest-frame UV luminosity using the bolometric corrections in \citet{Shen2020_quasar}. We neglect dust extinction here and the AGN luminosity should be interpreted as an upper limit.

As previously shown in section~\ref{sec:correlation}, SMBH accretion rates are found higher in massive compact galaxies. However, even neglecting dust extinction, we find that the UV luminosities of AGN in \thesanone are at least one order of magnitude smaller than the UV luminosities of galaxies at all stellar masses. If the AGN luminosity is subdominant compared to the galaxy luminosity, it should not have any impact on the half-light radius of the galaxy. But this statement strongly depends on the physics model for SMBH growth and feedback employed in \thesan, and is subjected to great uncertainties. The actual impact of unresolved AGN contamination could be significant. For example, in the case study of GN-z11, the two-component (a point source plus an extended component) fitting leads to about three times larger effective radius of the galaxy~\citep{Tacchella2023b}. 

We can estimate the potential impact of unresolved sources by computing the radius containing $45$ ($25$) percent of the total light, which corresponds to the contamination of a point source at the centre of the galaxy with $10$ ($50$) percent of the total galaxy light. We show the comparison in the bottom panel of Figure~\ref{fig:sb_profile}. We find that $50$ percent of contamination can result in about a factor of two smaller galaxy sizes across the whole stellar mass range. However, even with this ambitious assumption of point source contamination, this is not enough to explain the discrepancy with observations.

\section{Conclusions}
\label{sec:conclusion}

In this paper, we provide a thorough analysis of galaxy intrinsic and apparent sizes at $z\gtrsim 6$ based on the \thesan simulations. We study the correlation of galaxy size with various galaxy properties and the large-scale environments they reside in. We investigate the physical mechanisms that determine the angular momentum transport in galaxies and drive the compaction in massive galaxies. Through comparison with observations, we find a discrepancy between the galaxy size predictions of \thesan and observations at the low-mass/faint end. Our key findings can be summarized below. 
\begin{itemize}[leftmargin=0.05\linewidth]
    
    \item \textbf{Intrinsic size--mass relation:} The intrinsic half-mass radius increases (decreases) with galaxy stellar mass in low-mass (massive) galaxies. The breaking point of the two trends is around $10^{8}\msun$ and is insensitive to redshift. The redshift evolution at the low-mass end is mild. The most massive galaxies in the simulation volume continuously become more compact from $z\simeq 12$ to $z\simeq 6$ and eventually reach $r^{\ast}_{1/2}\lesssim 200-300\,{\rm pc}$. In the $(95.5\,{\rm cMpc})^{3}$ volume of \thesanone, four of these compact massive galaxies at $z\simeq 6$ are found to be quenched. Although the limitation of numerical resolution of the \thesanone simulation could make the simulation miss a subpopulation of extremely compact galaxies, the median size mass relations derived from our simulations show good agreement with the high-resolution \thesanhr simulation using identical physics and other cosmological simulations with different physics input and resolution. 

    \vspace{0.2cm}

    \item \textbf{Correlation with galaxy properties:} More compact galaxies tend to have lower halo spin at fixed stellar mass, which qualitatively agrees with the classical disk formation theory. In addition, more compact galaxies have lower sSFR, stronger AGN activity, lower gas abundance, and larger gas disk fraction. This suggests stronger feedback regulation after a period of centralized star-formation and rapid gas depletion, likely driven by the gravitational instability in gas disks. By tracking the evolution of the main progenitors of galaxies selected at $z\simeq 6$, we find that compact (quenched) galaxies typically undergo three phases of size evolution, the initial growth of mass and size driven by mergers and accretion, the rapid gas depletion and compaction phase, and quenching after reaching a characteristic stellar surface density. 

    \vspace{0.2cm}

    \item \textbf{Correlation with environmental properties:} Galaxies in denser environments tend to have more extended stellar distributions. The evolution track of massive galaxies all show an early phase of size growth with frequent minor or major mergers, and later more compact galaxies have quieter merger histories.

    \vspace{0.2cm}

    \item \textbf{Comparison to analytical models:} A deeper analysis of the size-spin correlation shows that the sizes of massive galaxies agree roughly with the prediction of the analytical disk formation theory. Though the most massive compact galaxy population experiences substantial loss in specific angular momentum. However, the sizes of low-mass galaxies barely correlate with the halo spin and are more consistent with a spherical shell model with galaxy sizes set by the shock radius of cold gas inflow and feedback-driven outflow. Therefore, the hump-like feature in the intrinsic size--mass relation can be explained by a merger-driven, spherical shock model for size growth of low-mass galaxies and a disk instability-driven compaction track of massive galaxies.

    \vspace{0.2cm}

    \item \textbf{Gas angular momentum profile:} We investigate the specific angular momentum profiles of dark matter, gas, and stars of simulated galaxies, We find a clear decoupling of gas and dark matter at $0.1\,R_{\rm vir}\lesssim r \lesssim 0.3\,R_{\rm vir}$ and the slow alignment of gas angular momentum with the central stellar disk. This phenomenon likely corresponds to the inflowing cold gas streams that roughly conserve angular momentum before mixing with the lower-angular momentum gas accreted earlier at radii below the impact parameter. The stellar specific angular momentum is about $3$ times smaller than that of gas, which is a consequence of both feedback-driven outflow and disk instability. Low-mass galaxies have poorer spin alignment with the overall spin of the galaxy and this is consistent with their sizes better determined by the spherical shell model.

    \vspace{0.2cm}

    \item \textbf{Discrepancy with observations:} We compute the apparent sizes of galaxies based on mock images generated using the Monte-Carlo radiative transfer method. The negative slope of the intrinsic size--mass relation is overturned by dust attenuation and we find a mild positive correlation between the UV effective radius and galaxy stellar mass as found in observations. A combination of projection effects, tracer effects, dust attenuation, and band selection can lead to order unity differences between the measured galaxy size and the intrinsic one, with non-trivial stellar mass dependence. The median apparent sizes of galaxies in rest-frame UV are at least three times larger than the observed ones at $10^{7}\msun \lesssim  M_{\ast}\lesssim 10^{9}\msun$ or $-20 \lesssim M_{\rm UV}\lesssim -15$ at $6 \lesssim z \lesssim 10$. Such a discrepancy is robust against numerical uncertainties and exists in many other cosmological simulations with various physics inputs, simulation techniques, and numeric resolutions. Other than physical reasons, the selection effects and fitting biases associated with the surface brightness limit of existing galaxy surveys could be the major observational factors driving this discrepancy.
    
\end{itemize}

The size--mass and size--luminosity relations during the epoch of reionization in the \thesan simulations show distinct features compared to those observed at lower redshifts. These relations reveal a two-phase behavior where the sizes of low-mass galaxies are significantly influenced by the strength of feedback-driven outflows, whereas the sizes of more massive galaxies are affected by internal disk instability. These findings are especially relevant in the era of JWST, providing an opportunity to enhance our understanding of galaxy formation through observations of galaxy sizes or morphology in general at high redshifts.

\section*{Acknowledgements}
The computations in this paper were run on the Engaging cluster at Massachusetts Institute of Technology (MIT). MV acknowledges support through the National Aeronautics and Space Administration (NASA) Astrophysics Theory Program (ATP) 19-ATP19-0019, 19-ATP19-0020, 19-ATP19-0167, and the National Science Foundation (NSF) grants AST-1814053, AST-1814259, AST-1909831, AST-2007355, and AST-2107724. EG acknowledges support from the CANON Foundation Europe through the Canon Fellowship program during part of the work presented in this paper.\\

\noindent Software citations:
\vspace{-0.2cm}
\begin{itemize}
    \item {\sc Numpy}:    \citet{Harris2020}
    \item {\sc Scipy}:    \citet{Virtanen2020}
    \item {\sc Astropy}:  \citet{Astropy2013,Astropy2018,Astropy2022}
    \item {\sc Matplotlib}: \citet{Hunter2007}
    \item {\sc Swiftsimio}: \citet{Borrow2020, Borrow2021}
    \item {\sc Arepo-rt}: \citet{Springel2010, Kannan2019, Weinberger2020}
\end{itemize}

%%%%%%%%%%%%%%%%%%%%%%%%%%%%%%%%%%%%%%%%%%%%%%%%%%
\section*{Data Availability}
The data underlying this paper can be shared upon request to the corresponding author of this paper. The \thesan simulation data is publicly available at \href{https://www.thesan-project.com/}{https://www.thesan-project.com/}. The post-processed data and analysis scripts are stored on the Engaging cluster at MIT.



%\bibliographystyle{mnras}
%\bibliography{ref} % if your bibtex file is called example.bib

%\begin{thebibliography}{99}
%\end{thebibliography}

%%%%%%%%%%%%%%%%%%%%%%%%%%%%%%%%%%%%%%%%%%%%%%%%%%

%%%%%%%%%%%%%%%%% APPENDICES %%%%%%%%%%%%%%%%%%%%%

\appendix

\section{Numerical convergence of galaxy size}
\label{sec:app-numeric}

\subsection{Gravitational softening}

The choice of gravitational softening imposes an effective limit on the accelerations that may be adequately reproduced in the system. The density and circular velocity profiles can deviate from the true solution when the specific radial acceleration of particles, $V^{2}_{\rm c}(r)/r$, drops below a characteristic value of $a_{\epsilon}\sim 0.5\, G\, V^{2}_{\rm vir}/\epsilon$~\citep{Power2003}. Therefore, an arbitrarily large $\epsilon$ can break the convergence of mass density distribution at larger radii. This typically happens at a radius of an order unity factor times $\epsilon$.

However, for typical total mass density profiles of the halo (e.g. NFW, Hernquist), a well-defined maximum value exists for radial acceleration. If $\epsilon$ is chosen such that $a_{\epsilon}$ is always larger than this maximum value, the softening will not further affect the convergence of the mass density profile. For example, for the NFW profile, this condition is satisfied when
\begin{equation}
    \epsilon \lesssim \dfrac{\ln{(1+c)-c/(1+c)}}{c^2}\, R_{\rm vir} \simeq 0.05\, R_{\rm vir} \, ,
\end{equation}
where we assume $c=4$ for typical high-redshift haloes. We obtain the halo mass limit where this effect becomes important as
\begin{equation}
    M^{\epsilon}_{\rm vir} = 2.4 \times 10^{7} \msun \times \left(\dfrac{\epsilon}{0.3 \pkpc}\right)^{3} \, .
\end{equation}
For the $\epsilon$ of \thesanone and \thesanhr, all the haloes relevant to our analysis in the main text are safely above this threshold.

The softening itself acts as a way of ``smoothing'' mass distribution and ``puffing up'' structures. A useful reference scale is $\epsilon_{\rm sp} \simeq 2.8 \epsilon$, which is the radius of the compact support given the cubic spline kernel chosen by the simulations. This is where the gravitational acceleration becomes purely Newtonian. However, decreasing softening length does not lead to decreasing (and converging) galaxy sizes given the same particle mass resolution. First, an arbitrarily small softening length will inflate the integration error given the same time-stepping choice~\citep{Power2003}. Secondly, small softening lengths exacerbate the segregation of stars and dark matter particles in halo centers and can counterintuitively increase galaxy sizes as softening length reduced~\citep{Ludlow2019-size,Ludlow2020,Ludlow2023}, which will be discussed in sections below. In simulations with sufficient mass resolution~\citep{Ludlow2020,Ludlow2023}, galaxy sizes can become smaller than the canonical $\epsilon_{\rm sp}$ limit and converge in relatively massive galaxies ($M_{\ast}\gtrsim 100\,m_{\rm b}$, where $m_{\rm b}$ is the baryonic mass resolution) when $\epsilon$ is chosen to be slightly smaller than $r_{\rm conv}$ (introduced below).

\subsection{Two-body relaxation}

When a finite number of particles is used to represent a system, individual particle accelerations will inevitably deviate from the mean-field value when particles pass close to each other. Even when orbits are integrated with perfect accuracy, these ``collisions'' lead to changes of the order of unity in energy on the relaxation time scale~\citep{Binney1987}
\begin{equation}
    t_{\rm relax}(r) \simeq t_{\rm circ}(r)\, \dfrac{N(<r)}{\ln(r/\epsilon)} \, ,
\end{equation}
where $t_{\rm circ}\equiv 2\pi r / V_{\rm c}(r)$ is the circular orbital time scale. This is more sensitive to the particle number than the gravitational softening length. \citet{Power2003} suggests the following convergence criterion for halo density profile against two-body relaxation
\begin{equation}
    \dfrac{t_{\rm relax}(r)}{t_{\rm circ}(R_{\rm vir})} \equiv \dfrac{\sqrt{200}}{8}\,\dfrac{N(<r)}{\ln{N(<r)}}\,\left(\dfrac{\bar{\rho}}{\rho_{\rm crit}}\right)^{-1/2} \gtrsim \kappa \, ,
\end{equation}
given optimized choices of softening, force accuracy, and time stepping. %Assuming an NFW profile with $c=4$, for $r = 0.01, 0.03, 0.05\, R_{\rm vir}$, this translates to a minimum number of particle of $N_{\rm min} \simeq 1000, 500, 350$.
$\kappa$ is an order unity constant that has been studied in detail in \citet{Ludlow2019-convergence}, and a simpler expression of $r_{\rm conv}$ was found as $r_{\rm conv}\simeq 0.055\,l/(1+z)$ with little halo mass dependence, where $l$ is the comoving mean interparticle separation in cosmological simulations. Taking the \thesanone simulation as an example, $r_{\rm conv}$ takes the value of about $0.36\pkpc$ at $z\simeq 6$. 

\subsection{Energy transfer between particle species}

In addition to two-body relaxation, close encounters between collisionless particles in simulations can also transfer energy between different species with different particle mass~\citep{Ludlow2019-size,Ludlow2020,Ludlow2023}. In the \thesan simulations, the dark matter particle mass is set five times larger than the initial mass of gas cells (and roughly speaking the typical mass of stellar particles). Energy equipartition of the system will tend to transfer energy from dark matter to baryonic particles and spuriously increase the size of galaxies. This can lead to a flattening in galaxy size mass relation at a scale similar to $r_{\rm conv}$~\citep[see Figure 7 in][]{Ludlow2020} independent of $\epsilon$.

\citet{Ludlow2023} provided a thorough study of the typical time scale for spurious heating. By comparing this time scale with the age of the Universe, they derived convergence criteria for a range of galaxy properties including $r^{\ast}_{1/2}$. We choose $N^{\rm lim}_{\rm 200}=30$ as defined in Table 2 in \citet{Ludlow2023} as a conservative estimate. We also scale it according to the change in Hubble time $1/H(z)$, which is a proxy for the age of the Universe at $z$. We obtain the minimum halo mass scale that simulated galaxies are safe against spurious heating and convert it to stellar mass with a $M_{\ast}/M_{\rm vir}$ ratio of $0.03$ (conservatively large), which will be referred to as $M^{\rm lim}_{\rm spur}$. This mass scale is marked with arrows in Figure~\ref{appfig:res-rhalf} below and Figure~\ref{fig:size-mass} in the main text.

\begin{figure}
    \centering
    \includegraphics[width=1\linewidth]{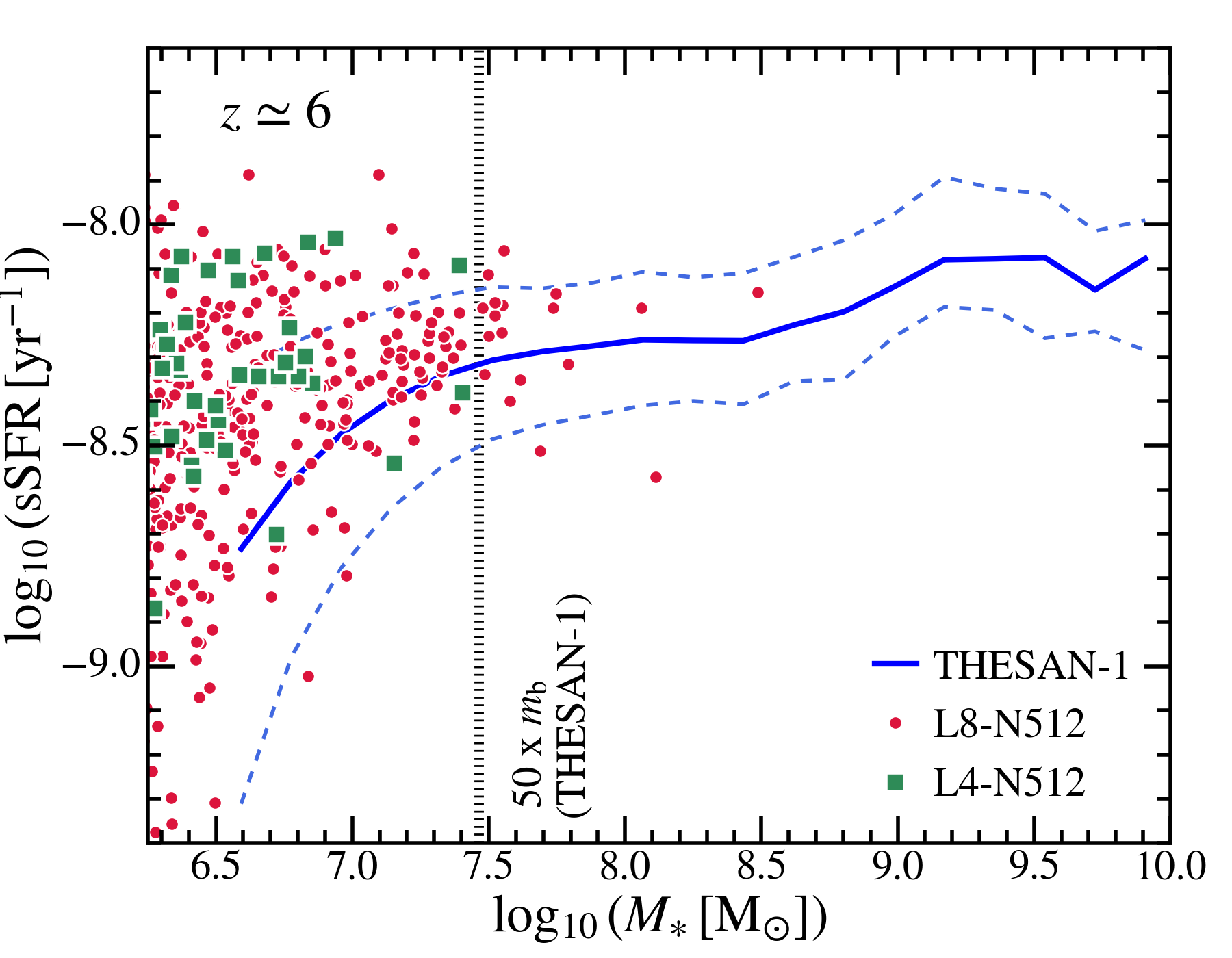}
    \caption{The relation between galaxy sSFR and stellar mass in simulations with different resolutions. The solid line shows the median relation while the dashed lines show 1-$\sigma$ scatter. The vertical lines show the mass limits corresponding to $50$ times the baryonic mass resolution of \thesanone. The sSFR converges above the mass limit we choose. At lower masses, sSFR can be artificially and systematically suppressed in low-resolution simulations.}
    \label{appfig:res-sfr}
\end{figure}

\begin{figure}
    \centering
    \includegraphics[width=1\linewidth]{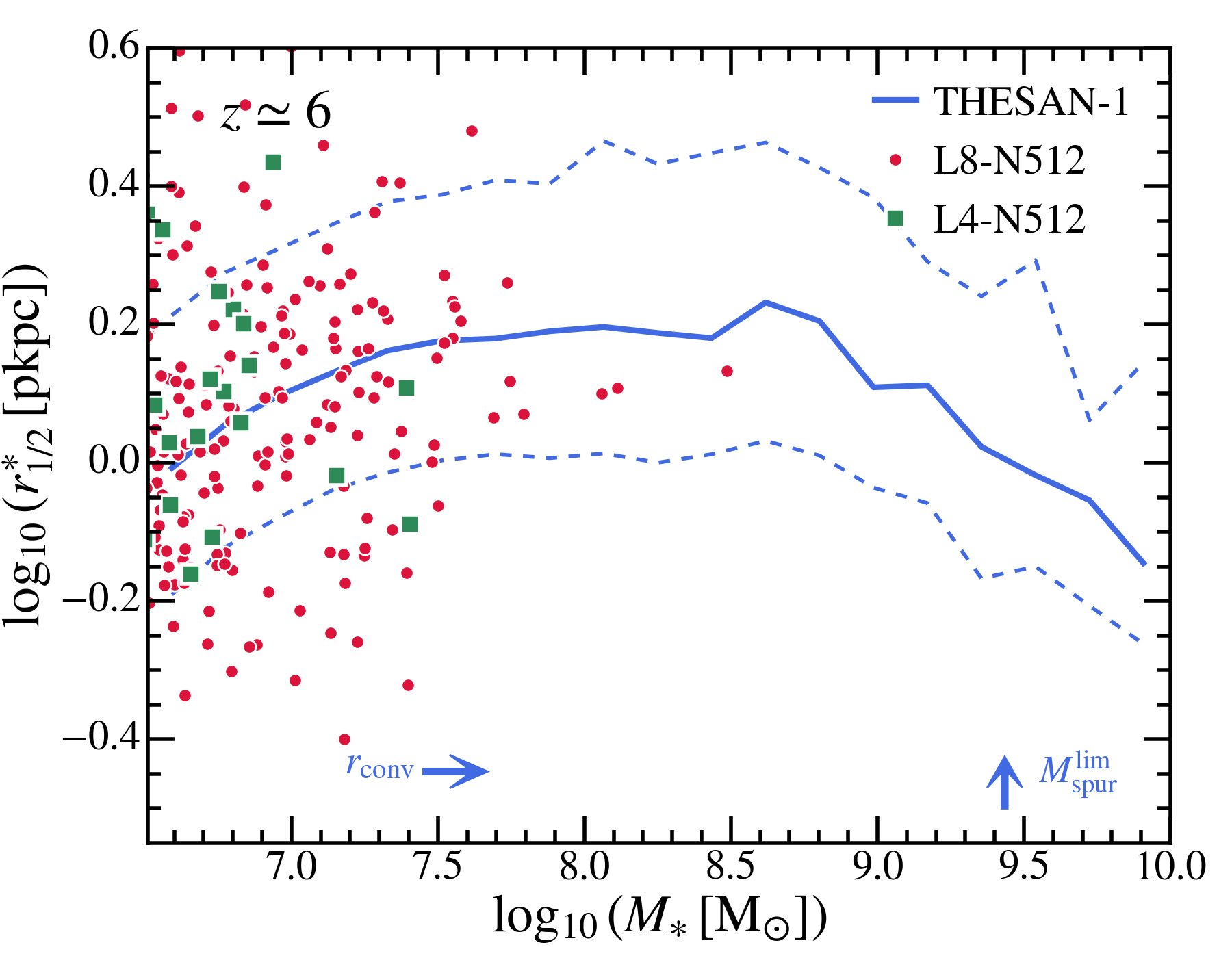}
    \caption{Size--mass relation in simulations with different resolutions at $z\simeq 6$. The plotting style is the same as Figure~\ref{appfig:res-sfr}. The two blue arrows indicate the convergence radius and the spurious heating mass limit of \thesanone, respectively. We find good convergence of the size--mass relation at the resolution level of \thesanone.}
    \label{appfig:res-rhalf}
\end{figure}

\subsection{Convergence of \thesanone and \thesanhr simulations}

In Figure~\ref{appfig:res-sfr}, we show the galaxy sSFR versus stellar mass relation at $z\simeq 6$ in \thesanone and two \thesanhr simulations (L8-N512 and L4-N512). The numerical details of these simulations are listed in Table~\ref{tab:sims}. We find that the sSFR converges at around $50$ times the baryonic mass resolution of the simulation. Below this mass, the sSFRs of galaxies are systematically suppressed in low-resolution simulations and a population of artificially quenched galaxies can show up. Therefore, in the galaxy classification step described in Section~\ref{sec:method:def-ms}, we have specifically removed this population from our analysis. In Figure~\ref{appfig:res-rhalf}, we show the galaxy size--mass relation at $z\simeq 6$ in \thesanone and \thesanhr simulations. We find that both the median and 1-$\sigma$ scatter of the relation converge already at the resolution level of \thesanone. The sizes and stellar masses of simulated galaxies are safely above the spurious heating regime defined by $r_{\rm conv}$ and $M^{\rm lim}_{\rm spur}$ calculated above. 

\section{Central versus satellite galaxies}
\label{sec:app-satellite}

\begin{figure}
    \centering
    \includegraphics[width=1\linewidth]{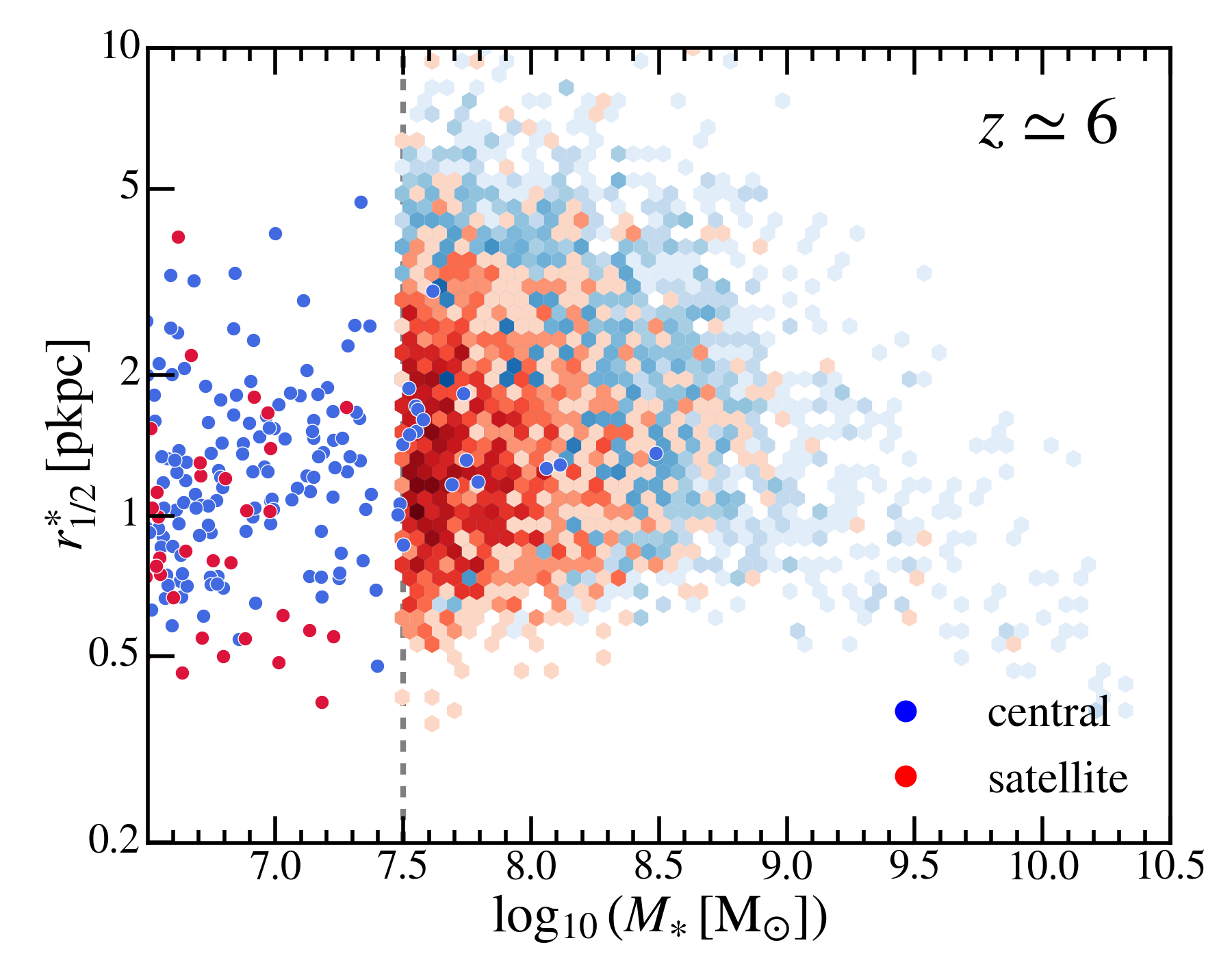}
    \caption{Size--mass relation of central versus satellite galaxies at $z\simeq 6$ in \thesanone (colored distribution) and \thesanhr (solid points). Although the satellites are slightly smaller than the centrals, the difference is too small to cause any notable shift in the median size of central galaxies versus all galaxies.}
    \label{appfig:central}
\end{figure}

In Figure~\ref{appfig:central}, we show the size--mass relation of central versus satellite galaxies at $z\simeq 6$ in \thesan. At fixed stellar mass, the satellites are slightly smaller than the centrals. However, this difference is small enough and it results in almost no differences in the median size of central galaxies and that of all galaxies. We also note that the small region around $M_{\ast}\sim 10^{7.5}-10^{8.5}\msun$, $r^{\ast}_{1/2}\sim 0.5\pkpc$ is purely dominated by satellite galaxies. This explains the low-mass compact galaxy population at the same position in Figure~\ref{fig:size-mass-correlation-env}, which resides in dense environments and is likely affected by the tidal fields and ram pressure of their hosts.

%\section{Galaxy $M_{\ast}$-$M_{\rm UV}$ relation}
%\begin{figure*}
%\centering
%\includegraphics[width=\linewidth]{figures/apparent_size/mass_lum_2d.png}
%\caption{.}
%\label{appfig:mass-lum}
%\end{figure*}

\section{Size-mass relation in different environments}
\label{sec:app-env}

\begin{figure}
    \centering
    \includegraphics[width=1\linewidth]{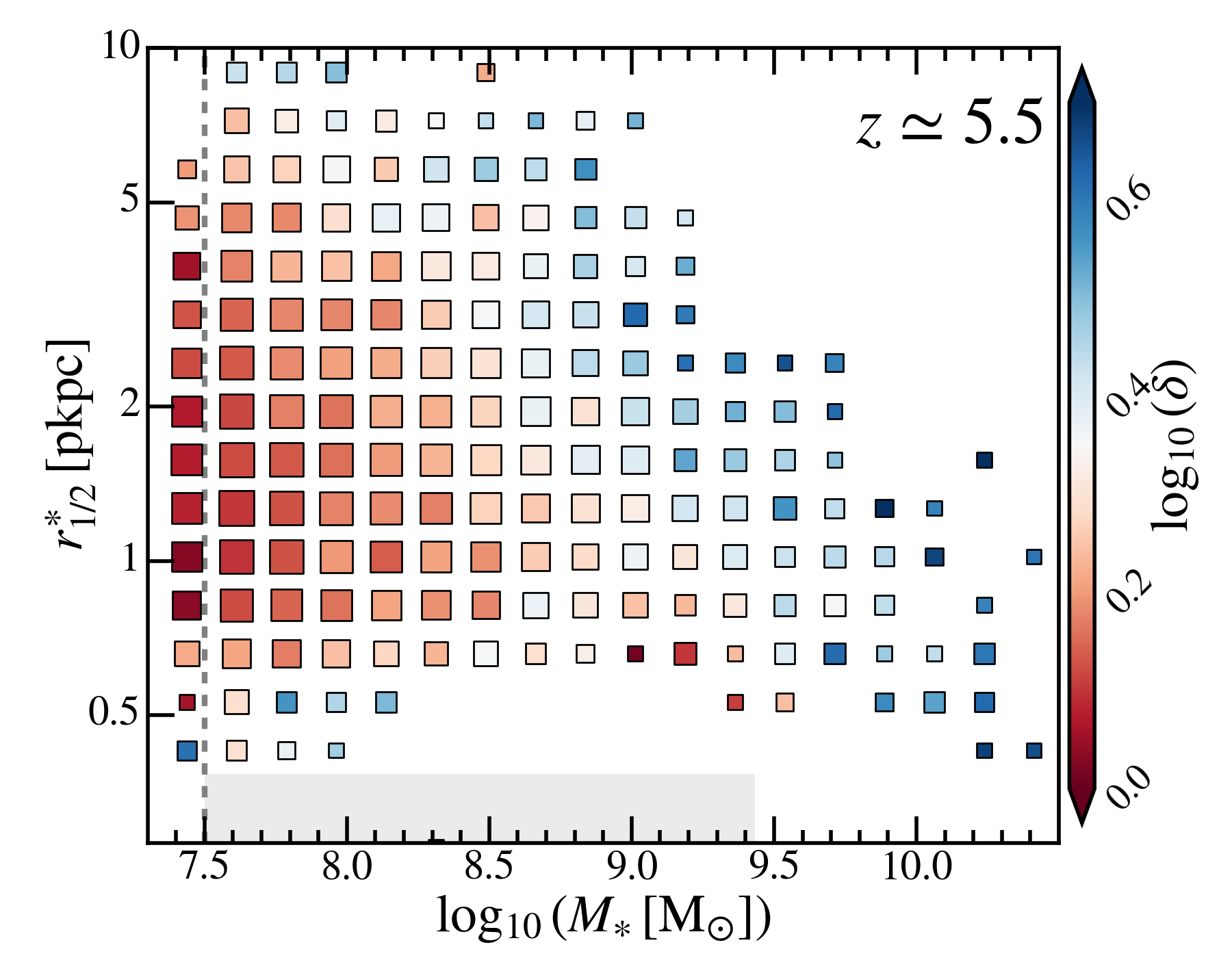}
    \includegraphics[width=1\linewidth]{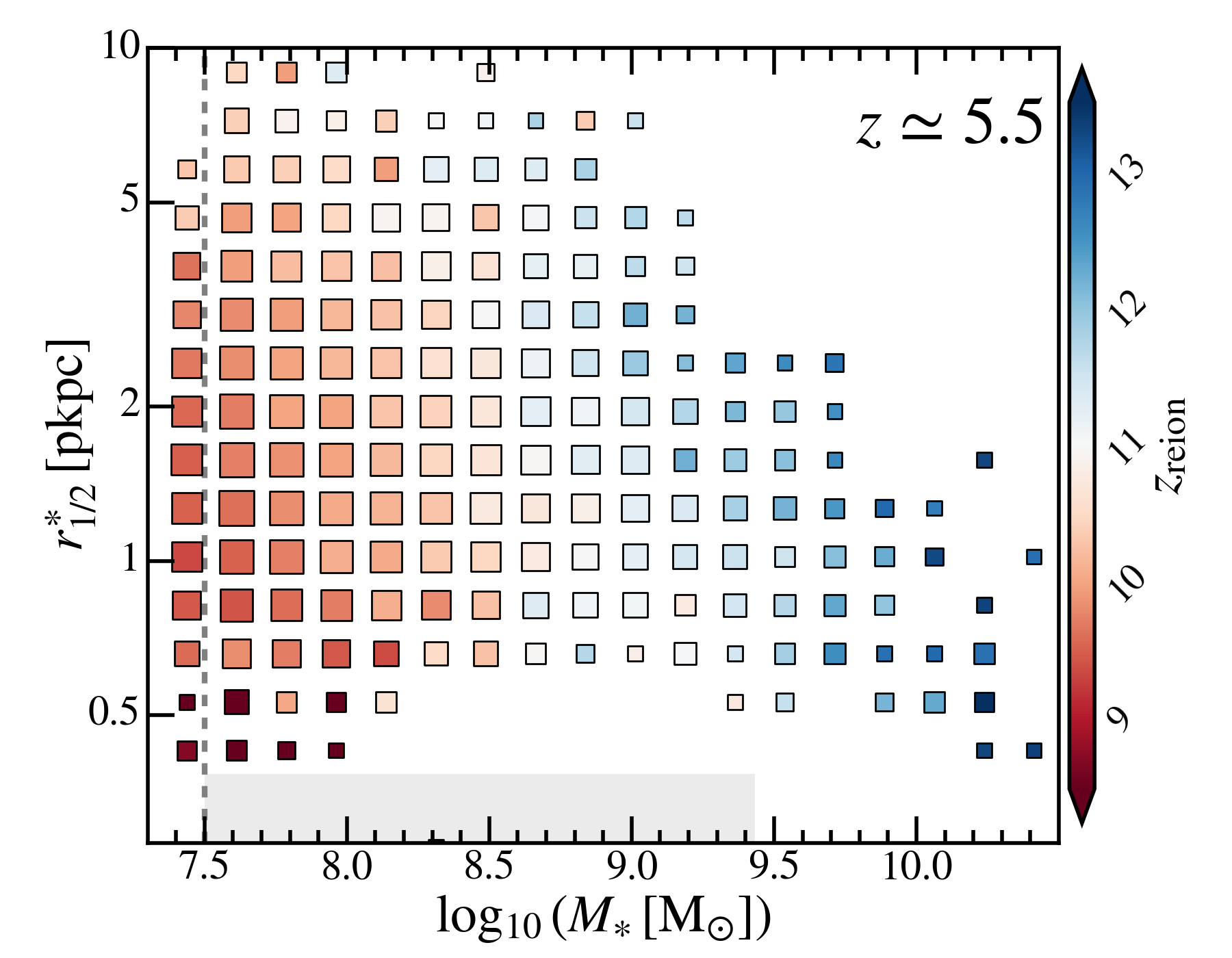}
    \caption{Galaxy intrinsic size--mass relation at $z\simeq 5.5$ color-coded by the matter overdensity (top), and redshift of reionization (bottom) in the $1\,{\rm cMpc}$ environment of galaxies. This is the same as Figure~\ref{fig:size-mass-correlation-env} except for larger smoothing length when computing overdensity of $z_{\rm reion}$.}
    \label{appfig:env}
\end{figure}

In Section~\ref{sec:correlation-env} in the main text, we present the galaxy size-mass relation color-coded by the total matter overdensity and redshift of reionization, smoothed over $250$ ckpc scale. Here in Figure~\ref{appfig:env}, we show the results when the smoothing length is increased to the protocluster scale (1 cMpc). Regarding overdensity, we find the same trend as in Figure~\ref{fig:size-mass-correlation-env}. Galaxies in underdense regions tend to be more compact at fixed stellar mass. However, the dependence on $z_{\rm reion}$ is reversed and is completely controlled by the overdensity. The redshift of reionization on cMpc scales is strongly correlated with the amount of radiation sources in the environment and thus mass overdensity. Denser regions are reionized earlier. As summarized in the main text, the difference in the trends identified at $250$ ckpc versus 1 cMpc scales reflects the transition of large-scale reionization (which depends more on overdensity and the ionizing photon budget in the environment) to small-scale reionization (where local ionizing sources and gas abundance start to be more relevant).

%%%%%%%%%%%%%%%%%%%%%%%%%%%%%%%%%%%%%%%%%%%%%%%%%%

% Don't change these lines
\bsp	% typesetting comment
\label{lastpage}
\end{document}